# HERB: a unified framework for the evaluation of Hydrogen Embrittlement mechanisms driven by the Rice-Beltz concept

Kai Zhao[1*]

*1. School of Mechanical Engineering, Jiangnan University, Wuxi 214122, China*

## Abstract

The multiscale picture of hydrogen embrittlement (HE) mechanisms has been under controversy for a long time. Here I report a thermomechanically-consistent HERB framework driven by the Rice-Beltz concept meanwhile incorporating the hydrogen transport near the crack-tip and void growth within the plastic zone. Triggered solely by dislocation emission from the crack tip, the HERB theory unifies multiple HE mechanisms, such as HEDE, HELP, NVC and HESIV within a single framework. Specifically, a generalized model for predicting the hydrogen-informed dislocation emission is established by incorporating the Rice-Beltz model with the transition state theory. Accounting for the dynamic variation of the trapping energy of spherical inclusions, hydrogen transport is modeled in the dislocation free zone in front of the crack tip. Semi-analytical expressions of the density of geometrically necessary dislocations are obtained by incorporating the Hutchinson-Rice-Rosengren solution with the conventional theory of mechanism-based strain gradient plasticity model. By exploring the feasibility of stochastic analysis, the present theory demonstrates that the hydrogen-informed void dynamics is dominated by the dislocation density between the limits of Lifshitz-Allen-Cahn and Lifshitz-Slyozov-Wagner laws, even though individual events remain unpredictable. These insights fundamentally reshape hydrogen/dislocation interactions across multiple scales, including the core width, short-range and long-range levels.

## 1. Introduction

Hydrogen, the lightest and most abundant element in the universe, is capable of permeating the matrix of solid metals, and reducing their fracture toughness, fatigue resistance and ductility, *i.e*. the so-called hydrogen embrittlement (***HE***) [1-6]. As presently the most mature hydrogen storage approach, high-pressure gaseous hydrogen ($HPGH_2$) storage boasts advantages such as rapid charging and discharging speed, simple structure of the hydrogen storage equipment, low energy consumption for preparing compressed hydrogen, and a wide temperature adaptation range [7]. However, during the process of $HPGH_2$ storage and transportation, hydrogen atoms will penetrate the wall of the hydrogen storage container and accumulate at the crack tip, which promotes crack propagation and increases the risk of failure of $HPGH_2$ storage tanks [8, 9]. It is thus of both theoretical and practical significance to reveal the underlying physics of the hydrogen-enhanced failure process.

[*] Corresponding author: kai.zhao@jiangnan.edu.cn (K. Z.)

Since the observation of ***HE*** by Johnson in the 19th century [10], numerous mechanisms, *e.g.*, the hydrogen-enhanced decohesion (***HEDE***) [11-14], hydrogen-enhanced localized plasticity (***HELP***) [15-18], adsorption-induced dislocation emission (***AIDE***) [19], nanovoid coalescence (***NVC***) [20, 21], and hydrogen-enhanced stress-induced vacancy (***HESIV***) [22] have been proposed to illustrate how hydrogen embrittles metals. Among the most frequently called upon mechanisms of ***HE***, ***HEDE*** refers to the decohesion associated with the decrease of the atomic bond strength due to hydrogen segregation at grain boundaries (GBs) or other interfaces [23]. However, it is noted that there has been little direct experimental evidence for the action of ***HEDE***, although cleavage at subgrain boundaries and transgranular fracture were observed in advanced high strength steels recently [24]. Indeed, as argued by Song and Curtin [25], it is physically unrealistic to force the stress intensity at the crack tip must be lower than the value $K_{Ie}$ required for dislocation emission. Rather, the primary role of hydrogen is to suppress dislocation emission from the crack tip, leaving cleavage as the only energy dissipation mechanism only [24].

Since the pioneering studies by Birnbaum [16], Sofronis [26], Robertson [27] *et al*., the dislocation-dominated ***HELP*** has been well known as one of mainstream ***HE*** mechanisms [28]. The essential physics of the ***HELP*** model is that, by making dislocation motion easier, hydrogen promotes slip localization, resulting in the accumulation of local atomic misfit and potentially dangerous stress concentrations [29]. However, the question of whether the ***HELP*** mechanism is responsible for ***HE*** remains open [29], since the evolution of the hydrogen-enhanced stress concentration into a catastrophic failure is governed by weak-link statistics and notoriously difficult to predict and control. In the continuum picture of the ***HELP*** mechanism [16], the localized plasticity (*i.e*., the plastic strain $\varepsilon_p$) is deterministic. However, previous studies [30-32] have revealed that small-scale plasticity is characterized by intermittent strain-rate fluctuations and strain avalanches that follow robust power-law statistics. The question thus comes as whether the continuum concept of the ***HELP*** mechanism is valid for small-scale plasticity, for which I would discuss on three different levels hierarchically.

On the dislocation core width level, the Peierls-Nabarro (***PN***) model [33-39] and its modifications [40-43] have been widely applied to describe singularity fields around dislocation. Although numerous simulations [44-47] have been dedicated to describe dislocation core structures under hydrogen environment, continuum models of ***HELP*** mechanism have not considered the effect of hydrogen on dislocation core energies yet [48]. Grand canonical Monte Carlo (GCMC) simulations conducted by Yu *et al*. [45] reveal that hydrogen increases the core radii and decreases the core energies, leading to the reduction of dislocation line energy. Based on the density functional theory, Li *et al*. [44] revealed a hydrogen-induced transition of screw dislocation core structure in bcc W. Specifically, at low H concentrations, dislocation maintains the intrinsic easy-core structure, and H atoms are attached to the "periphery" of dislocation to enhance dislocation motion; in contrast, at high H concentrations, dislocation transforms into a hard-core, metal hydride-like structure, while H atoms are implanted into the "body" of dislocation to significantly reduce the dislocation mobility. Leyson *et al*. [49] developed a multiscale approach to study the interaction of hydrogen with the core and the strain field of edge dislocations, and revealed a sharp transition between hydride forming and non-hydride forming regimes, with respect to a critical hydrogen chemical potential $\mu_H^c$ related to the nano-hydride nucleus of the system. Recently, Zhang *et al*. [50] proposed a stochastic Peierls-Nabarro (***sPN***) model to account for the standard deviation of the perturbation in the interplanar potential of high-entropy alloys (HEA), leading to stochastic variations in the dislocation width, Peierls stress and Peierls energy. Considering the similarity between the compositional randomness in an HEA and the stochastic distribution of hydrogen atoms within the core width, one might ask whether the ***sPN*** model can be applied for the hydrogen/dislocation interaction, or not?

On the level of short-range interactions, *i.e*., how hydrogen atoms (and derivative hydrogen-vacancy complexes) affect the nucleation [51, 52], and motion of a dislocation, and interactions between two neighboring dislocations, which has been evaluated by Sofronis and Birnbaum [26] actually in the initial version of the ***HELP*** mechanism, *i.e*., the hydrogen-induced "elastic shielding" results in a decrease in the elastic force between two dislocations. As proved by nanoindentation experiments [52, 53], the "defactants" concept [27, 54-57] can be applied to understand the hydrogen effect on dislocation nucleation by reducing the formation energy of dislocations and stacking faults. Leyson *et al*. [58] developed a multiscale model to describe the onset of homogeneous nucleation of a dislocation loop with the presence of hydrogen, with faithfully predicting the experimentally observed drop of the pop-in force. While most studies favor to believe that hydrogen atoms (or hydrogenated vacancies) either enhance [29, 59-61] or impede [62-64] dislocation motion solely, recent *in-situ* scanning electron microscopy experiments [65] reveal that hydrogen can both move or pin dislocation in bcc metals. The kinetic Monte Carlo simulations by Katzarov *et al*. [66] also demonstrate that the steady-state dislocation velocity does not change with the bulk hydrogen concentration monotonically, thus support the dual role effects of hydrogen on the dislocation mobility [67].

On the level of long-range interactions, *i.e.*, how hydrogen atoms affect the collective behavior of dislocation ensembles (as well as other types of defects, e.g., GBs, twin boundaries [68], precipitates [69, 70], vacancy clusters [21]), the underlying mechanisms are still not clear due to the unpredictability of the complex many-body system. While most studies are concerned about the hydrogen trapping capacity of dislocations [71], GBs [72, 73], and precipitates [74-79], few attentions are paid to resolve how these preexisting defects influence the plastic evolution under hydrogen environment [80-83]. Combining the *in-situ* electron channeling contrast imaging experiments, molecular dynamics (MD) and GCMC simulations, Koyama *et al*. [84] revealed that the driving force for dislocation movement is provided by the high stress concentration arising from hydrogen segregation at GBs. Based on the mean-field rate theory [85] and grouping numerical method [86], Li *et al*. [21] developed a cluster dynamics model to study the long-time evolution of hydrogen/vacancy system. However, in their model, the dislocation density was set as constant. Recently, Zhao *et al*. [87] proposed a mean-filed picture to account for the effect of hydrogen on the collective behavior of dislocations under nanoindentation. Although the plastic pop-in can be correctly captured by our model, the failure to reproduce the intermittent fluctuations of the force *v.s.* displacement curve implies that the stochastic feature of dislocation reaction kinetics has to be considered [88].

A major drawback of continuum models of the ***HELP*** mechanism [27] is that they could only describe the effect of hydrogen atmosphere on dislocations, but are not explicitly involved in the fracture process. In order to correlate the ***HELP*** mechanism with the hydrogen-induced fracture, phenomenological models, such as, ***NVC*** [20, 21], and ***HESIV*** [22] mechanisms, have been proposed. These studies thus strongly indicate that we should treat the ***HE*** as a global (*i.e.*, across multiple temporal and spatial scales) ductile-to-brittle transition (***DBT***) modulated by hydrogen, instead of being stuck on a tiny island of localized hydrogen-involved processes.

The ***DBT*** [89, 90] is referred to the change in fracture behavior of materials, where materials exhibit brittle characteristics at low temperatures, while become more ductile at elevated temperatures. Although found more than one century [90], the underlying mechanism of the ***DBT*** is still under controversy, *i.e.*, the transition is dominated by either the dislocation nucleation [91-93] or dislocation mobility [94-96]. By incorporating the ***PN*** concept into the dislocation nucleation description, Rice [92] showed that under the mode-II (in-plane shear) loading the dislocation emission is controlled by an energy criterion involving the unstable stacking fault (***USF***) energy $\gamma_{usf}$. Subsequent studies have advanced the Rice framework and the Rice-Thomson model by accounting for the elastic anisotropy [97], the effect of crack blunting [98, 99], three-dimensional dislocation nuclei [93], successive nucleation events [100], and surface steps [101] formed at the crack-tip [102]. In previous studies [103, 104], we have explored the feasibility to rationalize the dislocation emission within the Peierls-Rice-Beltz (***PRB***) framework with the presence of microstructural defects (*e.g.*, GBs) under hydrogen environment. However, as argued in previous studies [89, 105], a complete description of the hydrogen-informed ***DBT*** process also requires the knowledge of dislocation mobility, which would be considered via the ***HELP*** mechanism in present study.

In the conventional understanding of the ***DBT***, cleavage occurs prior to any dislocation emissions in the intrinsically brittle materials, but is prevented by dislocation emissions in the intrinsically ductile ones [105]. However, previous experiments [106-110] have evidenced that dislocation emissions from sharp cleavage cracks do not necessarily guarantee ductile behavior in a cleavage solid. Thus, a quasi-cleavage process emerges as an intermediate failure mechanism between the two extremes of cleavage and dislocation emission. Martin *et al*. [111, 112] have rationalized the quasi-cleavage in hydrogen embrittled steels by the growth and extension of voids nucleated at and along intersections of intense slip bands. Furthermore, Neeraj *et al*. [20] proposed an alternative nanovoid nucleation, growth and coalescence (*i.e.*, ***NVC***) micromechanism for ***HE***. Thus, the knowledge of nanocrack initiation and void kinetics would be evaluated to bridge the ***HELP*** mechanism with the fracture process in present study.

Starting from the ***PRB*** theory [93], here I attempt to construct a theoretical framework (named as ***HERB***) to describe the dislocation emission from the crack tip, hydrogen transport (diffusion and trapping), vacancy coalescence and void nucleation hierarchically, by involving various ***HE*** mechanisms including ***HEDE***, ***HELP***, ***NVC*** and ***HESIV*** simultaneously. It is demonstrated that the stochasticity across multiple temporal and spatial scales causes the individual event of void growth unpredictable, while the statistical trend of a vast number of events is physically clear.

## 2. Theoretical framework

In this section, I present the main content of the ***HERB*** framework. Initially, the dislocation emission from the crack-tip under mixed mode-I+II loading is derived from the classical ***Rice-Beltz*** theory in **Section 2.1**, by introducing the minimization of the strain energy density to renormalize the localized $K_I^*$ and $K_{II}^*$ fields. Then, the hydrogen transport is analyzed within the elastic core near the crack-tip in **Section 2.2**, by considering the trapping capability of hydrogen traps changes dynamically upon external loading. Finally, the stochastic kinetics of void growth within the plastic zone is established in **Section 2.3**, with the distribution of dislocation density derived by cooperating the Hutchinson-Rice-Rosengren (***HRR***) solution into the mechanism-based strain gradient (***MSG***) plasticity theory.

## 2.1 Phase- Ⅰ: The dislocation emission from the crack-tip

Based on the ***Rice-Thomson*** model [91] considering the dislocation shielding effect, the total force applied on the dislocation near the crack-tip under hydrogen environment is evaluated for the configuration shown in **Fig. 1**,

$$f = f_K + f_{d^2} + f_{dd'} + f_{gbe} + f_H \ , \tag{2.1.1}$$

where, the first term $f_K$ arises from the stress field of the blunt crack-tip. For the linear elastic analysis of plane problems, conventional approaches have been well established by Muskhelishvili [113], and later extended by Irwin [114] to obtain the well-known equations of the stress field near a sharp crack considering the first terms of series expansion. Creager and Paris [115] further derived the solutions for blunt cracks,

$$\begin{aligned}\sigma_{xx} =& \frac{K_I}{\sqrt{2\pi r}}\left(\cos\frac{\theta}{2} - \frac{1}{2}\sin\theta\sin\frac{3\theta}{2} - \frac{\rho}{2r}\cos\frac{3\theta}{2}\right) \\ &+ \frac{K_{II}}{\sqrt{2\pi r}}\left(-2\sin\frac{\theta}{2} - \frac{1}{2}\sin\theta\cos\frac{3\theta}{2} + \frac{\rho}{2r}\sin\frac{3\theta}{2}\right),\end{aligned} \tag{2.1.2}$$

$$\begin{aligned}\sigma_{yy} =& \frac{K_I}{\sqrt{2\pi r}}\left(\cos\frac{\theta}{2} + \frac{1}{2}\sin\theta\sin\frac{3\theta}{2} + \frac{\rho}{2r}\cos\frac{3\theta}{2}\right) \\ &+ \frac{K_{II}}{\sqrt{2\pi r}}\left(\frac{1}{2}\sin\theta\cos\frac{3\theta}{2} - \frac{\rho}{2r}\sin\frac{3\theta}{2}\right),\end{aligned} \tag{2.1.3}$$

$$\begin{aligned}\tau_{xy} =& \frac{K_I}{\sqrt{2\pi r}}\left(\frac{1}{2}\sin\theta\cos\frac{3\theta}{2} - \frac{\rho}{2r}\sin\frac{3\theta}{2}\right) \\ &+ \frac{K_{II}}{\sqrt{2\pi r}}\left(\cos\frac{\theta}{2} - \frac{1}{2}\sin\theta\sin\frac{3\theta}{2} - \frac{\rho}{2r}\cos\frac{3\theta}{2}\right),\end{aligned} \tag{2.1.4}$$

where, $K_I$ and $K_{II}$ are the generalized mode-I and mode-II stress intensity factors (SIFs) assuming that the blunt notch is replaced by a sharp crack with a translation of the crack-tip on the curvature center $C$ [116], $(r, \theta)$ is defined as the polar coordinate $x_1O'y_1$ in **Fig. 1**. $f_K$ is then evaluated as $b\tau_{r\theta}$, where $b$ is the Burgers vector, and the shear component $\tau_{r\theta}$ in the polar coordinate system is calculated by the coordinate transformation as,

$$\begin{aligned}\tau_{r\theta} &= \cos\theta\sin\theta\left(\sigma_{yy} - \sigma_{xx}\right) + (\cos^2\theta - \sin^2\theta)\tau_{xy} \\ &= \frac{K_I}{2\sqrt{2\pi r}}\sin\frac{\theta}{2}\left(1 + \cos\theta + \frac{\rho}{r}\right) + \frac{K_{II}}{\sqrt{2\pi r}}\left(\cos\frac{3\theta}{2} + \frac{1}{2}\sin\theta\sin\frac{\theta}{2} - \frac{\rho}{2r}\cos\frac{\theta}{2}\right). \\ &= \frac{1}{2\sqrt{2\pi r}}\left[K_I\sin\frac{\theta}{2}\left(1 + \cos\theta + \frac{\rho}{r}\right) + K_{II}\cos\frac{\theta}{2}\left(-1 + 3\cos\theta - \frac{\rho}{r}\right)\right]\end{aligned} \tag{2.1.5}$$

The second term $f_{d^2}$ in the **Eq.**(2.1.1) is the self-image force caused by the free surface effect. Initially, to distinguish the stress components $\sigma_{yy}$ and $\tau_{yx}$ along the crack surface, image dislocation distributions are introduced along the crack surface [117]. The image dislocation $(\zeta_i)$ distribution at a distance $\beta$ from the crack-tip, caused by the dislocation $\zeta$ emitted from either the crack-tip or the GB (see **Fig. 1**), is given by,

$$F_x(\beta) = \frac{-b}{\pi\rho_i}\sqrt{\frac{r}{|\beta|}}\left\{\cos\eta\cos\left(\phi - \frac{\theta}{2}\right) + \frac{1}{2}\sin\theta\sin\left(\phi - \eta + \frac{\theta}{2}\right) - \sin\phi\sin\left(2\phi - \eta - \frac{\theta}{2}\right)\right\}, \tag{2.1.6}$$

and

$$F_y(\beta) = \frac{-b}{\pi\rho_i}\sqrt{\frac{r}{|\beta|}}\left\{2\sin\eta\cos\left(\phi - \frac{\theta}{2}\right) + \cos\left(2\phi - \frac{\theta}{2}\right)\sin(\phi - \eta) - \frac{1}{2}\sin\theta\sin\left(\phi - \eta + \frac{\theta}{2}\right)\right\}. \tag{2.1.7}$$

Here $F_x(\beta)d\beta$ represents the sum of the $x$-directional components of the Burgers vector of the image dislocations between $\beta$ and $\beta + d\beta$, $b$ is the magnitude of the Burgers vector, and $\phi$ and $\eta$ are angles defined in **Fig. 1**. Substituting the **Eqs**.(2.1.6) and (2.1.7) into the stress field equations of a discrete dislocation [118] and integrating the functions over $\beta$ along the crack surface, the stress fields imposed by all image dislocations are given by,

$$\sigma_{xx} = \frac{\mu}{2\pi(1-\upsilon)}\left\{-\int_{-\infty}^{0}\frac{y[3(x-\beta)^2+y^2]}{[(x-\beta)^2+y^2]^2}F_x(\beta)\mathrm{d}\beta + \int_{-\infty}^{0}\frac{(x-\beta)[(x-\beta)^2-y^2]}{[(x-\beta)^2+y^2]^2}F_y(\beta)\mathrm{d}\beta\right\}, \tag{2.1.8}$$

$$\sigma_{yy} = \frac{\mu}{2\pi(1-\upsilon)}\left\{-\int_{-\infty}^{0}\frac{y[(x-\beta)^2-y^2]}{[(x-\beta)^2+y^2]^2}F_x(\beta)\mathrm{d}\beta + \int_{-\infty}^{0}\frac{(x-\beta)[(x-\beta)^2+3y^2]}{[(x-\beta)^2+y^2]^2}F_y(\beta)\mathrm{d}\beta\right\}, \tag{2.1.9}$$

and

$$\tau_{xy} = \frac{\mu}{2\pi(1-\upsilon)}\left\{-\int_{-\infty}^{0}\frac{(x-\beta)[(x-\beta)^2-y^2]}{[(x-\beta)^2+y^2]^2}F_x(\beta)\mathrm{d}\beta + \int_{-\infty}^{0}\frac{y[(x-\beta)^2-y^2]}{[(x-\beta)^2+y^2]^2}F_y(\beta)\mathrm{d}\beta\right\}. \tag{2.1.10}$$

Considering the practice of numerical solving, Shimokawa and Tsuboi [119] set the interval of the integration as -100 < $\beta$ < 0 nm, and calculated the $f_{d^2}$ in the same manner as $f_K$. Besides, some geometric constraints should be declared,

$$\eta = \arctan\left(\frac{r\sin\theta}{r\cos\theta-\alpha}\right), \tag{2.1.11}$$

$$\rho_r = \frac{r\sin\theta}{\sin\eta}, \tag{2.1.12}$$

$$\phi = \arctan\left(\frac{r\sin\theta}{r\cos\theta+\beta}\right), \tag{2.1.13}$$

and

$$\rho_i = \frac{r\sin\theta}{\sin\phi}\,, \tag{2.1.14}$$

where, the subscripts '$r$' and '$i$' in the **Eqs**.(2.1.12) and (2.1.14) represent real and imaginary (dislocations), respectively.

The third term $f_{dd'}$ in the **Eq**.(2.1.1) can be decomposed into the direct and indirect interaction force $f_{dd',d}$ and $f_{dd',i}$ caused by the residual dislocation $\zeta'$ after the dislocation $\zeta$ emission from the GB, as shown in **Fig. 1**. In other words, the residual dislocation $\zeta'$ exerts two types of stress fields to the dislocation $\zeta$: the direct and indirect stress fields. The direct force $f_{dd',d}$ can be evaluated as,

$$f_{dd',d} = -\frac{\mu b^2}{2\pi(1-\upsilon)\rho_r}\,. \tag{2.1.15}$$

On the other hand, the indirect force $f_{dd',i}$ can be calculated in the same manner as $f_{d^2}$, where the image dislocation distribution caused by the residual dislocation $\zeta'$ is obtained by setting $b = -b$, $\rho_i = \alpha + \beta$, $r = \alpha$, and $\phi = \eta = \theta = 0$. Then, substituting the distributions into the **Eqs**.(2.1.8)-(2.1.10), the value of $f_{dd',i}$ can be determined.

The fourth term $f_{gbe} = -4\Delta E_{gb}\rho_0/\pi(\rho_0^2 + \rho_r^2)$ (where, $\Delta E_{gb}$ is the energy change due to the dislocation emission, $\rho_0$ is the cutoff distance determined by the Rice-Thomson model [119]) is hypothesized to be caused by the local GB structural transition after dislocation emission from the GB, with details found in our previous study [104].

The fifth term $f_H$ is the shear force exerted on the dislocation induced by the hydrogen atmosphere [103],

$$f_H = -\tau_H b\,, \tag{2.1.16}$$

Revisit **Fig. 1** and consider a Cartesian coordinate system $x_2O''y_2$ centered at the core of the emitted dislocation, and a single hydrogen dilatation line at a point with the polar coordinate ($r'$, $\varphi$). The shear stress exerted at the core of the emitted dislocation along the slip plane by the hydrogen dilatation line is,

$$\frac{\sigma_{r'r'} - \sigma_{\varphi\varphi}}{\pi r'^2}\sin 2\varphi = -\frac{\mu\Delta a}{\pi r'^2}\sin 2\varphi \tag{2.1.17}$$

where, $\sigma_{r'r'}$ and $\sigma_{\varphi\varphi}$ are nonzero stress components of the plane strain axisymmetric field at a distance $r'$ from the dilatation line, while $\Delta a = \Delta a'/2(1-\upsilon)$ with the "unconstrained area of expansion" $\Delta a' = V_H/N_A h$, where $V_H$ ($= N_A\Omega$) is the partial molar volume of hydrogen in solution and $h$ is the distance between two successive hydrogen atoms along the dilatation line [26]. By applying the principle of linear superposition, the shear stress $d\tau_H$ due to dilatation lines in an infinitesimal area d$S$ at the coordinate ($r'$, $\varphi$) is written as,

$$d\tau_H = n dS \left(-\frac{\mu \Delta a}{\pi r'^2}\right) \sin 2\varphi \tag{2.1.18}$$

where, $n$ ($= c_H h$) is the in-plane concentration of dilatation lines (also denoting the number of hydrogen atoms per unit area in the plane normal to the dilatation line), with $c_H$ is the hydrogen concentration near the dislocation core [16],

$$c_H(r', \varphi) = \frac{c_0}{1-c_0} \exp\left(-\frac{W_{int}}{k_B T}\right) \Big/ \left(1 + \frac{c_0}{1-c_0} \exp\left(-\frac{W_{int}}{k_B T}\right)\right), \tag{2.1.19}$$

where, $W_{int}$ is the interaction energy between the hydrogen cluster and the dislocation in the semi-infinite plane,

$$W_{int} = -\frac{\mu b (1+\upsilon) \sin\varphi}{3\pi (1-\upsilon) r'} \Omega , \tag{2.1.20}$$

Integrating the above **Eq**.(2.1.18) over the entire area $S$ occupied by the hydrogen atmosphere, the hydrogen-induced net shear stress $\tau_H$ is,

$$\tau_H = \frac{\mu \Omega}{2\pi(1-\upsilon)} \int_{r_c}^{r_H} \int_0^{2\pi} c_H \sin \frac{2\varphi}{r'} d\varphi dr' \tag{2.1.21}$$

where, $r_c$ is the core radius of the dislocation, $r_H$ is the radius of the hydrogen cluster around the dislocation.

While the stress field near the crack-tip under hydrogen environment has been configured above, the 2D Peierls model developed by Rice [92] and Rice and Beltz [93] (*i.e.*, the ***PRB*** framework) is introduced to describe the temperature-dependent activation energy barriers for dislocation nucleation,

$$U[\vec{\delta}(r)] = U_0 + \int_0^\infty \Phi[\vec{\delta}(r)] dr + \frac{1}{2} \int_0^\infty s[\delta(r)] \cdot \delta(r) dr - \int_0^\infty \frac{K_{II}^{eff}}{\sqrt{2\pi r}} \delta(r) dr , \tag{2.1.22}$$

with

$$s[\delta(r)] = \frac{\mu}{2\pi(1-\upsilon)} \int_0^\infty \sqrt{\frac{\xi}{r}} \frac{d\delta(\xi)/d\xi}{r-\xi} d\xi , \tag{2.1.23}$$

where, the first term $U_0$ is the elastic strain energy of the loaded cracked solid without any slip; the potential $\Phi$ is the change in atomic stacking energy due to a slip discontinuity $\vec{\delta}$; the third term accounts for the elastic interaction energy between the infinitesimal increments of slip; the fourth term represents the elastic interaction energy between the slip and the crack surface.

The activation energy per unit length under applied loading is the difference between the energies for the stable equilibrium slip distribution $\vec{\delta}_{stable}(r)$ and the saddle-point slip distribution $\vec{\delta}_{saddle}(r)$,

$$Q_{2d} = U\left[\vec{\delta}_{saddle}(r)\right] - U\left[\vec{\delta}_{stable}(r)\right] . \tag{2.1.24}$$

Proceeding via a perturbation analysis of the shear distribution, Rice and Beltz [93] obtained a closed-form approximation of the activation energy as,

$$\Theta_{2d} = \frac{(1-\upsilon)Q_{2d}}{\mu b^2} = m\left(1 - \sqrt{\frac{G}{\gamma_{usf}}}\right)^{3/2} , \tag{2.1.25}$$

where, the dimensionless factor $m$ is approximated as 0.287 due to the extremely weak dependence on $\gamma_{usf}/\mu b$ [120, 121], the applied energy release rate $G$ can be correlated with the effective mode-II SIF $K_{II}^{eff}$ via $G = (1-\upsilon){K_{II}^{eff}}^2/2\mu$ [121].

By defining the localized mode-I SIF $K_I^*$ and mode-II SIF $K_{II}^*$ nominally (this is the key step from the continuum to atomistic mechanics, since atomistic details are considered), we can rewrite the total force in the hydrogen-informed Rice-Thomson-Shimokawa-Tsuboi (***RTST***) model as,

$$f = b\tau_{r\theta}^* , \tag{2.1.26}$$

where, $\tau_{r\theta}^*$ is the nominal shear component within the polar coordinate system, and can be related to the nominal SIFs similar as **Eq.**(2.1.5),

$$\tau_{r\theta}^* = f/b = \frac{1}{2\sqrt{2\pi r}}\left[K_I^* \sin\frac{\theta}{2}\left(1 + \cos\theta + \frac{\rho}{r}\right) + K_{II}^* \cos\frac{\theta}{2}\left(-1 + 3\cos\theta - \frac{\rho}{r}\right)\right] , \tag{2.1.27}$$

Meanwhile, the local strain energy density (*i.e.*, the first-order derivative of $U_0$ in **Eq.**(2.1.22) with respect to the volume) can be written as,

$$V_\varepsilon = \frac{1}{2}[\tilde{\sigma}]^T[\tilde{\varepsilon}] \tag{2.1.28}$$

with the stress and strain matrix are,

$$[\tilde{\sigma}] = \begin{matrix} \sigma_{xx}^* \\ \sigma_{yy}^* \\ \tau_{xy}^* \end{matrix}, [\tilde{\varepsilon}] = \begin{matrix} \varepsilon_{xx}^* \\ \varepsilon_{yy}^* \\ \gamma_{xy}^* \end{matrix} \tag{2.1.29}$$

with the strain matrix is,

$$[\tilde{\varepsilon}] = \begin{matrix} \varepsilon_{xx}^* = \left(\sigma_{xx}^* - \upsilon\sigma_{yy}^*\right)/E \\ \varepsilon_{yy}^* = \left(\sigma_{yy}^* - \upsilon\sigma_{xx}^*\right)/E \\ \gamma_{xy}^* = 2(1+\upsilon)\tau_{xy}^*/E \end{matrix} \tag{2.1.30}$$

for plane stress, and,

$$[\tilde{\varepsilon}] = \begin{matrix} \varepsilon_{xx}^* = (1+\upsilon)\left((1-\upsilon)\sigma_{xx}^* - \upsilon\sigma_{yy}^*\right)/E \\ \varepsilon_{yy}^* = (1+\upsilon)\left((1-\upsilon)\sigma_{yy}^* - \upsilon\sigma_{xx}^*\right)/E \\ \gamma_{xy}^* = 2(1+\upsilon)\tau_{xy}^*/E \end{matrix} \tag{2.1.31}$$

for plane strain. Explicitly, the components of the nominal stress field within the Cartesian coordinates are,

$$\begin{aligned} \sigma_{xx}^* = &\frac{K_I^*}{\sqrt{2\pi r}}\left(\cos\frac{\theta}{2} - \frac{1}{2}\sin\theta\sin\frac{3\theta}{2} - \frac{\rho}{2r}\cos\frac{3\theta}{2}\right) \\ &+ \frac{K_{II}^*}{\sqrt{2\pi r}}\left(-2\sin\frac{\theta}{2} - \frac{1}{2}\sin\theta\cos\frac{3\theta}{2} + \frac{\rho}{2r}\sin\frac{3\theta}{2}\right) \end{aligned} \tag{2.1.32-a}$$

$$\begin{aligned} \sigma_{yy}^* = &\frac{K_I^*}{\sqrt{2\pi r}}\left(\cos\frac{\theta}{2} + \frac{1}{2}\sin\theta\sin\frac{3\theta}{2} + \frac{\rho}{2r}\cos\frac{3\theta}{2}\right) \\ &+ \frac{K_{II}^*}{\sqrt{2\pi r}}\left(\frac{1}{2}\sin\theta\cos\frac{3\theta}{2} - \frac{\rho}{2r}\sin\frac{3\theta}{2}\right), \end{aligned} \tag{2.1.32-b}$$

$$\begin{aligned} \tau_{xy}^* = &\frac{K_I^*}{\sqrt{2\pi r}}\left(\frac{1}{2}\sin\theta\cos\frac{3\theta}{2} - \frac{\rho}{2r}\sin\frac{3\theta}{2}\right) \\ &+ \frac{K_{II}^*}{\sqrt{2\pi r}}\left(\cos\frac{\theta}{2} - \frac{1}{2}\sin\theta\sin\frac{3\theta}{2} - \frac{\rho}{2r}\cos\frac{3\theta}{2}\right). \end{aligned} \tag{2.1.32-c}$$

By minimizing the strain energy density with respect to the nominal SIFs ($K_I^*$, $K_{II}^*$) under the linear constraint from shear stress (**Eq**.(2.1.27)), one can find values of two scalar variables, *i.e*. the nominal SIFs. Mathematically, this is a constrained optimization problem in finite-dimensional space, we can have the following solutions,

$$\begin{cases} K_I^* = \frac{D(A_{22}B - A_{12}C)}{A_{11}C^2 - 2A_{12}BC + A_{22}B^2} \\ K_{II}^* = \frac{D(A_{11}C - A_{12}B)}{A_{11}C^2 - 2A_{12}BC + A_{22}B^2} \end{cases} \tag{2.1.33}$$

where:

$$\begin{aligned} B &= \sin(\theta/2)(1 + cos\theta + \rho/r) \\ C &= \cos(\theta/2)(-1 + 3cos\theta - \rho/r) \\ D &= 2\sqrt{2\pi r}(f/b) \end{aligned} \tag{2.1.34}$$

The coefficients $A_{11}$, $A_{12}$, $A_{22}$ and more details about the derivation can be found in **Appendix A**.

To involve the effect of GBs and hydrogen atmosphere on the dislocation nucleation, here we relate the effective mode-II SIF $K_{II}^{eff}$ in **Eq**.(2.1.22) with the above-defined localized mode-I SIF $K_I^*$ and localized mode-II SIF $K_{II}^*$ in **Eq**.(2.1.33) of the hydrogen-informed ***RTST*** model [92, 104],

$$K_{II}^{eff} = K_I^* \cos^2\frac{\theta}{2}\sin\frac{\theta}{2} + K_{II}^* \cos\frac{\theta}{2}\left(1 - 3\sin^2\frac{\theta}{2}\right), \tag{2.1.35}$$

Thus, the ***PRB*** framework now can be extended to consider the dislocation emission from a GB crack via,

$$\frac{(1-\upsilon)Q_{2d}}{\mu b^2} = m\left(1 - \sqrt{\frac{1-\upsilon}{2\mu\gamma_{usf}}}\left|K_{II}^{eff}\right|\right)^{3/2}. \tag{2.1.36}$$

The above activation energy derived as an implicit function of the applied mode-I SIF $K_I$ and applied mode-II SIF $K_{II}$, however, cannot be directly employed in the transition-state-theory-based analysis of dislocation nucleation in following steps. Inspired by previous studies [122, 123], and considering the temperature dependence, the activation Gibbs free energy can be rewritten as,

$$Q_{2d}(K_I, K_{II}, T) = (1 - T/T_m)Q_{2d}(K_I, K_{II}), \tag{2.1.37}$$

where, $T_m$ is the surface disordering temperature (to a first approximation, can be taken as the melting temperature) [122], and the activation enthalpy under zero temperature $Q_{2d}(K_I, K_{II}, T = 0)$ is given by,

$$Q_{2d}(K_I, K_{II}) = C\left(1 - \frac{K_r}{K_r^0}\right)^n, \tag{2.1.38}$$

where, $C$, $n$, $K_r^0$ are fitting parameters obtained by fitting the zero-temperature activation enthalpy $Q_{2d}$ values calculated from **Eq**.(2.1.36) to **Eq**.(2.1.38). The reduced (or equivalent) SIF $K_r$ is a function of the applied Mode-I SIF $K_I$ and applied Mode-II SIF $K_{II}$ [124],

$$K_r \equiv K_r(K_I, K_{II}) = \sqrt{K_I^2 + K_{II}^2} \tag{2.1.39}$$

In practice, we use the normalized form of **Eq**.(2.1.38), *i.e.*, $\Theta_{2d}(K_I, K_{II}) = \tilde{C}\left(1 - \frac{K_r}{K_r^0}\right)^n$ with the dimensionless parameter $\tilde{C} = C\frac{1-\upsilon}{\mu b^2}$. The average rate of the dislocation nucleation can be described as,

$$\omega = \omega_0 N \exp\left(-\frac{Q_{3d}(K_I, K_{II}, T)}{k_B T}\right), \tag{2.1.40}$$

where, $\omega_0$ is the attempt frequency (to a first approximation, can be estimated as the Debye frequency, *i.e.*, $k_B T_D/\hbar$, where $T_D$ is the Debye temperature). Under the ***transition state theory*** (***TST***) framework, a dislocation nucleation event will occur once the reduced $K_r$ (of the applied mode-I SIF $K_I$ and mode-II SIF $K_{II}$) is equal to the most probable SIF $K_{Id}^p$,

$$\left.\frac{Q_{3d}(K_I,K_{II},T)}{k_B T}\right|_{K_r=K_r^p} = \ln\left.\left(\frac{k_B T N\omega_0}{\dot{K}_r\Omega(K_I,K_{II},T)}\right)\right|_{K_r=K_r^p}, \tag{2.1.41}$$

where, $N\omega_0$ represents the number of potential nucleation sites in the vicinity of the crack-tip multiplied by the attempt frequency $\omega_0$ [125]. The 3D energy barrier $Q_{3d}$ is estimated from the 2D energy barrier $Q_{2d}$ with a scaling factor $s_0$, *i.e.*, $Q_{3d} = s_0 Q_{2d}$. The activation volume-like term is defined as,

$$\Omega(K_I,K_{II},T) = -\frac{\partial Q_{3d}}{\partial K_r} = \left(1-\frac{T}{T_m}\right)\frac{s_0 C n}{K_r^0}\left(1-\frac{K_r}{K_r^0}\right)^{n-1}. \tag{2.1.42}$$

By numerically solving the **Eq**.(2.1.41) with the material constants listed in **Table 1**, the most probable mixed-mode SIF $K_r^p$ for dislocation nucleation from a GB crack can be obtained. Comparing the values of $K_r^p$ and the critical SIF required for intergranular crack cleavage $K_{IG}$ [104, 126], the fracture patterns can be determined as, *i.e.*, either the brittle cleavage or the ductile blunting via dislocation emission.

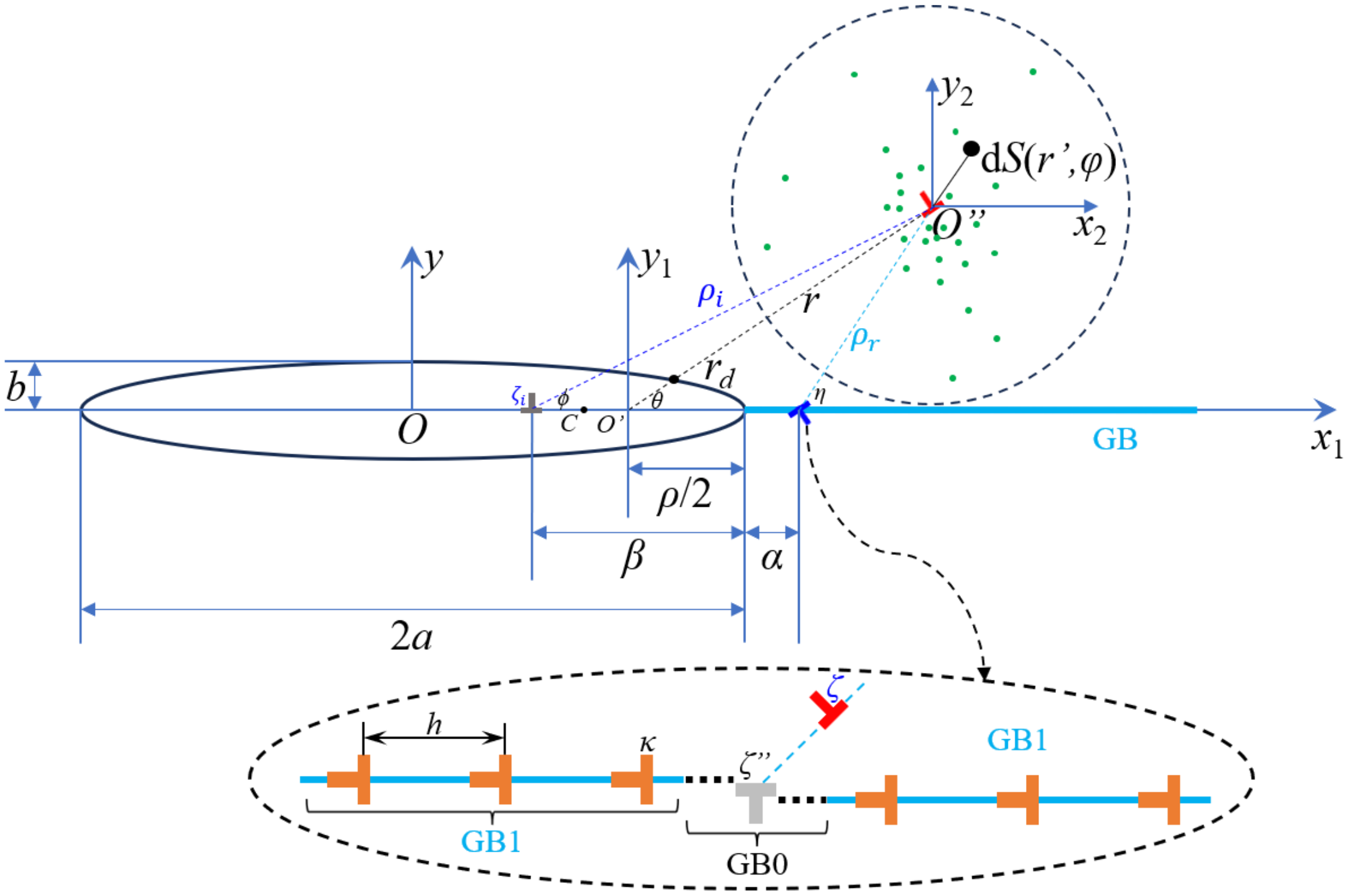

**Fig. 1**. Schematics of an edge dislocation emitted from the intergranular cack-tip under hydrogen environment. The green points represent hydrogen atoms.

## 2.2 Phase- Ⅱ: The brittle cleavage in the DFZ via the HEDE mechanism

After the emitted dislocations glide away from the crack-tip, they come to rest on the slip plane to form a plastic zone. With the simultaneous crack blunting and stress relaxation, a dislocation-free zone (***DFZ***) is formed between the crack-tip and the plastic zone [127], as shown in **Fig. 2** schematically. Similar to the concept of Thomson's elastic core region [128], the ***DFZ*** size $R_c$ can be determined as [129],

$$R_C = \frac{\xi}{\lambda^2}\frac{E(\alpha b)^2}{W_{ad}} \tag{2.2.1}$$

where, the parameter $\xi$ (taking on the value 1 for plane stress and $(1-\upsilon^2)$ for plane strain) characterizes the state of constraint illustrated in the ***HRR*** theory [130-134], $\sqrt{2\pi}\lambda$ is unity for plane stress and $(1-2\upsilon)$ for plane strain, $\alpha$ is an experimentally measurable material constant that reflects the strength of interaction between moving dislocations in a plastically deforming body [135], and $W_{ad}$ is simply the work of adhesion. Recent theoretical studies by Tehranchi *et al*. [12] suggest that the ***HEDE*** mechanism is operative by reducing the SIF $K_{Ic}$ for cleavage below the value $K_{Ie}$ required for ductile dislocation emission and blunting under hydrogen environment. Therefore, let us consider the hydrogen transport within the ***DFZ***, the ***HEDE*** concept could be realized by tuning the adhesion work as a function of the hydrogen concentration [128, 136], the hydrogen-segregation-induced change in the work of adhesion can be predicted using well-established thermodynamics relations [129, 137],

$$W_{ad} = W_{ad}^0 - \int_0^\Gamma [\mu_b(\Gamma_b) - \mu_s(\Gamma_s/2)]\mathrm{d}\Gamma \tag{2.2.2}$$

where, $W_{ad}^0$ is the work of adhesion for the pure interface, $\Gamma$ is the surface excess of the segregant (*i.e.*, hydrogen), and $\mu_b(\Gamma_b)$ and $\mu_s(\Gamma_s/2)$ are the boundary and surface adsorption isotherms, respectively.

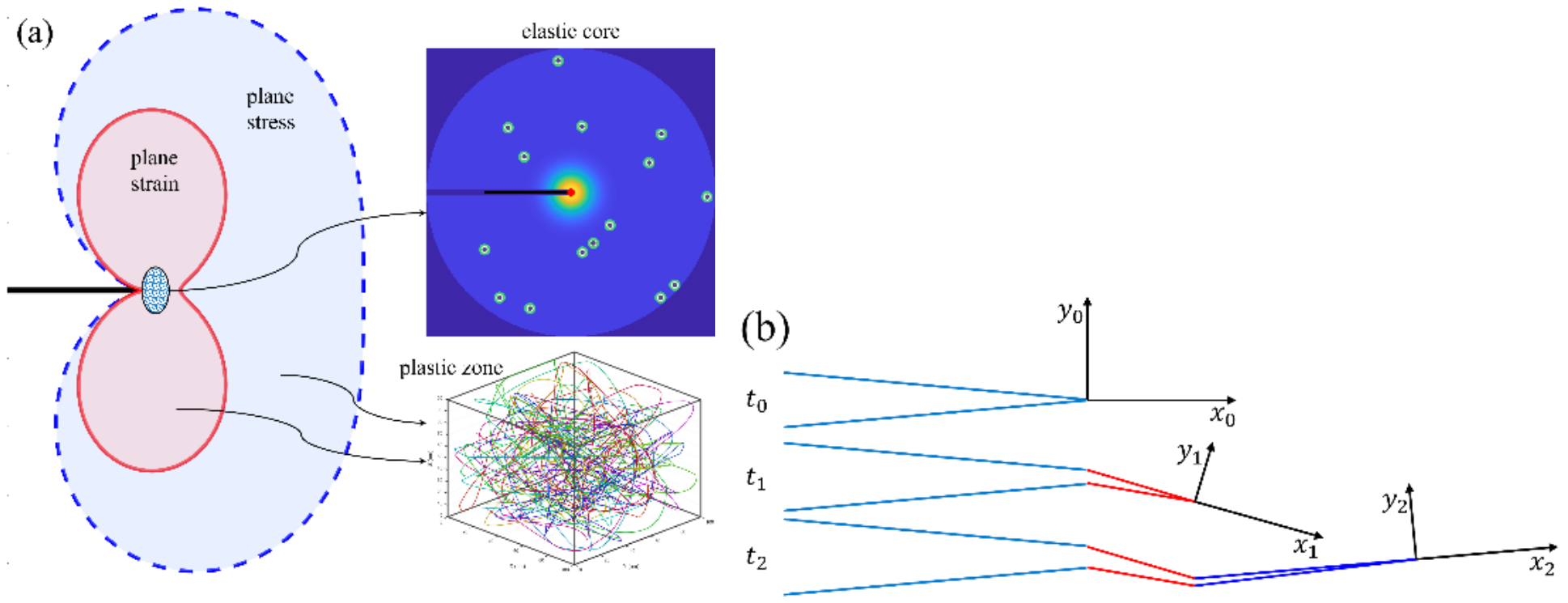

**Fig. 2**. (a) Schematics of hydrogen transport within the elastic core in the vicinity of the crack-tip. The red solid and blue dash curves represent the plastic zone shape estimated by the *von Mises* criterion in plane strain and plane stress under mode-I loading, respectively. (b) Schematics of the dynamics rebuilding of the crack-tip coordinate system at different moments.

Considering the existence of various microstructures near the crack tip and its influence on the transgranular cleavage cracking, the anisotropic crack extension behaviors would be evaluated using the linear elastic fracture mechanics (***LEFM***) approach. Established by Lieberman and Zirinsky [138, 139], the elastic constants of arbitrary oriented single crystals can be determined by the following procedures,

1) determine the direction cosines connecting the two unitary orthogonal axis systems as:

$$\begin{bmatrix} x_1' \\ x_2' \\ x_3' \end{bmatrix} = \begin{bmatrix} \beta_{11} & \beta_{12} & \beta_{13} \\ \beta_{21} & \beta_{22} & \beta_{23} \\ \beta_{31} & \beta_{32} & \beta_{33} \end{bmatrix} \begin{bmatrix} x_1 \\ x_2 \\ x_3 \end{bmatrix} \tag{2.2.3}$$

2) write the quadratic combinations of the direction cosines in the following 6×6 matrix:

$$\gamma = \begin{bmatrix} \beta_{11}^2 & \beta_{12}^2 & \beta_{13}^2 & \beta_{12}\beta_{13} & \beta_{13}\beta_{11} & \beta_{11}\beta_{12} \\ \beta_{21}^2 & \beta_{22}^2 & \beta_{23}^2 & \beta_{22}\beta_{23} & \beta_{23}\beta_{21} & \beta_{21}\beta_{22} \\ \beta_{31}^2 & \beta_{32}^2 & \beta_{33}^2 & \beta_{32}\beta_{33} & \beta_{33}\beta_{31} & \beta_{31}\beta_{32} \\ 2\beta_{21}\beta_{31} & 2\beta_{22}\beta_{32} & 2\beta_{23}\beta_{33} & (\beta_{22}\beta_{33}+\beta_{23}\beta_{32}) & (\beta_{21}\beta_{33}+\beta_{23}\beta_{31}) & (\beta_{22}\beta_{31}+\beta_{21}\beta_{32}) \\ 2\beta_{31}\beta_{11} & 2\beta_{32}\beta_{12} & 2\beta_{33}\beta_{13} & (\beta_{13}\beta_{32}+\beta_{12}\beta_{33}) & (\beta_{13}\beta_{31}+\beta_{11}\beta_{33}) & (\beta_{11}\beta_{32}+\beta_{12}\beta_{31}) \\ 2\beta_{11}\beta_{21} & 2\beta_{12}\beta_{22} & 2\beta_{13}\beta_{23} & (\beta_{12}\beta_{23}+\beta_{13}\beta_{22}) & (\beta_{13}\beta_{21}+\beta_{11}\beta_{23}) & (\beta_{11}\beta_{22}+\beta_{12}\beta_{21}) \end{bmatrix} \tag{2.2.4}$$

for the transformation of the elastic compliance matrix $S_{ij}$, and

$$\alpha = \begin{bmatrix} \beta_{11}^2 & \beta_{12}^2 & \beta_{13}^2 & 2\beta_{12}\beta_{13} & 2\beta_{13}\beta_{11} & 2\beta_{11}\beta_{12} \\ \beta_{21}^2 & \beta_{22}^2 & \beta_{23}^2 & 2\beta_{22}\beta_{23} & 2\beta_{23}\beta_{21} & 2\beta_{21}\beta_{22} \\ \beta_{31}^2 & \beta_{32}^2 & \beta_{33}^2 & 2\beta_{32}\beta_{33} & 2\beta_{33}\beta_{31} & 2\beta_{31}\beta_{32} \\ \beta_{21}\beta_{31} & \beta_{22}\beta_{32} & \beta_{23}\beta_{33} & (\beta_{22}\beta_{33}+\beta_{23}\beta_{32}) & (\beta_{21}\beta_{33}+\beta_{23}\beta_{31}) & (\beta_{22}\beta_{31}+\beta_{21}\beta_{32}) \\ \beta_{31}\beta_{11} & \beta_{32}\beta_{12} & \beta_{33}\beta_{13} & (\beta_{13}\beta_{32}+\beta_{12}\beta_{33}) & (\beta_{13}\beta_{31}+\beta_{11}\beta_{33}) & (\beta_{11}\beta_{32}+\beta_{12}\beta_{31}) \\ \beta_{11}\beta_{21} & \beta_{12}\beta_{22} & \beta_{13}\beta_{23} & (\beta_{12}\beta_{23}+\beta_{13}\beta_{22}) & (\beta_{13}\beta_{21}+\beta_{11}\beta_{23}) & (\beta_{11}\beta_{22}+\beta_{12}\beta_{21}) \end{bmatrix} \tag{2.2.5}$$

for the transformation of the elastic constant matrix $C_{ij}$. The elastic coefficients of bcc Fe in the basic coordinate system (*x*-[100], *y*-[010], *z*-[001]) are: $C_{11}$ = 243.36 GPa; $C_{12}$ = 145.01 GPa; $C_{44}$ = 116.04 GPa [140].

3) calculate the transformed matrix:

$$S' = \gamma S \gamma^T \tag{2.2.6-a}$$

$$C' = \alpha C \alpha^T \tag{2.2.6-b}$$

where, $\gamma^T$ and $\alpha^T$ are the transpose matrix of $\gamma$ and $\alpha$, respectively. The transformed compliance coefficients $S'_{ij}$ for 12 different orientations can be found in our previous studies [104].

As demonstrated by Zhang *et al*. [141], the displacement and stress fields of a half crack in an infinite anisotropic medium can be derived by the ***Lekhnitskii formalism***. Detailed descriptions and the application to cracks are given by Sih *et al*. [142]. Here we briefly review the main formulas, where the governing equation of the plane strain problem in an anisotropic linear elastic medium is,

$$b_{11}\frac{\partial^4 U}{\partial y^4} + b_{22}\frac{\partial^4 U}{\partial x^4} + (2b_{12} + b_{66})\frac{\partial^4 U}{\partial x^2 \partial y^2} - 2b_{16}\frac{\partial^4 U}{\partial x \partial y^3} - 2b_{26}\frac{\partial^4 U}{\partial x^3 \partial y} = 0 \tag{2.2.7}$$

where $U$ is the ***Airy stress function*** and $b_{ij}$ are correlated with the orientation-dependent compliance moduli $S'_{mn}$ determined by the **Eq**.(2.2.6-a),

$$b_{ij} = S'_{ij} - \frac{S'_{i3}S'_{j3}}{S'_{33}} \tag{2.2.8}$$

From a mathematical standpoint, the plane stress and plane strain problems are identical except for the values of elastic constants entering into the reduced strain-stress relations. Henceforth, it is understood that $a_{ij}$ ($= S'_{ij}$) and $b_{ij}$ for $i, j$ = 1, 2, 6 are the reduced elastic constants used in the plane stress and plane strain problems, respectively [142]. For brevity, we would not repeat almost the same derivation for the plane stress problem here. In previous study [104], we have calculated $b_{ij}$ coefficients for 12 different lattice orientations, covering the GB angle within the range of [20.05°, 148.4°]. The characteristic equation of the governing equation **Eq**.(2.2.7) is,

$$b_{11}\mu_j^4 - 2b_{16}\mu_j^3 + (2b_{12} + b_{66})\mu_j^2 - 2b_{26}\mu_j + b_{22} = 0 \tag{2.2.9}$$

By denoting the roots $\mu_i$ ($i = 1{\sim}4$, $\mu_1 = \bar{\mu}_3$ and $\mu_2 = \bar{\mu}_4$) and introducing two variables $s_1 = \mu_1$, $s_2 = \mu_2$, the ***Airy stress function*** for general plane strain problem can be expressed as,

$$U(x, y) = 2\Re[U_1(z_1) + U_2(z_2)] \tag{2.2.10}$$

where $z_i = x + s_i y$ ($i$ = 1, 2), $U_1$ and $U_2$ are arbitrary functions to be determined by the boundary conditions. Hence, the displacement fields can be expressed as,

$$\begin{aligned} u_x &= 2\Re[p_1\phi(z_1) + p_2\psi(z_2)], \\ u_y &= 2\Re[q_1\phi(z_1) + q_2\psi(z_2)]. \end{aligned} \tag{2.2.11}$$

where $p_i$ and $q_i$ are of the form,

$$p_1 = b_{11}s_1^2 + b_{12} - b_{16}s_1, \quad p_2 = b_{11}s_2^2 + b_{12} - b_{16}s_2,$$
$$q_1 = \frac{b_{12}s_1^2 + b_{22} - b_{26}s_1}{s_1}, \qquad q_2 = \frac{b_{12}s_2^2 + b_{22} - b_{26}s_2}{s_2}. \tag{2.2.12}$$

By considering the boundary condition of a crack in an infinite medium, the stress function $\phi$ and $\psi$ can be determined. Thus, the displacement fields at the crack-tip, expressed in polar coordinates ($\theta$, $r$), can be written as,

$$u_x = K_I\sqrt{\frac{2r}{\pi}}\,\Re\left[\frac{1}{s_1 - s_2}\left(s_1 p_2\sqrt{\cos\theta + s_2\sin\theta} - s_2 p_1\sqrt{\cos\theta + s_1\sin\theta}\right)\right],$$
$$u_y = K_I\sqrt{\frac{2r}{\pi}}\,\Re\left[\frac{1}{s_1 - s_2}\left(s_1 q_2\sqrt{\cos\theta + s_2\sin\theta} - s_2 q_1\sqrt{\cos\theta + s_1\sin\theta}\right)\right]. \tag{2.2.13}$$

for mode-I loading, and,

$$u_x = K_{II}\sqrt{\frac{2r}{\pi}}\,\Re\left[\frac{1}{s_1 - s_2}\left(p_2\sqrt{\cos\theta + s_2\sin\theta} - p_1\sqrt{\cos\theta + s_1\sin\theta}\right)\right],$$
$$u_y = K_{II}\sqrt{\frac{2r}{\pi}}\,\Re\left[\frac{1}{s_1 - s_2}\left(q_2\sqrt{\cos\theta + s_2\sin\theta} - q_1\sqrt{\cos\theta + s_1\sin\theta}\right)\right]. \tag{2.2.14}$$

for mode-II loading, respectively.

And the stress fields are,

$$\sigma_{xx} = \frac{K_I}{\sqrt{2\pi r}}\,\Re\left[\frac{s_1 s_2}{s_1 - s_2}\left(\frac{s_2}{\sqrt{\cos\theta + s_2\sin\theta}} - \frac{s_1}{\sqrt{\cos\theta + s_1\sin\theta}}\right)\right],$$
$$\sigma_{yy} = \frac{K_I}{\sqrt{2\pi r}}\,\Re\left[\frac{1}{s_1 - s_2}\left(\frac{s_1}{\sqrt{\cos\theta + s_2\sin\theta}} - \frac{s_2}{\sqrt{\cos\theta + s_1\sin\theta}}\right)\right],$$
$$\tau_{xy} = \frac{K_I}{\sqrt{2\pi r}}\,\Re\left[\frac{s_1 s_2}{s_1 - s_2}\left(\frac{1}{\sqrt{\cos\theta + s_1\sin\theta}} - \frac{1}{\sqrt{\cos\theta + s_2\sin\theta}}\right)\right]. \tag{2.2.15}$$

for mode-I loading, and,

$$\sigma_{xx} = \frac{K_{II}}{\sqrt{2\pi r}}\,\Re\left[\frac{1}{s_1 - s_2}\left(\frac{s_2^2}{\sqrt{\cos\theta + s_2\sin\theta}} - \frac{s_1^2}{\sqrt{\cos\theta + s_1\sin\theta}}\right)\right],$$
$$\sigma_{yy} = \frac{K_{II}}{\sqrt{2\pi r}}\,\Re\left[\frac{1}{s_1 - s_2}\left(\frac{1}{\sqrt{\cos\theta + s_2\sin\theta}} - \frac{1}{\sqrt{\cos\theta + s_1\sin\theta}}\right)\right],$$
$$\tau_{xy} = \frac{K_{II}}{\sqrt{2\pi r}}\,\Re\left[\frac{1}{s_1 - s_2}\left(\frac{s_1}{\sqrt{\cos\theta + s_1\sin\theta}} - \frac{s_2}{\sqrt{\cos\theta + s_2\sin\theta}}\right)\right]. \tag{2.2.16}$$

for mode-II loading, respectively. Thus, the *von Mises* equivalent stress is,

$$\sigma_{vm} = \sqrt{\left[\left(\sigma_{xx} - \sigma_{yy}\right)^2 + \left(\sigma_{yy} - \sigma_{zz}\right)^2 + (\sigma_{zz} - \sigma_{xx})^2\right]/2 + 3\tau_{xy}^2} \tag{2.2.17}$$

where, $\sigma_{zz} = 0$ for plane stress and $\sigma_{zz} = \upsilon\left(\sigma_{xx} + \sigma_{yy}\right)$ for plane strain. Novak *et al*. [70] have demonstrated that whereas the ductile fracture in the uncharged steel is *strain-controlled*, the local fracture event for the initiation of brittle fracture in the presence of hydrogen is *stress-controlled*. Thus, by setting the microcrack initiation criterion as the *von Mises* stress reaches the minimum of the debonding strength of all planar and volumetric defects existed in the ***DFZ***, we have,

$$\sigma_{vm} = \min\{\Sigma_{db}^{GB,H}, \Sigma_{db}^{inc,H}, \Sigma_{db}^{PB,H}, \dots\} \tag{2.2.18}$$

where, $\Sigma_{db}^{GB,H}$, $\Sigma_{db}^{inclu,H}$ and $\Sigma_{db}^{PB,H}$ represent the debonding strength of GBs, inclusions and phase boundaries under hydrogen environment, which could be written as functions of local hydrogen concentration $c_H$ mathematically,

$$\begin{cases} \Sigma_{db}^{GB,H} = \Sigma_{db}^{GB}(c_H) \\ \Sigma_{db}^{inc,H} = \Sigma_{db}^{inc}(c_H) \\ \Sigma_{db}^{PB,H} = \Sigma_{db}^{PB}(c_H) \\ \cdots \end{cases} \tag{2.2.19}$$

Due to the existence of microstructural defects, *e.g*. GBs, second phase particles, the crack path would be usually deflected, even branched. Thus, the coordinate system should be rebuilt in real time with the development of the crack path as shown in **Fig. 2**-(b). However, for demonstration purposes only, here we consider the intergranular fracture as shown in **Fig. 1**. Specifically, if the crack propagates along a GB between two (upper and lower) grains (as shown in **Fig. 1**), the effect of GB energy $\gamma_{gb}$ has to be considered to calculate the energy required to create two new surfaces, *i.e*., the Dupré equation, which could be crystallographically different [143, 144],

$$G_I^{gb} = \gamma_s^1 + \gamma_s^2 - \gamma_{gb}\ , \tag{2.2.20}$$

where, $\gamma_s^1$ and $\gamma_s^2$ are the free surface energies of the adjoining grains (simply note the upper grain as grain-#1, while the lower one as grain-#2 in **Fig. 1**), respectively. In the case of negligible plastic deformation, the fracture along symmetrical GBs should always be more favorable than cleavage (inside the grain) along a plane parallel to the GB ($\gamma_s^1 = \gamma_s^2 = \gamma_s$) [145]. The calculated $G_I^{gb}$ and $K_{IG}$ for 12 different GBs can be found in our previous studies [104]. For GBs without hydrogen charged, $W_{ad}^0 = G_I^{gb}$. More generally, for the case of fast decohesion, as described by Hirth and Rice [137], the ideal work of fracture (*i.e*., **Eq**.(2.2.2)) can be reformulated as [70], $2\gamma_{int} = 2\gamma_s - \gamma_i - (\Delta g_i - \Delta g_s)\Gamma$. In the study of hydrogen-induced intergranular fracture in steels [70], $2\gamma_s$ and $\gamma_i$ are the free surface and carbide/matrix interface energies in the absence of hydrogen, respectively; $\Delta g_s$ and $\Delta g_i$ are the Gibbs free energy excesses when hydrogen is absorbed onto the free surface created upon separation and the particle/matrix interface, respectively; $\Gamma$ is the interface coverage by hydrogen measured in hydrogen solute atoms per unit area.

Under hydrogen environment, inspired by the classical Griffith theory and for a first approximation, we have the debonding strength of a general phase boundary as,

$$\Sigma_{db} = \beta\sqrt{\frac{E^*W_{ad}}{d_0}} \tag{2.2.21}$$

where, $\beta$ is a factor accounting for the interfacial defect geometry and loading patterns, $E^*$ is the reduced elastic modulus, and $d_0$ is the characteristic length of the defect. Specifically, the strength $\Sigma$ of carbides with the particle size $l$, can be reduced through the Smith model [70, 146], *viz*:

$$\frac{l}{d}\Sigma^2 + \tau_{eff}^2\left(1 + \frac{4}{\pi}\frac{\tau_0}{\tau_{eff}}\sqrt{\frac{l}{d}}\right)^2 = \frac{4E\gamma_{eff}}{\pi(1-\upsilon^2)d} \tag{2.2.22}$$

where, $d$ is the grain diameter, $\tau_0$ is the stress resisting dislocation motion, $\tau_{eff}$ is an effective stress equal to the difference between the applied shear stress $\tau_{xy}$ and the stress $\tau_0$, *i.e.*, $\tau_{eff} = \tau_{xy} - \tau_0$, and $\gamma_{eff}$ is the effective decohesion work of the carbide/matrix interface as a function of the hydrogen solute atoms carried by the dislocations,

$$\gamma_{eff} = 2\gamma_{int} + \gamma_p \tag{2.2.23}$$

where, the plastic work $\gamma_p$ accompanying the decohesion initiation event is [147, 148],

$$\gamma_p = A(2\gamma_{int})^q \tag{2.2.24}$$

The Sofronis-McMeeking-Krom-Koers-Bakker (***SMKKB***) hydrogen transport equation is given as [70, 149],

$$\frac{C_L+C_T(1-\theta_T)}{C_L}\frac{\partial C_L}{\partial t} - \nabla\cdot(D_L\nabla C_L) + \nabla\cdot\left(\frac{D_LC_L\tilde{V}_H}{RT}\nabla\sigma_H\right) + \theta_T\frac{\mathrm{d}N_T}{\mathrm{d}\varepsilon_p}\frac{\partial\varepsilon_p}{\partial t} = 0 \tag{2.2.25}$$

where, $C_L$ and $C_T$ are the hydrogen concentration in lattice sites and trap sites, respectively; $D_L$ is the concentration independent diffusivity in normal interstitial lattice sites (NILS) $D_L = M_LRT$, where $M_L$ is the mobility of the hydrogen in lattice sites; $\tilde{V}_H$ is the partial molar volume of hydrogen, *i.e.*, $2.0\times10^{-6}$ m$^3$ for $\alpha$-iron at 293 K [150]; $\sigma_H$ is the hydrostatic stress $\sigma_H = \frac{1}{3}\sum_{i=1}^{3}\sigma_{ii}$; $\theta_T$ is the occupancy of the trap sites and $N_T$ is the trap density. The last term in **Eq**.(2.2.25), named as the (plastic) strain rate factor, was not taken into account in the original model of Sofronis and McMeeking [151].

However, numerous experiments [152] have indicated that hydrogen-induced facture might occur within the elastic strain regime, the validation of **Eq**.(2.2.25), *i.e*. in which the hydrogen trapping is dominated by the last plastic term, is thus questioned. Let us consider a spherical inclusion in the matrix, the hydrogen trapping capability is actually determined by the trap binding energy $\Delta E_T$, which is the sum of the interfacial energy $E_i$ and elastic strain energy $E_s$ [153]. Assuming the elastic constants of the matrix and inclusion are equal and denoting the shear modulus as $G$, Poisson's ratio as $v$, the magnitude of the Burgers vector of the misfit dislocations as $b$, and the unstrained radius of the spherical inclusion as $r$, we have,

$$E_i = 4\pi r^2 Gb\left\{\frac{1}{10} + \left[1 + \beta - \sqrt{1+\beta^2} - \beta\ln\left(2\beta\left(\sqrt{1+\beta^2} - \beta\right)\right)\right]/2\pi^2\right\} \quad \text{(2.2.26-a)}$$

$$E_s = 4\pi r^2 G\delta_0^2(1+v)/3(1-v) \quad \text{(2.2.26-b)}$$

where, $\beta = \pi f_0/(1-v)$ with the misfit $f_0$ defined as $f_0 = |a_m - a_i|/a_m$ ($a_m$ and $a_i$ are lattice constants of the matrix and inclusion, respectively), the parameter $\delta_0$ is given in terms of the dilatation strain $\varepsilon_0$ (for small strains) by Nabarro [154, 155],

$$\delta_0 = (1 + 4G/3K)\varepsilon_0 = \sqrt[3]{\Omega_i/\Omega_m} - 1 \quad \text{(2.2.27)}$$

where, $K$ is the bulk modulus of the inclusion, $\Omega_i$ and $\Omega_m$ are the atomic volumes of the inclusion and matrix, respectively (assuming $\Omega_i > \Omega_m$ for brevity). With the subscript 0 representing the relaxed situation, Jesser [153, 156] gives that,

$$f = |\delta - \delta_0| = (1 + 4G/3K)|\varepsilon_v - \varepsilon_0| \quad \text{(2.2.28)}$$

where, $\varepsilon_v$ is the volumetric strain due to applied external loading. The total energy $E_i + E_s$ (*i.e*., the trap binding energy $\Delta E_T$) nonlinearly scales with the misfit $f$, thus the volumetric strain $\varepsilon_v$. This means that the trapping capability of hydrogen traps evolves dynamically during the loading process, but was assumed unchanged in most simulation studies previously. Assuming that the normalized trap density ($N_T^{(i)}/N_T^{total}\big|_{t=0}$) follows a given distribution with respect to the trap binding energy $\Delta E_T$ in unstrained status (*i.e*., the moment $t = 0$),

$$N_T^{(i)}/N_T^{total}\Big|_{t=0} = \varphi(\Delta E_T)|_{t=0} \quad \text{(2.2.29)}$$

Upon loading, the surrounding matrix of a spherical inclusion experiences elastic deformation, leading to the variation of the lattice misfit, thus inducing the change of the trap binding energy. Therefore, the trap density distribution $\varphi(\Delta E_T)$ evolves dynamically, and the **Eq**.(2.2.25) can be rewritten as,

$$\left(1+(1-\theta_T)\frac{C_T}{C_L}\right)\frac{\partial C_L}{\partial t}-D_L\nabla\cdot(\nabla C_L)+\frac{D_L\widetilde{V}_H}{RT}\nabla\cdot(C_L\nabla\sigma_H)+\theta_T\frac{\mathrm{d}N_T}{\mathrm{d}(\Delta E_T)}\frac{\partial(\Delta E_T)}{\partial\varepsilon_v}\frac{\partial\varepsilon_v}{\partial t}=0 \qquad (2.2.30)$$

For plane problem loaded under mode-I or mode-II situation, within the elastic core in the vicinity of the crack-tip, the strain components are given in terms of the stress components calculated in **Eq**.(2.2.15) or **Eq**.(2.2.16),

$$\begin{cases}\varepsilon_{xx}=\frac{1}{2G}\left[\sigma_{xx}-\frac{\lambda^*}{2(\lambda^*+G)}\left(\sigma_{xx}+\sigma_{yy}\right)\right]\\ \varepsilon_{yy}=\frac{1}{2G}\left[\sigma_{yy}-\frac{\lambda^*}{2(\lambda^*+G)}\left(\sigma_{xx}+\sigma_{yy}\right)\right]\\ \varepsilon_{xy}=\frac{1}{2G}\tau_{xy}\end{cases} \qquad (2.2.31)$$

where, $\lambda^*=\frac{3-\kappa}{\kappa-1}G$, in which, the Kolosov constant is,

$$\kappa=\begin{cases}3-4\upsilon, & \text{for plane strain}\\ \frac{3-\upsilon}{1+\upsilon}, & \text{for plane stress}\end{cases} \qquad (2.2.32)$$

Thus, for present study in Cartesian coordinates, the volumetric strain $\varepsilon_v=\varepsilon_{xx}+\varepsilon_{yy}+\varepsilon_{zz}$ ($\varepsilon_{zz}\neq 0$ for plane stress), leads to the partial difference $\partial\varepsilon_v/\partial t$ in the last term in **Eq.**(2.2.25) be proportional to the applied SIF rate $\dot{K}_I$ (or $\dot{K}_{II}$), and,

$$\partial(\Delta E_T)/\partial\varepsilon_v=\partial E_i/\partial\varepsilon_v+\partial E_s/\partial\varepsilon_v \qquad (2.2.33)$$

with,

$$\begin{aligned}\partial E_i/\partial\varepsilon_v&=(\partial E_i/\partial\beta)\cdot(\partial\beta/\partial\varepsilon_v)\\ &=-\frac{2r^2Gb}{\pi}\ln\left(2\beta\left(\sqrt{1+\beta^2}-\beta\right)\right)\cdot\frac{\pi(1+4G/3K)}{1-\upsilon}\mathrm{sgn}(\varepsilon_v-\varepsilon_0)\end{aligned} \qquad (2.2.34\text{-a})$$

$$\partial E_s/\partial\varepsilon_v=\frac{8\pi r^2G(1+4G/3K)^2(1+\upsilon)}{3(1-\upsilon)}\varepsilon_v \qquad (2.2.34\text{-b})$$

where, the quantity $\beta$ is already updated as $\beta=\pi f/(1-\upsilon)$, the sign function (or signum function) of a real number $x$ is defined as,

$$\mathrm{sgn}(x):=\begin{cases}-1 & \text{if } x<0\\ 0 & \text{if } x=0\\ 1 & \text{if } x>0\end{cases} \qquad (2.2.35)$$

To compute $\mathrm{d}N_T/\mathrm{d}(\Delta E_T)$ in numerical implementation, one has to reevaluate the trap density distribution $\varphi(\Delta E_T)$ in every timestep iteratively. However, in semi-analytical study, commonly assuming a Gaussian distribution (*e.g.*, $N_T^{(i)} = \frac{N_{Tm}^{(i)}}{\sqrt{2\pi}\sigma}\exp\left[-\frac{\left(\Delta E_T^{(i)}-\Delta E_{Tm}^{(i)}\right)^2}{2\sigma^2}\right]$, where, $N_{Tm}^{(i)}$ is the peak density of the *i-th* type of trapping sites, $\Delta E_{Tm}^{(i)}$ is the mean energy of the distribution, $\sigma$ is the standard deviation), or Weibull distribution, etc., is not sufficient to approximate the real distribution of trapping sites in experiments, meanwhile the following estimation was used in Ref.s[157-159],

$$N_T = N_L\left(\frac{D_L}{D_{eff}} - 1\right)\exp\left(-\frac{\Delta E_T}{RT}\right) \tag{2.2.36}$$

where, $D_{eff}$ is the effective diffusion coefficient related to the NILS diffusion coefficient $D_L$ through,

$$D_{eff} = \frac{D_L}{1+\partial C_T/\partial C_L} :\rightarrow \frac{D_L}{D_{eff}} = 1 + (1-\theta_T)\frac{C_T}{C_L} \tag{2.2.37}$$

Thus, we can have $\mathrm{d}N_T/\mathrm{d}(\Delta E_T)$ in **Eq**.(2.2.30) as,

$$\frac{\mathrm{d}N_T}{\mathrm{d}(\Delta E_T)} = -\frac{N_L}{RT}\frac{(1-\theta_T)C_T}{C_L}\exp\left(-\frac{\Delta E_T}{RT}\right) \tag{2.2.38}$$

It is noted that we only demonstrate the dynamic variation of density distribution of the ideally spherical inclusion in above theoretical framework, while the hydrogen trapping capability of other defects, such as, GBs and twin boundaries, would also change dynamically with the local strain field.

In this context, from a statistical perspective [70, 160], we can evaluate the fracture probability $\mathrm{d}\phi$ in an element (subjected to a stress $\sigma$) in the numerical domain as,

$$\mathrm{d}\phi = 1 - \exp\left(-\mathrm{d}V\int_0^\sigma g(\Sigma)\mathrm{d}\Sigma\right) \tag{2.2.39}$$

where, $g(\Sigma)\mathrm{d}\Sigma$ is the number of particles per unit volume having strengths within the range of $[\Sigma, \Sigma + \mathrm{d}\Sigma]$, for which a convenient and versatile expression is the three-parameter Weibull assumption [161],

$$\int_0^\sigma g(\Sigma)\mathrm{d}\Sigma = \left(\frac{\sigma-\Sigma_u}{\Sigma_0}\right)^m f N_T^{(i)} \tag{2.2.40}$$

where, $m$ is the shape factor, $\Sigma_0$ is the scale parameter, $\Sigma_u$ is the lower bound strength (of the largest feasible cracked particle), $f$ represents the fraction of "eligible" particles per unit volume that participate in the decohesion process, taking the value 5% as suggested by Lin *et al*. [160] and Novak *et al*. [70]. The total failure probability $\Phi$ of the entire structure is,

$$\Phi = 1 - \exp\left(-\int_0^V \int_0^\sigma g(\Sigma)\mathrm{d}\Sigma\mathrm{d}V\right) \tag{2.2.41}$$

Technically, we employ a coupled finite element (FE) framework to solve this boundary value problem for solid mechanics and mass transport sequentially at every time step. The domain is discretized using 4-node bilinear isoparametric quadrilateral elements (Q4) with 2×2 Gaussian quadrature for numerical integration. However, if one wants more precise simulations, such as the cellular automata and phase field techniques to describe 3D microstructure with GBs, inclusion particles *etc*. inside, the ***HEDE*** mechanism should be realized by implementing **Eq**.(2.2.2) and **Eq**.(2.2.21) in user-defined subroutines.

## 2.3 Phase-Ⅲ: Hydrogen-assisted void formation in the plastic zone

Now let us move into the plastic zone outside the elastic core, as shown in **Fig. 2**. Previous studies [162] have demonstrated that the void growth can be reached via the condensation of vacancies, which are primarily generated by the dragging of intersection jogs, intrinsically correlated with the dislocation density. Meanwhile, under hydrogen environment, it has been proposed that hydrogen-vacancy complexes could interact with dislocations [21] to promote the nano-void formation observed experimentally [20]. Thus, the dislocation density field would be evaluated theoretically as a top priority here by incorporating the ***HRR*** solution into the ***MSG*** plasticity theory.

Revisit the well-known ***HRR*** solution for the stress and deformation fields near a crack tip in an elastic-plastic material, which was initially established for the mode-I problem of a power-law hardening material under the framework of deformation plasticity. First, let us consider the following constitutions under uniaxial loading [163],

$$\frac{\varepsilon}{\varepsilon_0} = \alpha_0\left(\frac{\sigma}{\sigma_0}\right)^n \tag{2.3.1}$$

$$\frac{\varepsilon}{\varepsilon_0} = \begin{cases} \sigma/\sigma_0, & \sigma < \sigma_0 \\ \alpha_0(\sigma/\sigma_0)^n, & \sigma \geq \sigma_0 \end{cases} \tag{2.3.2}$$

$$\frac{\varepsilon}{\varepsilon_0} = \frac{\sigma}{\sigma_0} + \alpha_0\left(\frac{\sigma}{\sigma_0}\right)^n \tag{2.3.3}$$

where, $\sigma_0$ and $\varepsilon_0$ are reference stress and strain, respectively; $\alpha_0$ is a dimensionless material parameter and $n$ is the hardening exponent. In the vicinity of the crack tip, all the three-above stress-strain correlations could be approximated as,

$$\frac{\varepsilon}{\varepsilon_0} \approx \alpha_0 \left(\frac{\sigma}{\sigma_0}\right)^n \tag{2.3.4}$$

Assuming the material obeys the ***J2*** plasticity, thus [164],

$$\varepsilon_{ij} = \frac{1+\upsilon}{E} s_{ij} + \frac{1-2\upsilon}{3E} \sigma_{kk}\delta_{ij} + \frac{3}{2}\alpha_0\varepsilon_0 \left(\frac{\sigma_e}{\sigma_0}\right)^{n-1} \frac{s_{ij}}{\sigma_0} \tag{2.3.5}$$

where, $\sigma_e = \sqrt{\frac{3}{2} s_{ij}s_{ij}}$ is the effective stress, and the deviatoric stress tensor is,

$$s_{ij} = \sigma_{ij} - \frac{1}{3}\sigma_{kk}\delta_{ij} \tag{2.3.6}$$

Then, by defining the elastic far-field mixity parameter as [165, 166],

$$M_e = \frac{2}{\pi}\arctan\left|\lim_{r\to\infty} \frac{\sigma_{\theta\theta}(\theta=0)}{\sigma_{r\theta}(\theta=0)}\right| = \frac{2}{\pi}\arctan\left|\frac{K_I}{K_{II}}\right| \in [0,1] \tag{2.3.7}$$

while the plastic near-field mixity parameter is,

$$M_p = \frac{2}{\pi}\arctan\left|\lim_{r\to 0} \frac{\sigma_{\theta\theta}(\theta=0)}{\sigma_{r\theta}(\theta=0)}\right| = \frac{2}{\pi}\arctan\left|\frac{\tilde{\sigma}_{\theta\theta}(\theta=0,M_p)}{\tilde{\sigma}_{r\theta}(\theta=0,M_p)}\right| \in [0,1] \tag{2.3.8}$$

for plane problems under mixed mode-I+II loading, we have:

***I) Plane strain***

By introducing the Airy stress function and deriving the partial differential equation governing the stress function from the compatibility equation, the problem reduces to a fourth-order nonlinear differential equation,

$$\begin{gathered}\nabla^2\nabla^2\phi + \frac{3\alpha_0}{4}\left\{\left(\frac{1}{r^2}\frac{\partial^2}{\partial\theta^2} - \frac{1}{r}\frac{\partial}{\partial r} - \frac{\partial^2}{\partial r^2}\right)\left[\sigma_e^{n-1}\left(\frac{1}{r^2}\frac{\partial^2\phi}{\partial\theta^2} + \frac{1}{r}\frac{\partial\phi}{\partial r} - \frac{\partial^2\phi}{\partial r^2}\right)\right]\right. \\ \left. + \frac{4}{r^2}\frac{\partial^2}{\partial r\partial\theta}\left[r\sigma_e^{n-1}\frac{\partial}{\partial r}\left(\frac{1}{r}\frac{\partial\phi}{\partial\theta}\right)\right]\right\} = 0\end{gathered} \tag{2.3.9}$$

where, the bi-harmonic term is,

$$\begin{gathered}\nabla^2\nabla^2\phi = \frac{\partial^4\phi}{\partial r^4} + \frac{2}{r^2}\frac{\partial^4\phi}{\partial r^2\partial\theta^2} + \frac{1}{r^4}\frac{\partial^4\phi}{\partial\theta^4} + \frac{2}{r}\frac{\partial^3\phi}{\partial r^3} \\ - \frac{2}{r^3}\frac{\partial^3\phi}{\partial r\partial\theta^2} - \frac{1}{r^2}\frac{\partial^2\phi}{\partial r^2} + \frac{4}{r^4}\frac{\partial^2\phi}{\partial\theta^2} + \frac{1}{r^3}\frac{\partial\phi}{\partial r}\end{gathered} \tag{2.3.10}$$

The approximate solution of **Eq.**(2.3.9) focuses on the crack tip by means of a series expansion on the radius,

$$\phi(r,\theta) \sim r^s \tilde{\phi}(\theta) + r^t \tilde{\phi}_2(\theta) + \cdots \tag{2.3.11}$$

and the solution can be restricted to only the dominant first term,

$$\phi(r,\theta) = K_M^p r^s \tilde{\phi}(\theta) \tag{2.3.12}$$

Since the elastic energy is an arbitrarily small portion of the total energy, the bi-harmonic term in the **Eq**.(2.3.9) can be omitted, thus the resolved **Eq**.(2.3.9) becomes homogeneous with respect to $\theta$ and can be expressed in the form of an eigenvalue equation for $s$,

$$\begin{aligned} \left(\frac{\mathrm{d}^2}{\mathrm{d}\theta^2} - n(s-2)(n(s-2)+2)\right)\left\{\tilde{\sigma}_e^{n-1}\left[s(2-s)\tilde{\phi} + \frac{\mathrm{d}^2\tilde{\phi}}{\mathrm{d}\theta^2}\right]\right\} \\ +4(s-1)(n(s-2)+1)\frac{\mathrm{d}}{\mathrm{d}\theta}\left(\tilde{\sigma}_e^{n-1}\frac{\mathrm{d}\tilde{\phi}}{\mathrm{d}\theta}\right) = 0 \end{aligned} \tag{2.3.13}$$

We can rewrite the above **Eq**.(2.3.13) taking into account the adopted denotation,

$$\left(\frac{\mathrm{d}^2}{\mathrm{d}\theta^2} - a_1\right)\left[\tilde{\sigma}_e^{n-1}\left[a_2\tilde{\phi} + \frac{\mathrm{d}^2\tilde{\phi}}{\mathrm{d}\theta^2}\right]\right] + a_3\frac{d}{d\theta}\left(\tilde{\sigma}_e^{n-1}\frac{\mathrm{d}\tilde{\phi}}{\mathrm{d}\theta}\right) = 0 \tag{2.3.14}$$

with,

$$\tilde{\sigma}_e^2 = \frac{3}{4}\left(a_2\tilde{\phi} + \frac{\mathrm{d}^2\tilde{\phi}}{\mathrm{d}\theta^2}\right)^2 + a_4\left(\frac{\mathrm{d}\tilde{\phi}}{\mathrm{d}\theta}\right)^2 \tag{2.3.15}$$

where,

$$\begin{cases} a_1 = n(s-2)[n(s-2)+2] \\ a_2 = s(2-s) \\ a_3 = 4(s-1)[n(s-2)+1] \\ a_4 = 3(1-s)^2 \end{cases} \tag{2.3.16}$$

The nonlinear fourth-order differential eigenvalue equation **Eq**.(2.3.14) is solved by means of transformation to the system of first-order differential equations,

$$\begin{cases} \frac{d\tilde{\phi}}{d\theta} = \tilde{\phi}_1 \\ \frac{d\tilde{\phi}_1}{d\theta} = \tilde{\phi}_2 \\ \frac{d\tilde{\phi}_2}{d\theta} = \tilde{\phi}_3 \\ \frac{d\tilde{\phi}_3}{d\theta} = \varphi(\tilde{\phi}, \tilde{\phi}_1, \tilde{\phi}_2, \tilde{\phi}_3) \end{cases} \tag{2.3.17}$$

where,

$$
\begin{aligned}
\varphi(\tilde{\phi},\tilde{\phi}_1,\tilde{\phi}_2,\tilde{\phi}_3) = &-\left[\tfrac{3}{4}n\left(a_2\tilde{\phi}+\tilde{\phi}_2\right)^2 + a_4\tilde{\phi}_1^2\right]^{-1} \\
&\times\left\{\tilde{\sigma}_e^2\left[(a_2-a_1+a_3)\tilde{\phi}_2 - a_1a_2\tilde{\phi}\right]\right. \\
&+\tfrac{n-3}{2}\varphi_1\left(a_2\tilde{\phi}_1+\tilde{\phi}_3\right)+\tfrac{n-1}{2}\left(a_2\tilde{\phi}+\tilde{\phi}_2\right) \\
&\times\left[\tfrac{3}{2}\left(a_2\tilde{\phi}_1+\tilde{\phi}_3\right)^2+\tfrac{3}{2}\left(a_2\tilde{\phi}+\tilde{\phi}_2\right)a_2\tilde{\phi}_2+2a_4\left(\tilde{\phi}_2^2+\tilde{\phi}_1\tilde{\phi}_3\right)\right] \\
&+\tfrac{n+1}{2}\varphi_1\left(a_2\tilde{\phi}_1+\tilde{\phi}_3\right)+a_3\tfrac{n-1}{2}\varphi_1\tilde{\phi}_1 \\
&\left.+\tfrac{1}{\tilde{\sigma}_e^2}\tfrac{(n-3)(n-1)}{4}\varphi_1^2\left(a_2\tilde{\phi}+\tilde{\phi}_2\right)\right\}
\end{aligned}
\tag{2.3.18}
$$

and,

$$
\varphi_1(\tilde{\phi},\tilde{\phi}_1,\tilde{\phi}_2,\tilde{\phi}_3)=\tfrac{3}{2}\left(a_2\tilde{\phi}+\tilde{\phi}_2\right)\left(a_2\tilde{\phi}_1+\tilde{\phi}_3\right)+2a_4\tilde{\phi}_1\tilde{\phi}_2 \tag{2.3.19}
$$

The components of the dimensionless displacement field are,

$$
\begin{cases}
\tilde{u}_r(\theta)=\tfrac{3}{4}(n+1)\tilde{\sigma}_e^{n-1}\left[a_2\tilde{\phi}+\tilde{\phi}_2\right] \\
\tilde{u}_\theta(\theta)=\tfrac{n+1}{n}\left[\tfrac{d\tilde{u}_r}{d\theta}-3\tilde{\sigma}_e^{n-1}(1-s)\cdot\tilde{\phi}_1\right]
\end{cases}
\tag{2.3.20}
$$

where,

$$
\begin{cases}
\tfrac{d\tilde{u}_r}{d\theta}=\tfrac{3(n+1)}{4}\tilde{\sigma}_e^{n-3}\left[\tfrac{n-1}{2}\varphi_1\left(a_2\tilde{\phi}+\tilde{\phi}_2\right)+\tilde{\sigma}_e^2\left(a_2\tilde{\phi}_1+\tilde{\phi}_3\right)\right] \\
\tilde{u}_\theta-\tfrac{d\tilde{u}_r}{d\theta}=\tfrac{1}{n}\tfrac{d\tilde{u}_r}{d\theta}-3\,\tfrac{n+1}{n}\tilde{\sigma}_e^{n-1}(1-s)\tilde{\phi}_1 \\
\tilde{u}_r+\tfrac{d\tilde{u}_\theta}{d\theta}=(s-2)\tilde{u}_r
\end{cases}
\tag{2.3.21}
$$

***II) Plane stress***

Analogous to the analysis for the plane strain condition, the equation governing the stress function for the plane stress case is,

$$
\begin{aligned}
&\nabla^2\nabla^2\phi+\tfrac{\alpha_0}{2}\left\{r^{-1}\tfrac{\partial^2}{\partial r^2}\left[\sigma_e^{n-1}\left(2r\tfrac{\partial^2\phi}{\partial r^2}-\tfrac{\partial\phi}{\partial r}-r^{-1}\tfrac{\partial^2\phi}{\partial\theta^2}\right)\right]\right. \\
&+6r^{-2}\tfrac{\partial^2}{\partial r\partial\theta}\left[\sigma_e^{n-1}r\tfrac{\partial}{\partial r}\left(r^{-1}\tfrac{\partial\phi}{\partial\theta}\right)\right] \\
&+r^{-1}\tfrac{\partial}{\partial r}\left[\sigma_e^{n-1}\left(-2r^{-1}\tfrac{\partial\phi}{\partial r}-2r^{-2}\tfrac{\partial^2\phi}{\partial\theta^2}+\tfrac{\partial^2\phi}{\partial r^2}\right)\right] \\
&\left.+r^{-2}\tfrac{\partial^2}{\partial\theta^2}\left[\sigma_e^{n-1}\left(-\tfrac{\partial^2\phi}{\partial r^2}+2r^{-1}\tfrac{\partial\phi}{\partial r}+2r^{-2}\tfrac{\partial^2\phi}{\partial\theta^2}\right)\right]\right\}=0
\end{aligned}
\tag{2.3.22}
$$

where, the bi-harmonic term $\nabla^4\phi$ (which represents the elastic strain contribution) can be omitted, since it has an exponent $r^{s-2}$, whereas the first term of the plastic part contains $r^{ns-2}$. If $n > 0$, we have $s-2 >$ $ns-2$. Thus, the first term of the plastic component has higher singularity than that of the elastic component, meaning that the bi-harmonic term can be reasonably omitted when $r \to 0$,

$$\begin{aligned}\left[n(s-2)-\frac{d^2}{d\theta^2}\right]\cdot\left[\tilde{\sigma}_e^{n-1}\left(s(s-3)\tilde{\phi}-2\frac{d^2\tilde{\phi}}{d\theta^2}\right)\right]\\ +n(s-2)[n(s-2)+1]\cdot\tilde{\sigma}_e^{n-1}\left[s(2s-3)\tilde{\phi}-\frac{d^2\tilde{\phi}}{d\theta^2}\right]\\ +6(s-1)[n(s-2)+1]\frac{d}{d\theta}\left(\tilde{\sigma}_e^{n-1}\cdot\frac{d\tilde{\phi}}{d\theta}\right)=0\end{aligned} \tag{2.3.23}$$

where,

$$\tilde{\sigma}_e^2=s^2(s^2-3s+3)\tilde{\phi}^2+3(s-1)^2\tilde{\phi}_1^2+\tilde{\phi}_2^2+s(3-s)\tilde{\phi}\tilde{\phi}_2 \tag{2.3.24}$$

The above eigenvalue equation for $s$ can be transformed into a system of linear ordinary differential equations that are functions of $s$, $\tilde{\phi}$ and its derivatives, *i.e.*,

$$\begin{cases}\frac{d\tilde{\phi}}{d\theta}=\tilde{\phi}_1\\ \frac{d\tilde{\phi}_1}{d\theta}=\tilde{\phi}_2\\ \frac{d\tilde{\phi}_2}{d\theta}=\tilde{\phi}_3\\ \frac{d\tilde{\phi}_3}{d\theta}=\psi(\tilde{\phi},\tilde{\phi}_1,\tilde{\phi}_2,\tilde{\phi}_3)\end{cases} \tag{2.3.25}$$

where,

$$\begin{aligned}\psi(\tilde{\phi},\tilde{\phi}_1,\tilde{\phi}_2,\tilde{\phi}_3)=\left(b_{10}\tilde{\phi}^2+b_{11}\tilde{\phi}_1^2+b_{12}\tilde{\phi}_2^2+b_{13}\tilde{\phi}\tilde{\phi}_2\right)^{-1}\\ \times\left\{\tilde{\sigma}_e^2\left(b_1\tilde{\phi}+b_2\tilde{\phi}_2\right)+\left(b_3\tilde{\phi}+b_4\tilde{\phi}_2\right)[b_5\varphi_1^2/\tilde{\sigma}_e^2\right.\\ +b_6\left(b_1\tilde{\phi}_1^2+\tilde{\phi}\tilde{\phi}_2\right)+b_7\tilde{\phi}_2^2+2\tilde{\phi}_3^2+b_8\tilde{\phi}_1\tilde{\phi}_3]\\ \left.+\psi_1\left[b_9\tilde{\phi}_1-2(n-1)\tilde{\phi}_3\right]\right\}\end{aligned} \tag{2.3.26}$$

with,

$$\begin{aligned}\psi_1=2s^2(s^2-3s+3)\tilde{\phi}\tilde{\phi}_1+(5s^2-9s+6)\tilde{\phi}_1\tilde{\phi}_2\\ +2\tilde{\phi}_2\tilde{\phi}_3+s(3-s)\tilde{\phi}\tilde{\phi}_3\end{aligned} \tag{2.3.27}$$

where,

$$
\begin{aligned}
&b_1 = -ns(s-2)^2[3+n(2s-3)],\\
&b_2 = s(s-3)+n(s-3)[3+n(s-2)]-6(s-1)[1+n(s-2)]\\
&b_3 = s(s-3)(n-1)/2, b_4 = 1-n,\\
&b_5 = (n-3)/2, b_6 = 2s^2(s^2-3s+3),\\
&b_7 = 5s^2-9s+6, b_8 = 4s^2-6s+6,\\
&b_9 = (n-1)\{s(s-3)-3(s-1)[n(s-2)+1]\},\\
&b_{10} = 2s^2(s^2-3s+3)+s^2(s-3)^2(n-1)/2,\\
&b_{11} = 6(1-s)^2, b_{12} = 2n, b_{13} = 2ns(3-s)
\end{aligned}
\tag{2.3.28}
$$

The dimensionless radial and tangential displacements are,

$$
\begin{cases}
\tilde{u}_r(\theta) = \frac{1}{n(s-2)+1}\tilde{\sigma}_e^{n-1}\left\{s(3-s)\frac{\tilde{\phi}}{2}+\tilde{\phi}_2\right\}\\
\tilde{u}_\theta(\theta) = \frac{n+1}{n}\left[\frac{d\tilde{u}_r(\theta)}{d\theta}-3\tilde{\sigma}_e^{n-1}(1-s)\tilde{\phi}_1\right]
\end{cases}
\tag{2.3.29}
$$

where,

$$
\begin{cases}
\frac{d\tilde{u}_r(\theta)}{d\theta} = \frac{\tilde{\sigma}_e^{n-1}}{n(s-2)+1}\left\{\frac{(n-1)\psi_1}{2\tilde{\sigma}_e^2}\left[\frac{s(3-s)}{2}\tilde{\phi}+\tilde{\phi}_2\right]+\frac{s(3-s)}{2}\tilde{\phi}_1+\tilde{\phi}_3\right\}\\
\frac{d\tilde{u}_\theta(\theta)}{d\theta} = \tilde{\sigma}_e^{n-1}\left[s\left(s-\frac{3}{2}\right)\tilde{\phi}-\frac{1}{2}\tilde{\phi}_2\right]-\tilde{u}_r(\theta)
\end{cases}
\tag{2.3.30}
$$

By substituting the above components of the dimensionless displacements into the following scaling integral, one can have,

$$
I_n(\theta,\theta^*,n) = \int_{-\pi}^{\pi}\Omega(\theta,\theta^*,n)\mathrm{d}\theta
\tag{2.3.31}
$$

and,

$$
\begin{aligned}
\Omega(\theta,\theta^*,n) = \frac{n}{n+1}\tilde{\sigma}_e^{n+1}\cos\theta\\
-\left[\tilde{\sigma}_{rr}\left(\tilde{u}_\theta-\frac{\mathrm{d}\tilde{u}_r}{\mathrm{d}\theta}\right)-\tilde{\sigma}_{r\theta}\left(\tilde{u}_r+\frac{\mathrm{d}\tilde{u}_\theta}{\mathrm{d}\theta}\right)\right]\sin\theta\\
-\frac{1}{n+1}(\tilde{\sigma}_{rr}\tilde{u}_r+\tilde{\sigma}_{r\theta}\tilde{u}_\theta)\cos\theta
\end{aligned}
\tag{2.3.32}
$$

Shih [166] has demonstrated that plastic SIF $K_M^p$ in pure mode-I (or pure mode-II) can be expressed directly in terms of the corresponding elastic SIF (*i.e.*, $K_I$ and $K_{II}$) using Rice's ***J***-integral. Similarly, in mixed mode-I+II small-scale yielding, $K_M^p$ can be also expressed in terms of the ***J***-integral,

$$
J = \frac{(K_I^2+K_{II}^2)}{E'} = \frac{\alpha_0\sigma_0^2}{E'}I_n(\theta^*)\left(K_M^p\right)^{n+1}
\tag{2.3.33}
$$

and,

$$K_M^p = \left[\frac{(K_I^2 + K_{II}^2)}{\alpha_0 \sigma_0^2 I_n(\theta^*)}\right]^{1/(n+1)} \tag{2.3.34}$$

where, $E' = E$ for plane stress, $E' = E/(1-\upsilon^2)$ for plane strain, $\theta^*$ is the crack growth direction angle and $\alpha$ and $n$ are hardening parameters defined in **Eq**.(2.3.1)~(2.3.4). Elastic SIFs $K_I$ and $K_{II}$ are defined as,

$$\begin{cases} K_I = \frac{\sigma\sqrt{\pi a}}{2}[1 + \eta - (1-\eta)\cos 2\Theta] \\ K_{II} = \frac{\sigma\sqrt{\pi a}}{2}(1-\eta)\sin 2\Theta \end{cases} \tag{2.3.35}$$

for the macro configuration under stress-controlled loading schematically shown in **Fig. 3** by Shlyannikov [165], while the $K_I$ and $K_{II}$ are constrained by the solution of $K_r^p$ in Phase-I in present study. Thus, we can have the plastic SIF, $K_M^p = \left[\frac{K_r^2}{\alpha_0 \sigma_0^2 I_n(\theta^*)}\right]^{1/(n+1)}$, by substituting **Eq**.(2.1.39) into **Eq**.(2.3.34). However, the configuration depicted in **Fig. 3** remains the most commonly employed when exploring more generalized cases in both scientific and engineering contexts.

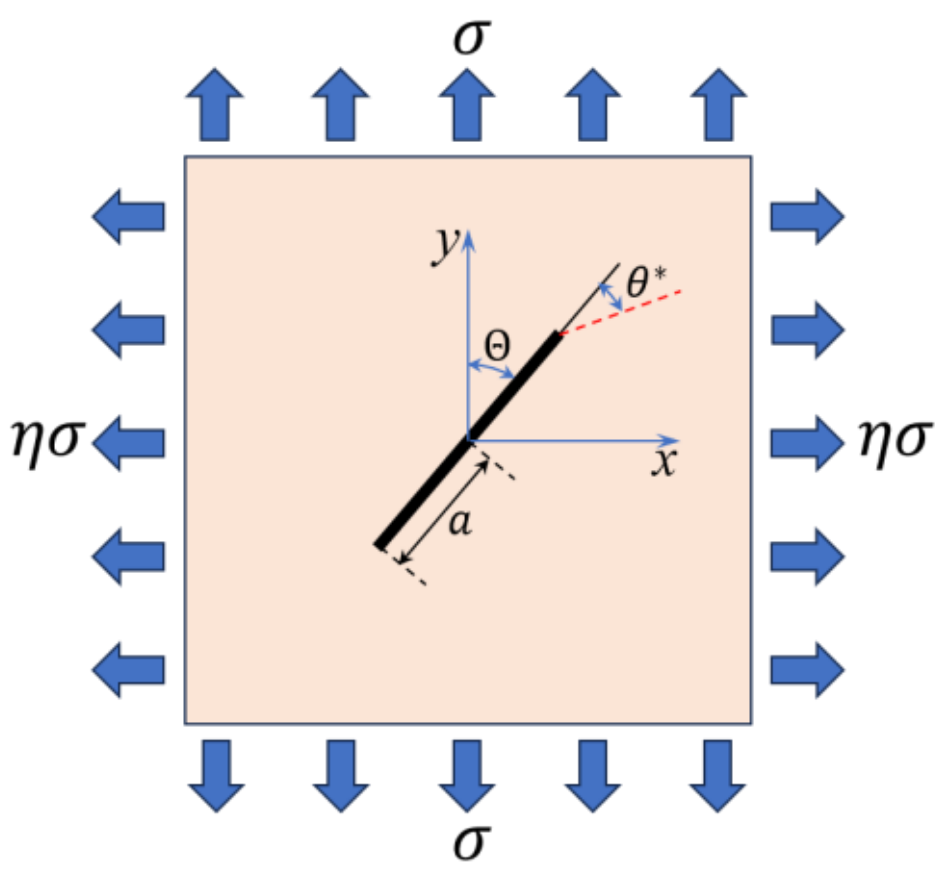

**Fig. 3**. Global coordinate system and loading pattern [165].

***III) Numerical iterative method for solving the nonlinear eigenvalue problems***

To solve the nonlinear eigenvalue equations, the shooting and finite-difference methods are applied iteratively. However, the shooting method and its variants were inadequate for mixed mode analysis. A more correct method employing the finite difference procedure was proposed by Shlyannikov [167] and highly accurate solutions with very rapid convergence were obtained. The finite-difference counterpart for **Eq**s.(2.3.17) and **Eq**s.(2.3.25) as well is introduced as,

$$\begin{cases} \tilde{\phi}_{i+1} - \tilde{\phi}_i = \frac{\Delta\theta_i}{2}(\tilde{\phi}_{1,i+1} + \tilde{\phi}_{1,i}) \\ \tilde{\phi}_{1,i+1} - \tilde{\phi}_{1,i} = \frac{\Delta\theta_i}{2}(\tilde{\phi}_{2,i+1} + \tilde{\phi}_{2,i}) \\ \tilde{\phi}_{2,i+1} - \tilde{\phi}_{2,i} = \frac{\Delta\theta_i}{2}(\tilde{\phi}_{3,i+1} + \tilde{\phi}_{3,i}) \\ \tilde{\phi}_{3,i+1} - \tilde{\phi}_{3,i} = \frac{\Delta\theta_i}{2}(\varphi_{i+1} + \varphi_i) \end{cases} \tag{2.3.36}$$

where, $i = 1, \ldots, i_{max}$ with $i_{max}$ is the number of node points within the interval $-\pi \le \theta \le \pi$. The step size of integration $\Delta\theta_i = \theta_{i+1} - \theta_i$ was varied in an appropriate manner. The solution of both the nonlinear equation systems **Eq**s.(2.3.17) and **Eq**s.(2.3.25) thus transfers to find the roots for the algebraic system **Eq**s.(2.3.36).

Before describing the solution of **Eq**s.(2.3.17) and **Eq**s.(2.3.25), we would formulate the Cauchy problem for the same system. Assuming an approximate solution for the *v-th* iteration to be known, the last equation of systems **Eq**s.(2.3.17) and **Eq**s.(2.3.25) can be treated as one nonlinear equation of the relative variable, *i.e.*, we consider the *i*+1 node in the (*v*+1)-*th* iteration,

$$\begin{gathered} \Phi(\tilde{\phi}_{3,i+1}^{v+1}) \equiv \tilde{\phi}_{3,i+1}^{v+1} - \tilde{\phi}_{3,i} - \{\varphi_{i+1}(\tilde{\phi}_{i+1}(\tilde{\phi}_{3,i+1}^{v+1}), \tilde{\phi}_{1,i+1}(\tilde{\phi}_{3,i+1}^{v+1}), \tilde{\phi}_{2,i+1}(\tilde{\phi}_{3,i+1}^{v+1}), \tilde{\phi}_{3,i+1}^{v+1}) \\ + \varphi_i(\tilde{\phi}_i, \tilde{\phi}_{1,i}, \tilde{\phi}_{2,i}, \tilde{\phi}_{3,i})\} \frac{\Delta\theta_i}{2} = 0 \end{gathered} \tag{2.3.37}$$

where, it should be noted that at the *i*+1 node,

$$\begin{cases} \varphi_{i+1} \equiv \varphi_{i+1}(\tilde{\phi}_{i+1}, \tilde{\phi}_{1,i+1}, \tilde{\phi}_{2,i+1}, \tilde{\phi}_{3,i+1}^{v+1}) \\ \tilde{\phi}_{i+1} \equiv \tilde{\phi}_{i+1}(\tilde{\phi}_{3,i+1}^{v+1}) \\ \tilde{\phi}_{1,i+1} \equiv \tilde{\phi}_{1,i+1}(\tilde{\phi}_{3,i+1}^{v+1}) \\ \tilde{\phi}_{2,i+1} \equiv \tilde{\phi}_{2,i+1}(\tilde{\phi}_{3,i+1}^{v+1}) \end{cases} \tag{2.3.38}$$

where, the values $\tilde{\phi}_{i+1}$, $\tilde{\phi}_{1,i+1}$, $\tilde{\phi}_{2,i+1}$ are expressed through the **Eq**.(2.3.17) or **Eq**.(2.3.25),

$$\begin{aligned} \tilde{\phi}_{2,i+1}^{(v+1)} &= \tilde{\phi}_{2,i} + \frac{\Delta\theta_i}{2}\left(\tilde{\phi}_{3,i+1}^{(v+1)} + \tilde{\phi}_{3,i}\right) \\ \tilde{\phi}_{1,i+1}^{(v+1)} &= \tilde{\phi}_{1,i} + \frac{\Delta\theta_i}{2}\left[2\tilde{\phi}_{2,i} + \frac{\Delta\theta_i}{2}\left(\tilde{\phi}_{3,i+1}^{(v+1)} + \tilde{\phi}_{3,i}\right)\right] \\ \tilde{\phi}_{i+1}^{(v+1)} &= \tilde{\phi}_i + \frac{\Delta\theta_i}{2}\left\{2\tilde{\phi}_{1,i} + \frac{\Delta\theta_i}{2}\left[2\tilde{\phi}_{2,i} + \frac{\Delta\theta_i}{2}\left(\tilde{\phi}_{3,i+1}^{(v+1)} + \tilde{\phi}_{3,i}\right)\right]\right\} \end{aligned} \tag{2.3.39}$$

The above **Eq**.(2.3.39) is solved by the Newton method with details demonstrated in Ref. [167]. Then, inspired by the ***MSG*** plasticity theory [168], we can derive the dislocation density field from the strain field near the crack tip as follows:

***I) Plane strain***

The strain field is given by,

$$\begin{cases}\varepsilon_{rr} = (\sigma_{rr} - \upsilon\sigma_{\theta\theta})/E + \frac{3}{4}\alpha_0 \left(\frac{\sigma_e}{\sigma_0}\right)^{n-1} (\sigma_{rr} - \sigma_{\theta\theta})/E \\ \varepsilon_{\theta\theta} = (\sigma_{\theta\theta} - \upsilon\sigma_{rr})/E + \frac{3}{4}\alpha_0 \left(\frac{\sigma_e}{\sigma_0}\right)^{n-1} (\sigma_{\theta\theta} - \sigma_{rr})/E \\ \varepsilon_{r\theta} = (1+\upsilon)\sigma_{r\theta}/E + \frac{3}{2}\alpha_0 \left(\frac{\sigma_e}{\sigma_0}\right)^{n-1} \sigma_{r\theta}/E \end{cases} \tag{2.3.40}$$

In the asymptotic sense, the elastic component can be omitted compared with the plastic one in the strain field, thus, we have,

$$\begin{cases}\varepsilon_{rr} \cong \varepsilon_{rr}^{p} = \frac{3}{4}\alpha_0 \left(\frac{\sigma_e}{\sigma_0}\right)^{n-1} (\sigma_{rr} - \sigma_{\theta\theta})/E \\ \varepsilon_{\theta\theta} \cong \varepsilon_{\theta\theta}^{p} = \frac{3}{4}\alpha_0 \left(\frac{\sigma_e}{\sigma_0}\right)^{n-1} (\sigma_{\theta\theta} - \sigma_{rr})/E \\ \varepsilon_{r\theta} \cong \varepsilon_{r\theta}^{p} = \frac{3}{2}\alpha_0 \left(\frac{\sigma_e}{\sigma_0}\right)^{n-1} \sigma_{r\theta}/E \end{cases} \tag{2.3.41}$$

where, the dominant singularity solution for the mixed-mode elastic-plastic stress distribution is,

$$\begin{cases}\sigma_{rr} = K_M^p r^{s-2}\tilde{\sigma}_{rr}(\theta) = K_M^p r^{s-2}\left(s\tilde{\phi} + \tilde{\phi}_2\right) \\ \sigma_{\theta\theta} = K_M^p r^{s-2}\tilde{\sigma}_{\theta\theta}(\theta) = K_M^p r^{s-2} \cdot s \cdot (1-s)\tilde{\phi} \\ \sigma_{r\theta} = K_M^p r^{s-2}\tilde{\sigma}_{r\theta}(\theta) = K_M^p r^{s-2}(1-s)\tilde{\phi}_1 \end{cases} \tag{2.3.42}$$

and, the effective stress is,

$$\sigma_e = K_M^p r^{s-2}\tilde{\sigma}_e(\theta) \tag{2.3.43}$$

where, the angular distribution function $\tilde{\sigma}_e(\theta)$ was already defined in **Eq.**(2.3.15) previously. Thus, the plastic strain field can be rewritten as,

$$\begin{cases}\varepsilon_{rr}^{p} = \frac{3\alpha_0 K_M^p}{4E}\left(\frac{K_M^p}{\sigma_0}\right)^{n-1} r^{n(s-2)} f_{rr}(\theta) \\ \varepsilon_{\theta\theta}^{p} = \frac{3\alpha_0 K_M^p}{4E}\left(\frac{K_M^p}{\sigma_0}\right)^{n-1} r^{n(s-2)} f_{\theta\theta}(\theta) \\ \varepsilon_{r\theta}^{p} = \frac{3\alpha_0 K_M^p}{2E}\left(\frac{K_M^p}{\sigma_0}\right)^{n-1} r^{n(s-2)} f_{r\theta}(\theta) \end{cases} \tag{2.3.44}$$

where, the angular distribution functions are,

$$\begin{cases}f_{rr}(\theta) = \tilde{\sigma}_e^{n-1}\left(s^2\tilde{\phi} + \tilde{\phi}_2\right) \\ f_{\theta\theta}(\theta) = -f_{rr}(\theta) \\ f_{r\theta}(\theta) = \tilde{\sigma}_e^{n-1}(1-s)\tilde{\phi}_1 \end{cases} \tag{2.3.45}$$

***II) Plane stress***

Similarly, we have the strain field as follows,

$$\begin{cases} \varepsilon_{rr} = (\sigma_{rr} - \upsilon\sigma_{\theta\theta})/E + \alpha_0 \left(\frac{\sigma_e}{\sigma_0}\right)^{n-1} \left(\sigma_{rr} - \frac{\sigma_{\theta\theta}}{2}\right)/E \\ \varepsilon_{\theta\theta} = (\sigma_{\theta\theta} - \upsilon\sigma_{rr})/E + \alpha_0 \left(\frac{\sigma_e}{\sigma_0}\right)^{n-1} \left(\sigma_{\theta\theta} - \frac{\sigma_{rr}}{2}\right)/E \\ \varepsilon_{r\theta} = \sigma_{r\theta}/2\mu + \frac{3}{2}\alpha_0 \left(\frac{\sigma_e}{\sigma_0}\right)^{n-1} \sigma_{r\theta}/E \end{cases} \tag{2.3.46}$$

where, the plastic component is,

$$\begin{cases} \varepsilon_{rr}^p = \alpha_0 \left(\frac{\sigma_e}{\sigma_0}\right)^{n-1} \left(\sigma_{rr} - \frac{\sigma_{\theta\theta}}{2}\right)/E \\ \varepsilon_{\theta\theta}^p = \alpha_0 \left(\frac{\sigma_e}{\sigma_0}\right)^{n-1} \left(\sigma_{\theta\theta} - \frac{\sigma_{rr}}{2}\right)/E \\ \varepsilon_{r\theta}^p = \frac{3}{2}\alpha_0 \left(\frac{\sigma_e}{\sigma_0}\right)^{n-1} \sigma_{r\theta}/E \end{cases} \tag{2.3.47}$$

where, the stress field take the same form as in **Eq.**(2.3.42), while its $\tilde{\sigma}_e(\theta)$ was defined in **Eq.**(2.3.24) previously. thus, the plastic strain field can be rewritten as,

$$\begin{cases} \varepsilon_{rr}^p = \frac{\alpha_0 K_M^p}{E} \left(\frac{K_M^p}{\sigma_0}\right)^{n-1} r^{n(s-2)} g_{rr}(\theta) \\ \varepsilon_{\theta\theta}^p = \frac{\alpha_0 K_M^p}{E} \left(\frac{K_M^p}{\sigma_0}\right)^{n-1} r^{n(s-2)} g_{\theta\theta}(\theta) \\ \varepsilon_{r\theta}^p = \frac{3\alpha_0 K_M^p}{2E} \left(\frac{K_M^p}{\sigma_0}\right)^{n-1} r^{n(s-2)} g_{r\theta}(\theta) \end{cases} \tag{2.3.48}$$

where, the angular distribution functions are,

$$\begin{cases} g_{rr}(\theta) = \tilde{\sigma}_e^{n-1} \left[\frac{s(1+s)}{2}\tilde{\phi} + \tilde{\phi}_2\right] \\ g_{\theta\theta}(\theta) = \tilde{\sigma}_e^{n-1} \left[s\left(\frac{1}{2} - s\right)\tilde{\phi} - \frac{\tilde{\phi}_2}{2}\right] \\ g_{r\theta}(\theta) = \tilde{\sigma}_e^{n-1}(1-s)\tilde{\phi}_1 \end{cases} \tag{2.3.49}$$

***III) Analytical evaluation of dislocation density***

In the ***MSG*** plasticity theory [168, 169], the effective plastic strain gradient is defined as [170],

$$\eta^p = \sqrt{c_1\eta_{iik}^p\eta_{jjk}^p + c_2\eta_{ijk}^p\eta_{ijk}^p + c_3\eta_{ijk}^p\eta_{kji}^p} \tag{2.3.50}$$

where, the coefficients are determined to be equal to $c_1 = 0$, $c_2 = 1/4$ and $c_3 = 0$. Thus, the above **Eq**.(2.3.50) can be recast as,

$$\eta^p = \frac{1}{2}\sqrt{\boldsymbol{\eta}^p \boldsymbol{\eta}^p} \tag{2.3.51}$$

where, the components of the third-order strain gradient tensor are obtained as,

$$\eta^p_{ijk} = \nabla\varepsilon^p_{ik,j} + \nabla\varepsilon^p_{jk,i} - \nabla\varepsilon^p_{ij,k} \text{ ; with } i,j,k \in \{r,\theta\} \tag{2.3.52}$$

where, $\nabla$ denotes the covariant derivative appropriate for physical components in the cylindrical coordinates system ($r$, $\theta$, $z$), which is a generalization of two-dimensional polar coordinates system ($r$, $\theta$). Fleck and Hutchinson have given the strain gradient tensor for the ***HRR*** fields as follows [169],

$$\eta_{ijk} = \ell^{-1}\mathscr{Y}_0\left(\frac{J}{\Sigma_0\mathscr{Y}_0 C_n r}\right)^{n/(n+1)}\tilde{\eta}_{ijk}(\theta,n) \tag{2.3.53}$$

However, they did not present the details of each component. Here, referring to the formulation for the gradient of a tensor field detailed in **Appendix B**, I present the 6 in-plane components of the plastic strain gradient tensor as follows,

$$\begin{cases} \eta^p_{rrr} = \dfrac{\partial\varepsilon^p_{rr}}{\partial r} \\ \eta^p_{\theta\theta r} = 2\left(\dfrac{\partial\varepsilon^p_{r\theta}}{r\partial\theta} + \dfrac{\varepsilon^p_{rr}-\varepsilon^p_{\theta\theta}}{r}\right) - \dfrac{\partial\varepsilon^p_{\theta\theta}}{\partial r} \\ \eta^p_{r\theta r} = \dfrac{\partial\varepsilon^p_{rr}}{r\partial\theta} - \dfrac{2\varepsilon^p_{r\theta}}{r} \\ \eta^p_{rr\theta} = 2\dfrac{\partial\varepsilon^p_{r\theta}}{\partial r} - \left(\dfrac{\partial\varepsilon^p_{rr}}{r\partial\theta} - \dfrac{2\varepsilon^p_{r\theta}}{r}\right) \\ \eta^p_{\theta\theta\theta} = \dfrac{\partial\varepsilon^p_{\theta\theta}}{r\partial\theta} + \dfrac{2\varepsilon^p_{r\theta}}{r} \\ \eta^p_{r\theta\theta} = \dfrac{\partial\varepsilon^p_{\theta\theta}}{\partial r} \end{cases} \tag{2.3.54}$$

***a) Plane strain***

Explicitly, let $\beta_0 = \frac{3\alpha_0 K^p_M}{4E}\left(\frac{K^p_M}{\sigma_0}\right)^{n-1}$ and $m = n(s-2)$ for plane strain, we have,

$$\begin{cases} \eta^p_{rrr} = m\beta_0 r^{m-1} f_{rr}(\theta) \\ \eta^p_{\theta\theta r} = \beta_0 r^{m-1}[4f'_{r\theta}(\theta) + f_{rr}(\theta) - (m+1)f_{\theta\theta}(\theta)] \\ \eta^p_{r\theta r} = \beta_0 r^{m-1}[f'_{rr}(\theta) - 4f_{r\theta}(\theta)] \\ \eta^p_{rr\theta} = \beta_0 r^{m-1}[4(m+1)f_{r\theta}(\theta) - f'_{rr}(\theta)] \\ \eta^p_{\theta\theta\theta} = \beta_0 r^{m-1}[f'_{\theta\theta}(\theta) + 4f_{r\theta}(\theta)] \\ \eta^p_{r\theta\theta} = m\beta_0 r^{m-1} f_{\theta\theta}(\theta) \end{cases} \tag{2.3.55}$$

where, the first-order derivatives of angular functions are,

$$\begin{aligned} f'_{rr}(\theta) = {} & \left[\tfrac{3}{4}\left(a_2\tilde{\phi} + \tilde{\phi}_2\right)^2 + a_4\tilde{\phi}_1^2\right]^{(n-1)/2}\left(s^2\tilde{\phi}_1 + \tilde{\phi}_3\right) \\ & + \tfrac{n-1}{2}\left(s^2\tilde{\phi} + \tilde{\phi}_2\right)\left[\tfrac{3}{4}\left(a_2\tilde{\phi} + \tilde{\phi}_2\right)^2 + a_4\tilde{\phi}_1^2\right]^{(n-3)/2} \\ & \times\left[\tfrac{3}{2}\left(a_2\tilde{\phi} + \tilde{\phi}_2\right)\left(a_2\tilde{\phi}_1 + \tilde{\phi}_3\right) + 2a_4\tilde{\phi}_1\tilde{\phi}_2\right] \end{aligned} \tag{2.3.56}$$

$$f'_{\theta\theta}(\theta) = -f'_{rr}(\theta) \tag{2.3.57}$$

$$\begin{aligned} f'_{r\theta}(\theta) = {} & (1-s)\tilde{\phi}_2\left[\tfrac{3}{4}\left(a_2\tilde{\phi} + \tilde{\phi}_2\right)^2 + a_4\tilde{\phi}_1^2\right]^{(n-1)/2} \\ & + \tfrac{n-1}{2}(1-s)\tilde{\phi}_1\left[\tfrac{3}{4}\left(a_2\tilde{\phi} + \tilde{\phi}_2\right)^2 + a_4\tilde{\phi}_1^2\right]^{(n-3)/2} \\ & \times\left[\tfrac{3}{2}\left(a_2\tilde{\phi} + \tilde{\phi}_2\right)\left(a_2\tilde{\phi}_1 + \tilde{\phi}_3\right) + 2a_4\tilde{\phi}_1\tilde{\phi}_2\right] \end{aligned} \tag{2.3.58}$$

***b) Plane stress***

Similarly, let $\beta_0 = \frac{3\alpha_0 K_M^p}{4E}\left(\frac{K_M^p}{\sigma_0}\right)^{n-1}$ and $m = n(s-2)$ for plane stress, we have,

$$\begin{cases} \eta^p_{rrr} = m\beta_0 r^{m-1} g_{rr}(\theta) \\ \eta^p_{\theta\theta r} = \beta_0 r^{m-1}[3g'_{r\theta}(\theta) + g_{rr}(\theta) - (m+1)g_{\theta\theta}(\theta)] \\ \eta^p_{r\theta r} = \beta_0 r^{m-1}[g'_{rr}(\theta) - 3g_{r\theta}(\theta)] \\ \eta^p_{rr\theta} = \beta_0 r^{m-1}[3(m+1)g_{r\theta}(\theta) - g'_{rr}(\theta)] \\ \eta^p_{\theta\theta\theta} = \beta_0 r^{m-1}[g'_{\theta\theta}(\theta) + 3g_{r\theta}(\theta)] \\ \eta^p_{r\theta\theta} = m\beta_0 r^{m-1} g_{\theta\theta}(\theta) \end{cases} \tag{2.3.59}$$

where, the derivatives of the angular functions are,

$$\begin{aligned} g'_{rr}(\theta) = {} & \left[\tfrac{s(1+s)}{2}\tilde{\phi}_1 + \tilde{\phi}_3\right]\left(l_0\tilde{\phi}^2 + l_1\tilde{\phi}_1^2 + l_2\tilde{\phi}_2^2 + l_3\tilde{\phi}\tilde{\phi}_2\right)^{(n-1)/2} \\ & + \tfrac{n-1}{2}\left[\tfrac{s(1+s)}{2}\tilde{\phi} + \tilde{\phi}_2\right]\left(l_0\tilde{\phi}^2 + l_1\tilde{\phi}_1^2 + l_2\tilde{\phi}_2^2 + l_3\tilde{\phi}\tilde{\phi}_2\right)^{(n-3)/2} \\ & \times\left[2l_0\tilde{\phi}\tilde{\phi}_1 + (2l_1 + l_3)\tilde{\phi}_1\tilde{\phi}_2 + 2l_2\tilde{\phi}_2\tilde{\phi}_3 + l_3\tilde{\phi}\tilde{\phi}_3\right] \end{aligned} \tag{2.3.60}$$

$$g'_{\theta\theta}(\theta) = \left[s\left(\frac{1}{2}-s\right)\tilde{\phi}_1 - \frac{\tilde{\phi}_3}{2}\right]\left(l_0\tilde{\phi}^2 + l_1\tilde{\phi}_1^2 + l_2\tilde{\phi}_2^2 + l_3\tilde{\phi}\tilde{\phi}_2\right)^{(n-1)/2}$$
$$+\frac{n-1}{2}\left[s\left(\frac{1}{2}-s\right)\tilde{\phi} - \frac{\tilde{\phi}_2}{2}\right]\left(l_0\tilde{\phi}^2 + l_1\tilde{\phi}_1^2 + l_2\tilde{\phi}_2^2 + l_3\tilde{\phi}\tilde{\phi}_2\right)^{(n-3)/2} \tag{2.3.61}$$
$$\times\left[2l_0\tilde{\phi}\tilde{\phi}_1 + (2l_1 + l_3)\tilde{\phi}_1\tilde{\phi}_2 + 2l_2\tilde{\phi}_2\tilde{\phi}_3 + l_3\tilde{\phi}\tilde{\phi}_3\right]$$

$$g'_{r\theta}(\theta) = (1-s)\tilde{\phi}_2\left(l_0\tilde{\phi}^2 + l_1\tilde{\phi}_1^2 + l_2\tilde{\phi}_2^2 + l_3\tilde{\phi}\tilde{\phi}_2\right)^{(n-1)/2}$$
$$+\frac{n-1}{2}(1-s)\tilde{\phi}_1\left(l_0\tilde{\phi}^2 + l_1\tilde{\phi}_1^2 + l_2\tilde{\phi}_2^2 + l_3\tilde{\phi}\tilde{\phi}_2\right)^{(n-3)/2} \tag{2.3.62}$$
$$\times\left[2l_0\tilde{\phi}\tilde{\phi}_1 + (2l_1 + l_3)\tilde{\phi}_1\tilde{\phi}_2 + 2l_2\tilde{\phi}_2\tilde{\phi}_3 + l_3\tilde{\phi}\tilde{\phi}_3\right]$$

with the coefficients as,

$$\begin{cases} l_0 = s^2(s^2 - 3s + 3) \\ l_1 = 3(s-1)^2 \\ l_2 = 1 \\ l_3 = s(3-s) \end{cases} \tag{2.3.63}$$

Therefore, in the numerical implementation of plane problems, we can directly write the effective plastic strain gradient $\eta^p$ as,

$$\eta^p = \frac{1}{2}\sqrt{\left(\eta^p_{rrr}\right)^2 + \left(\eta^p_{\theta\theta r}\right)^2 + \left(\eta^p_{r\theta r}\right)^2 + \left(\eta^p_{rr\theta}\right)^2 + \left(\eta^p_{\theta\theta\theta}\right)^2 + \left(\eta^p_{r\theta\theta}\right)^2} \tag{2.3.64}$$

Then, based on the ***MSG*** plasticity theory, the density of geometrically necessary dislocations (***GND***s) $\rho_G$ (required for the compatible deformation of crystals) is related to the effective plastic strain gradient $\eta^p$ by,

$$\rho_G = \tilde{r}\eta^p/b \tag{2.3.65}$$

where, $\tilde{r}$ is the Nye-factor, which is assumed to be approximately 1.90 to account for the underlying crystalline anisotropy of fcc metals [170-172].

The density of statistically stored dislocations (***SSD***s) $\rho_S$ (trapping each other in a random way) is [170],

$$\rho_S = \left(\frac{\sigma_{ref} f(\varepsilon^p)}{M\alpha G b}\right)^2 \tag{2.3.66}$$

where, $\sigma_{ref}$ is a reference stress, $f(\varepsilon^p)$ is a non-dimensional function of the equivalent plastic strain $\varepsilon^p$, as given from the uniaxial stress-strain curve, $M$ is the Taylor factor, $\alpha$ is an empirical coefficient taking values of [0.3, 0.5], $G$ is the shear modulus, and $b$ is the Burgers vector. Hence, the total dislocation density is evaluated as,

$$\rho_D = \rho_G + \rho_S \tag{2.3.67}$$

It is noted that in another analytical work for pure mode-I loading by Thomson [128], the dislocation density field near the crack tip is given as,

$$\rho_D = \frac{\epsilon_0}{b}\left(\frac{K_I^2 \cos\theta}{\pi\sigma_0^2}\right)^{n/(1+n)} \frac{\cos 2\theta}{r^{1+n/(1+n)}} + \frac{3}{2b}\frac{n\epsilon_0}{1+n}\left(\frac{K_I^2}{\pi\sigma_0^2}\right)^{n/(1+n)} \frac{\sin\theta \sin 2\theta}{\cos^{1/(1+n)}\theta}\frac{1}{r^{1+n/(1+n)}} \tag{2.3.68}$$

where, $\sigma_0$, $\epsilon_0$, $n$ are parameters defined in the following strain hardening constitution,

$$\sigma/\sigma_0 = (\epsilon/\epsilon_0)^n, \epsilon > \epsilon_0 \tag{2.3.69}$$

Previous experiments [173-175] have suggested that the plastic deformation produces the vacancy supersaturation, with concentrations as large as $10^{-3}$. As reviewed by Noell *et al*. [176], there exit two general theories for how vacancies are produced during plastic deformation. While the first mechanism proposed by Saada [177, 178], focuses on the annihilation of non-screw dislocations of different glide planes, the second one involves the formation of Burgers-vector-sized edge dislocation segments on screw dislocations after they intersect and cut through forest dislocations [162], and was verified by large-scale atomistic simulations [179, 180].

As demonstrated by Militzer *et al*. [181], the net rate of the vacancy concentration $c_V$ is given by, $\mathrm{d}c_v/\mathrm{d}t = \Pi = \chi\Omega_0\sigma\dot{\varepsilon}/Q_f$, where, $\chi$ is a dimensionless constant, $\Omega_0$ is the atomic volume (of the matrix material), $\sigma$ is flow stress, $\dot{\varepsilon}$ is strain rate and $Q_f$ is the vacancy formation energy. While the Taylor relation $\sigma \sim \sqrt{\rho_D}$ [182, 183] and $\dot{\varepsilon} \sim \rho_D$ [21] are well known, it thus indicates that the vacancy generation is strongly dependent on the dislocation density $\rho_D$. However, such naïve considerations do not involve the kinematic annihilation of vacancies due to the interaction with other types of microstructures. According to the classical reaction rate theory for the dynamics of nonequilibrium point defects (vacancies and interstitials in crystalline matrix), the concentrations of vacancies and (hydrogen) interstitials evolve as follows [184],

$$\begin{cases} \dfrac{\partial c_v}{\partial t} = \mathcal{M}_v - D_v\left(S_v^\rho + S^R\right)c_v - \varkappa c_v c_i - \nabla \cdot J_v \\ \dfrac{\partial c_i}{\partial t} = \mathcal{M}_i - D_i\left(S_i^\rho + S^R\right)c_i - \varkappa c_v c_i - \nabla \cdot J_i \end{cases} \tag{2.3.70}$$

where, $\mathcal{M}_v = \mathcal{M}_i = \mathcal{M}$ is the defect production rate (~ $\Pi$ derived above); $D_{v,i} = D_{i,v}^0 e^{-E_{v,i}^m/k_B T}$ are corresponding diffusivities of vacancies and hydrogen interstitials, respectively; $E_{v,i}^m$ represent the defects (vacancy, interstitial) migration energy; $S_{v,i}^\rho$ are dislocation sink intensity for vacancies ($S_v^\rho \equiv Z_{vN}\rho_N + Z_{vV}\rho_V + Z_{vI}\rho_I$ ) and interstitials ( $S_i^\rho \equiv Z_{iN}\rho_N + Z_{iV}\rho_V + Z_{iI}\rho_I$ ), respectively, with $\rho_N$ is network dislocation density (approximately equal to $\rho_D$ defined in **Eq**.(2.3.67)), $\rho_V$ and $\rho_I$ denoting densities of vacancy and interstitial loops, respectively; here, $Z_{\{\cdot,\cdot\}}$ is the preference factor with properties: $Z_{vN} = Z_{vV} = Z_{vI} = 1$, $Z_{iN} = 1 + B$, $Z_{iI} \cong Z_{iV} \cong 1 + B'$, where, $B$ and $B' \geq B$ are excesses for network bias and loop bias ($B \ll 1$); the sink intensity $S^R = 4\pi NR$ is determined through the voids density $N$ and their radius $R$; $\varkappa = 4\pi r_0(D_v + D_i)/\Omega$ is the recombination coefficient defined by the interaction radius of $r_0$ (vacancy loop radius), the atomic volume $\Omega$ and corresponding diffusivities; $J_v$ and $J_i$ are fluxes of vacancies and interstitials, respectively, since these point defects are mobile species.

It should be noted here that the above kinetics equations were originally designed for the irradiation convention by Kharchenko *et al*. [184], where the vacancy and self-interstitial are generated simultaneously as the so-called Frenkel pair, thus have the same production rate $\mathcal{M}$. However, to a first approximation, considering the fact that vacancy and hydrogen concentrations are generally at the same order demonstrated by experiments [185], we can hypothesize that the vacancy concentration ($c_v$) and hydrogen interstitial concentration ($c_i$) in present study have more or less the same production rate. Nonetheless, it is still flexible to write the production rate $\mathcal{M}_i$ in the kinetics equation of hydrogen concentration as a function of the time $t$: $\mathcal{M}_i(t)$, if one knows its exact expression and wants more precise simulations.

Under external mechanical loadings, irradiation, etc., point defects accumulate into voids, growing with the system evolution and resulting in the failure or swelling. The void growth (swelling) rate is $\dot{S} \equiv \frac{\mathrm{d}}{\mathrm{d}t}\frac{\Delta V}{V}$, where at fixed void density $N$, one has $\dot{S} = 4\pi NR^2\dot{R}$. Physically, the void growth rate is attributed to the volume change due to vacancy absorption, $4\pi D_v RNc_v$, the volume change due to interstitial absorption, $-4\pi D_i RNc_i$, and the volume change due to thermal vacancy emission, $-4\pi D_v RN\left(c_v^e(R) - c_v^e(\infty)\right)$; $c_v^e(R) = c_{v0}e^{R_s/R}$ is the equilibrium vacancy concentration near the void of the radius $R$, where, $c_{v0} = e^{-E_v^f/k_BT}$ is the equilibrium vacancy concentration, $E_v^f$ (= 1.53 eV for $\alpha$-Fe [21]) is the vacancy formation energy, $R_s \equiv 2\gamma\Omega/k_BT$, $\gamma$ is the surface energy of the void, $c_v^e(\infty) = c_{v0}$ defines the vacancy concentration near a flat surface. Therefore, the dynamic evolution of the void radius can be described as,

$$\frac{\partial R}{\partial t} = \frac{1}{R}\left\{D_v\left[c_v - \left(c_v^e(R) - c_v^e(\infty)\right)\right] - D_i c_i\right\} \quad (2.3.71)$$

To a first approximation, we temporally do not consider the diffusion of voids. By hypothesizing constant values of dislocation densities, one can consider the **Eq**.(2.3.70) and **Eq**.(2.3.71) as a closed loop system.

Furthermore, by using sink intensities and corresponding diffusivities, one can estimate mean lifetime scales for vacancies and interstitials as $\tau_{v,i} = \left(D_{v,i}S_{v,i}^{\rho}\right)^{-1}$, respectively. Then, by introducing the diffusion length $L_d = \sqrt{D_v\tau_v}$, the scaling factor $\mu \equiv \varkappa\tau_v$, the renormalized density of void sinks $\theta \equiv 4\pi N\Omega^{1/3}/S_v^{\rho}$, one can define $x \equiv \mu c_v$ and $y \equiv \mu c_i$ as renormalized concentrations of point defects dynamically evolving in the dimensionless time $\tau \equiv t/\tau_v$ and space $\boldsymbol{r}' \equiv \boldsymbol{r}/L_d$ with a production rate $P \equiv \mathcal{M}\mu\tau_v$. Thus, the main equations can be reformulated as the following dimensionless form,

$$\begin{cases} \partial_t x = P - (1+\theta R)x - xy - \nabla\cdot J_v \\ \partial_t y = P - \epsilon(1+\kappa\theta R)y - xy - \epsilon\nabla\cdot J_i \\ v\partial_t R = [x - (x(R) - x_0) - \kappa\epsilon y]/R \end{cases} \tag{2.3.72}$$

where, it is noted that $\tau \to t$, $\tilde{R} \equiv R/a \to R$ ($a = \Omega^{1/3}$ is the lattice constant), $\tilde{J}_v \equiv \mu\tau_v J_v \to J_v$ and $\tilde{J}_i \equiv \mu\tau_i J_i \to J_i$, $\tilde{\nabla} \equiv L_d\nabla \to \nabla$. Here the dimensionless parameters are defined as: $\epsilon \equiv \tau_v/\tau_i \gg 1$, $\kappa \equiv S_v^\rho/S_i^\rho < 1$, $v \equiv S_v^\rho \mu a^2$, $x_0 \equiv \mu c_{v0}$, and the dimensionless variable $x(\bar{R}) \equiv \mu c_v^e(R) \to x(R)$.

Physically, the damage rate $P$ always fluctuates in a stochastic manner, *i.e.*, defect populations and therefore, the void radius evolution are controlled by the stochastic dynamics. Following this logic, we thus hypothesize that the defect production rate $P$ fluctuates around its average value $P_0$: $P \to P(t) = P_0 + \xi(t)$, where $\xi(t)$ is the white noise with statistical properties: $\langle\xi(t)\rangle = 0$, $\langle\xi(t)\xi(t')\rangle = 2P_0\Sigma\delta(t-t')$, where $\Sigma$ is the noise intensity, and $\delta(t-t')$ is the Dirac delta function. In present study for dislocation-mediated vacancy condensation, the average defect production rate is calculated as [181],

$$P_0 = \mathcal{M}_0\mu\tau_v = \chi\Omega_0\bar{\sigma}\dot{\bar{\varepsilon}}/E_v^f \tag{2.3.73}$$

where, $\mathcal{M}_0$ is the average value of $\mathcal{M}$, $\bar{\sigma}$ is the average flow stress, $\dot{\bar{\varepsilon}}$ is the average strain rate, and $\Omega_0$ (= $1.18\times10^{-29}$ m$^3$) is the atomic volume. Specifically, the coefficient $\chi$ represents the fraction of mechanical work of the applied stress stored in the generated vacancies, and its magnitude approaches to 0.1 at elevated temperature near $0.5T_m$, where $T_m$ is the melting point [186]. As mentioned above, considering $\bar{\sigma}\sim\sqrt{\bar{\rho}_D}$ and $\dot{\bar{\varepsilon}}\sim\bar{\rho}_D$, the vacancy production rate $P_0$ thus can be expressed as a function of the mean dislocation density $\bar{\rho}_D$: $P_0\sim\bar{\rho}_D^{3/2}$, where $\bar{\rho}_D$ can be evaluated by combining the ***HRR*** and ***MSG*** theories.

In above **Eq**.(2.3.72), it is found that the void radius $R$ is the slowest mode. Thus, by setting $\partial_t x \cong \partial_t y = 0$ and keeping only the most essential contributions, we have $x \cong P/(1+\theta R)$ and $y \cong P/\epsilon(1+\theta R)$ and insert them into the differential equation for $R$, the following equation for void dynamics is arrived,

$$v\partial_t R = \frac{1}{R}\left[\frac{P(1-\kappa)}{1+\theta R} - x_0\left(e^{R_s/R} - 1\right)\right] \tag{2.3.74}$$

Note that $P = P_0 + \xi(t)$, next we take into account that fluctuations are possible only at $P_0 \neq 0$, and put $P \equiv P_0$ for simplicity, *i.e.*,

$$v\partial_t R = \frac{1}{R}\left[\frac{(P_0+\xi(t))(1-\kappa)}{1+\theta R} - x_0\left(e^{R_s/R}-1\right)\right] \tag{2.3.75}$$

Then, we have the ***Langevin equation*** as,

$$v\partial_t R = \frac{1}{R}\left[\frac{P(1-\kappa)}{1+\theta R} - x_0\left(e^{R_s/R}-1\right)\right] + g(R)\xi(t) \tag{2.3.76}$$

where, it is noted that $\bar{R}_s \equiv R_s/a \to R_s$, and $g(R) = \sqrt{1-\kappa}/\big(R(1+\theta R)\big)$ relates to the noise amplitude, satisfying the fluctuation-dissipation theorem [184]. By treating the stochastic dynamics in the Stratonovich sense [184], we can obtain the averaged void size,

$$v\partial_t\langle R\rangle = \langle\frac{1}{R}\left(\frac{P(1-\kappa)}{1+\theta R} - x_0\left(e^{R_s/R}\right)\right)\rangle - \Sigma P(1-\kappa)\,\langle\frac{1+2\theta R}{\big(R(1+\theta R)\big)^3}\rangle \tag{2.3.77}$$

where, the last term relates to the so-called noise-induced drift. Under the macroscopic approximation, we can put $\langle A(\cdot)\rangle \cong A(\langle\cdot\rangle), \forall A$, to estimate the noise effect on the system dynamics.

$$v\partial_t\langle R\rangle = \frac{1}{\langle R\rangle}\left(\frac{P(1-\kappa)}{1+\theta\langle R\rangle} - x_0\left(e^{R_s/\langle R\rangle}\right)\right) - \Sigma P(1-\kappa)\frac{1+2\theta\langle R\rangle}{\big(\langle R\rangle(1+\theta\langle R\rangle)\big)^3} \tag{2.3.78}$$

Specifically, considering the homogeneous limit and excluding adiabatically fast modes *x* and *y*, the stationary equation for critical void radius $R_c$ can be obtained by letting $\partial_t\langle R\rangle = 0$ in the above **Eq.**(2.3.78),

$$P = \frac{x_0}{1-\kappa}\frac{\left(e^{R_s/R_c}-1\right)(1+\theta R_c)}{1-\Sigma\frac{1+2\theta R_c}{R_c^2(1+\theta R_c)^2}} \tag{2.3.79}$$

In the case of large void density limit ($\theta \to \infty$), the above **Eq.**(2.3.79) can be rewritten as,

$$P \approx \frac{x_0}{1-\kappa}\left(e^{R_s/R_c}-1\right)(1+\theta R_c) \tag{2.3.80}$$

which is fair for both deterministic ($\Sigma = 0$) and stochastic ($\Sigma \neq 0$) scenarios. By substituting the relation $x \cong P/(1+\theta R)$, using the expansion $e^{R_s/R_c} \cong 1 + R_s/R_c$ and introducing the vacancy supersaturation $\Delta_x \equiv x/x_0$, we can obtain the expression $R_c = R_s/(1-\kappa)\Delta_x$ for the early-stage evolution of critical void radius. It is emphasized again that the adiabatic elimination of fast modes (point defect concentration *x* and *y*) in a homogeneous system was employed, where both dislocations and voids are active sinks.

In order to explore the influence of an external stochastic source on the universality of the void size growth, the ***Langevin equation*** **Eq.**(2.3.76) is renormalized as,

$$\frac{\mathrm{d}u}{\mathrm{d}t} = v(u) + g(u)\xi(t) \tag{2.3.81}$$

where, $u \equiv R/R_{c0}$ is the dimensionless void radius (with $R_{c0} \equiv R_c(\beta = 0)$ as the critical radius under no external influence sense, *i.e.*, $\beta = 0$, where we introduce the control parameter $\beta \equiv (1-\kappa)P/x_0$), $v(u) = (1+\beta)/u - 1/u^2 - \chi u$, $g(u) = 1/u$, and $\xi(t)$ is the Gaussian white noise satisfying $\langle \xi(t) \rangle = 0$ and $\langle \xi(t)\xi(t') \rangle = 2\beta\Sigma\delta(t - t')$. As argued by Kharchenko *et al*. [184], the control parameter $\beta$ affects the special value of $\chi$, while $v(u)$ has a unique maximum $v(u^*) = 0$ at $u^* = \sqrt{(1+\beta)/3\chi}$. Thus, one can find a renormalized stationary value for $\chi(\beta)$: $\chi(\beta; t \to \infty) \equiv \chi^* = 4(1+\beta)^3/27$ [187], which would be used in following calculations.

It is noted that **Eq**.(2.3.81) is a ***stochastic differential equation*** (***SDE***) with multiplicative noise, which could be written as:

$$\mathrm{d}u = v(u)\mathrm{d}t + \frac{\sqrt{2\beta\Sigma}}{u} \circ \mathrm{d}W \tag{2.3.82}$$

where, $\circ$ denotes the Stratonovich product, $\mathrm{d}W$ is the standard ***Wiener process*** ($\langle \mathrm{d}W^2 \rangle = dt$), and the diffusion coefficient is $\sigma(u) = \sqrt{2\beta\Sigma}/u$. To transform the multiplicative noise into additive noise using the ***Lamperti transformation*** (suitable for the Stratonovich interpretation), we introduce a new variable $y$ such that the diffusion term becomes constant:

$$y = \int_0^u \frac{\mathrm{d}u'}{\sigma(u')} = \int_0^u \frac{u'}{\sqrt{2\beta\Sigma}} \mathrm{d}u' = \frac{u^2}{2\sqrt{2\beta\Sigma}} \tag{2.3.83}$$

Let $\sigma_0 = \sqrt{2\beta\Sigma}$ for brevity, so $y = u^2/2\sigma_0$. And solving for $u$, we can have $u = \sqrt{2\sigma_0 y}$. Since the ***SDE*** is in the Stratonovich sense, the chain rule applies as in ordinary calculus:

$$\mathrm{d}y = \frac{\partial y}{\partial u} \circ du = \left(\frac{u}{\sigma_0}\right) \circ \left(v(u)\mathrm{d}t + \frac{\sigma_0}{u} \circ \mathrm{d}W\right) = \frac{v(u)u}{\sigma_0}\mathrm{d}t + dW \tag{2.3.84}$$

Note that the Stratonovich interpretation ensures no additional Ito correction term appears, unlike in the Ito convention where an extra $-(1/2)\sigma'(u)/\sigma(u)$ term would be present in the drift.

By substituting $v(u)$ into **Eq**.(2.3.84), the transformed ***SDE*** with additive noise is:

$$\mathrm{d}y = \frac{1}{\sigma_0}[(1+\beta) - u^{-1} - \chi u^2]\mathrm{d}t + \mathrm{d}W \tag{2.3.85}$$

By solving the above **Eq**.(2.3.85) numerically, we can obtain the stochastic trajectory of void dynamics.

Next, we would consider the formation of spatial instabilities in the system of point defects caused by elastic deformation of the lattice. The flux of defects takes the form,

$$\begin{cases} J_v = -D_v[(1-\varepsilon c_v)\nabla c_v - \eta_v \nabla^3 c_v] \\ J_i \ = -D_i[(1-\varepsilon c_i)\nabla c_i - \eta_i \nabla^3 c_i] \end{cases} \tag{2.3.86}$$

where, the elastic interaction strength is $\varepsilon = K\Omega^{-1}\varpi_{v,i}^2/k_B T$, with $K = E/3(1-2v)$ is the bulk modulus, $|\varpi_{v,i}| \approx a^3 \approx \Omega$ is the dilatation parameter ($\varpi_v < 0$, $\varpi_i > 0$). Thus, $\varepsilon$ can be rewritten as $\varepsilon \approx K\Omega/k_B T$. By substituting $c_v \equiv x/\mu$ and $c_i \equiv y/\mu$ into the above equations, meanwhile multiplied by the parameter $\mu\tau_v$, respectively, one can obtain the renormalized flux as,

$$\begin{cases} J_v = -L_d[(1-\varepsilon' x)\nabla x - \eta_v' \nabla^3 x] \\ J_i \ = -L_d[(1-\varepsilon' y)\nabla y - \eta_i' \nabla^3 y]\kappa\epsilon \end{cases} \leftrightarrow \begin{cases} J_v \equiv L_d \tilde{J}_v \\ J_i \ \equiv L_d \tilde{J}_i \kappa\epsilon \end{cases} \tag{2.3.87}$$

where, $\varepsilon' = \varepsilon/\mu$, $\eta_{v,i}' = \eta_{v,i}/L_d^2$. Then, by transforming the operator $\nabla$ in the $\boldsymbol{r}$ space into the operator $\frac{1}{L_d}\widetilde{\nabla}$ in the renormalized $\boldsymbol{r}'$ space, we can have,

$$\begin{cases} \nabla \cdot J_v \leftrightarrow \widetilde{\nabla} \cdot \tilde{J}_v \\ \nabla \cdot J_i \ \leftrightarrow \kappa\epsilon\widetilde{\nabla} \cdot \tilde{J}_i \end{cases} \tag{2.3.88}$$

*i.e.*, the notation $\tilde{*}$ represents operations of a variable $*$ in the real space into the renormalized space. Therefore, we have the following renormalized main equations,

$$\begin{cases} \partial_t x = P - (1+\theta R)x - xy + \widetilde{\nabla} \cdot \left[(1-\varepsilon' x)\widetilde{\nabla} x - \eta_v' \widetilde{\nabla}^3 x\right] + \xi(t) \\ \partial_t y = P - \epsilon(1+\kappa\theta R)y - xy + \kappa\epsilon\widetilde{\nabla} \cdot \left[(1-\varepsilon' y)\widetilde{\nabla} y - \eta_i' \widetilde{\nabla}^3 y\right] + \xi(t) \\ v\partial_t R = [x - (x(R) - x_0) - \kappa\epsilon y]/R \end{cases} \tag{2.3.89}$$

where, the renormalized *nabla* operator $\widetilde{\nabla} = (\partial/\partial x, \partial/\partial y, \partial/\partial z)$ is a vector, the *Laplacian* $\tilde{\Delta} \equiv \widetilde{\nabla}^2 = \partial^2/\partial x^2 + \partial^2/\partial y^2 + \partial^2/\partial z^2$, thus the third order derivative of a scalar function $\varphi$ is,

$$\widetilde{\nabla}^3 \varphi \equiv \widetilde{\nabla}\left(\widetilde{\nabla}^2 \varphi\right) \equiv \widetilde{\nabla}\left(\tilde{\Delta}\varphi\right) = \begin{pmatrix} \partial\left(\tilde{\Delta}\varphi\right)/\partial x \\ \partial\left(\tilde{\Delta}\varphi\right)/\partial y \\ \partial\left(\tilde{\Delta}\varphi\right)/\partial z \end{pmatrix} \tag{2.3.90}$$

In order not to be confused with the *xOy* coordinate system in the 2D numerical implementation, we rewrite the *x* and *y* as $\tilde{c}_v$ and $\tilde{c}_i$, respectively,

$$\begin{cases} \partial_t \tilde{c}_v = P - (1+\theta R)\tilde{c}_v - \tilde{c}_v\tilde{c}_i + \widetilde{\nabla} \cdot \left[(1-\varepsilon' \tilde{c}_v)\widetilde{\nabla}\tilde{c}_v - \eta_v' \widetilde{\nabla}^3 \tilde{c}_v\right] + \xi(x,y,t) \\ \partial_t \tilde{c}_i \ = P - \epsilon(1+\kappa\theta R)\tilde{c}_i - \tilde{c}_v\tilde{c}_i + \kappa\epsilon\widetilde{\nabla} \cdot \left[(1-\varepsilon' \tilde{c}_i)\widetilde{\nabla}\tilde{c}_i - \eta_i' \widetilde{\nabla}^3 \tilde{c}_i\right] + \xi(x,y,t) \\ v\partial_t R \ = \left[\tilde{c}_v - \tilde{c}_{v0}\left(e^{\tilde{R}_s/R} - 1\right) - \kappa\epsilon\tilde{c}_i\right]/R \end{cases} \tag{2.3.91}$$

It is noted the fast modes $\tilde{c}_v(x,y,t)$ and $\tilde{c}_i(x,y,t)$ are functions of both the spatial coordinates ($x$, $y$) and time $t$, while the slowest mode $R$ takes the form,

$$v\partial_t\langle R\rangle \ = \left[\langle\tilde{c}_v\rangle - \tilde{c}_{v0}\left(e^{\tilde{R}_s/\langle R\rangle} - 1\right) - \kappa\epsilon\langle\tilde{c}_i\rangle\right]/\langle R\rangle \tag{2.3.92}$$

where, the ensemble average $\langle R\rangle = \langle R\rangle(t)$, $\langle\tilde{c}_v\rangle = \langle\tilde{c}_v\rangle(t)$ and $\langle\tilde{c}_i\rangle = \langle\tilde{c}_i\rangle(t)$ are purely the function of time $t$. Alternatively, on could take the form of ***Langevin equation*** as follows,

$$v\partial_t R = \frac{1}{R}\left[\frac{P(1-\kappa)}{1+\theta R} - \tilde{c}_{v0}\left(e^{\tilde{R}_s/R} - 1\right)\right] + g(R)\xi(t) \tag{2.3.93}$$

The above ***stochastic partial differential equations*** (*SPDEs*) are solved on the quadratic mesh of the size $128\ell\times128\ell$ with periodic boundary conditions, the mesh size $\ell$ = 0.5 and the timestep $\Delta t$ = 0.0001.

Based on the theoretical system built above, hereto we could draw a picture to demonstrate the cooperation of various ***HE*** mechanisms on multiple temporal and spatial scales involved in the present ***HERB*** scheme as shown in **Fig. 4**.

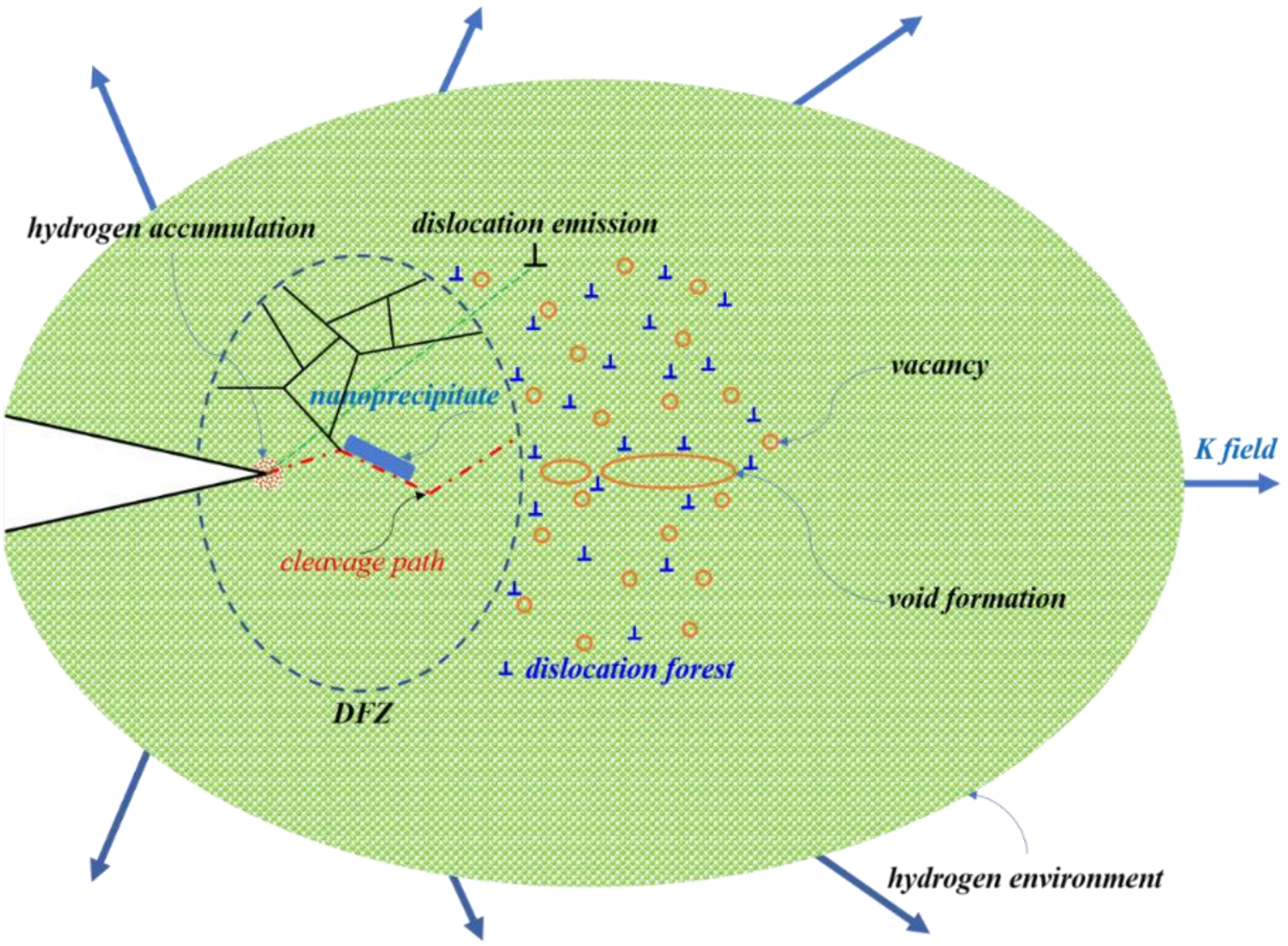

**Fig. 4**. Schematics of the whole ***HERB*** scheme.

**Table 1**. Material parameters of bcc Fe [104].

| | | |
|---|---|---|
| Burgers vector, $b$ (Å) | 2.4825 | Ref.[93] |
| shear modulus, $\mu$ (GPa) | 69.3 | Ref.[93] |
| Poisson's ratio, $\upsilon$ | 0.291 | Ref.[93] |
| surface energy, $\gamma_{surf}$ (J/m$^2$) | 2.37 | Ref.[92] |
| $\gamma_{surf}/\gamma_{usf}$ | 2.2 (Frenkel) | Ref.[92] |
| | 3.2 (EAM) | |
| Debye temperature, $T_D$ (K) | 477 (at 0 K) | Ref.[188] |
| | 373 (at 298 K) | Ref.[189] |
| surface disordering temperature, $T_m$ (K) | 1811 | Ref.[190] |

# 3. Results

Starting with the dislocation emission from a crack-tip under hydrogen environment, we would present in this section how multiple ***HE*** mechanisms were activated collectively in the whole ***HE*** process.

## *3.1 Dislocation emission under hydrogen environment*

Let us define a mixity parameter $M_e = K_{II}/K_I$, **Fig. 5** shows the energetics of the dislocation emission from the crack tip under mixed mode-I+II loading. As shown in **Fig. 5**a, the critical $K_I$ at which the total force exerted on the dislocation is equal to 0, increases with the mixity. However, the energy maximums (*i.e.*, at the critical $K_I$) of each case are almost the same. This result demonstrates the self-consistency of the present theory, since the activation energy required for dislocation emission is not dependent on the loading patterns. The material parameters of bcc Fe are listed in **Table 1**, the tip radius of a blunt crack is $\rho/b = 10$, and the distance $r = b$ is assumed for dislocation emission.

In order to access the rate dependence via the ***TST***, we further fit the part of energy curves where the force is larger than 0, and solve the most probable (reduced) SIF $K_r^p$, as shown in **Fig. 6**. The results reveal that the value of $K_r^p$ is significantly influenced by the mixity $M_e$. Generally, $K_r^p$ increases with the increasing hydrogen concentration $c_H$. However, it is not always the case, e.g. under pure mode-I loading (*i.e.*, $M_e = 0$), the reduced SIF $K_r^p$ of $c_H$ = 1000 appm scenario is larger than that of $c_H$ = 100 appm scenario under relatively lower rate, but shows the inverse trend under higher rate.

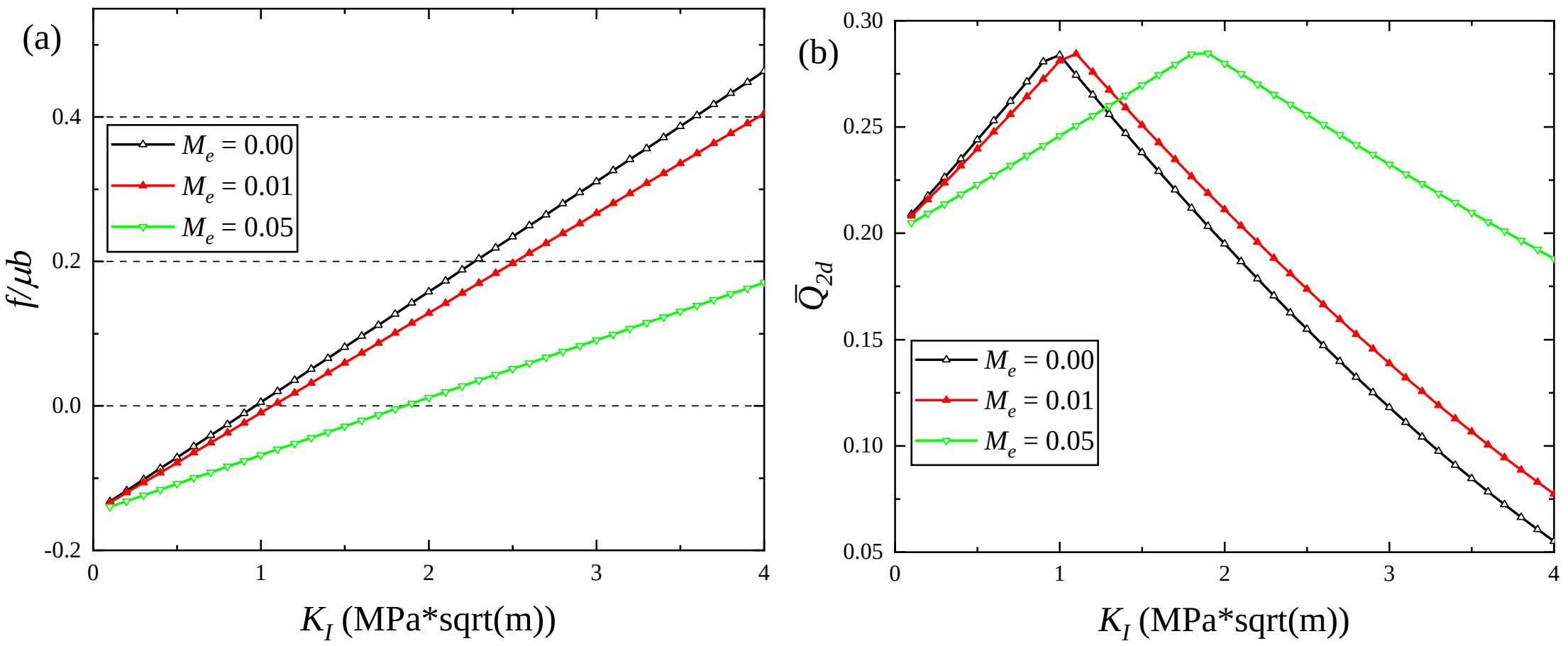

**Fig. 5**. Activation energy for mixed mode-I+II loading. (a) Normalized force $f/\mu b$ *v.s.* the applied mode-I SIF $K_I$; (b) Normalized activation energy $\tilde{Q}_{2d} = (1-\upsilon)Q_{2d}/\mu b^2$ *v.s.* $K_I$, where $\mu$ is the shear modulus, $b$ is the magnitude of the Burgers vector and $\upsilon$ is the Poisson's ratio. The hydrogen concentration is set as $c_H$ = 100 appm, and the slip angle is $\theta$ = 8° here.

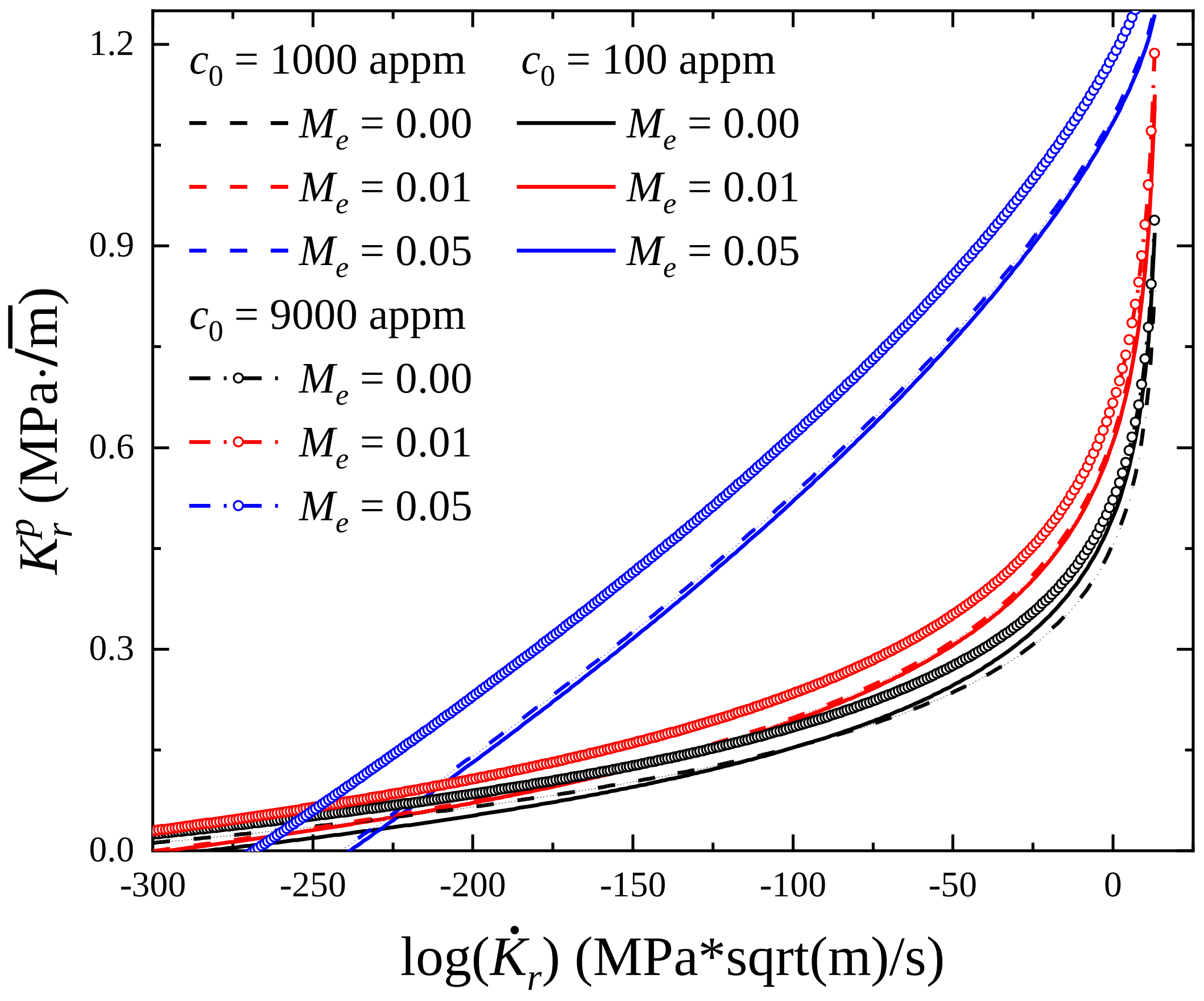

**Fig. 6**. The most probable (reduced) SIF $K_r^p$ as a function of the loading rate under different hydrogen concentrations. The slip angle is set as $\theta$ = 8° here.

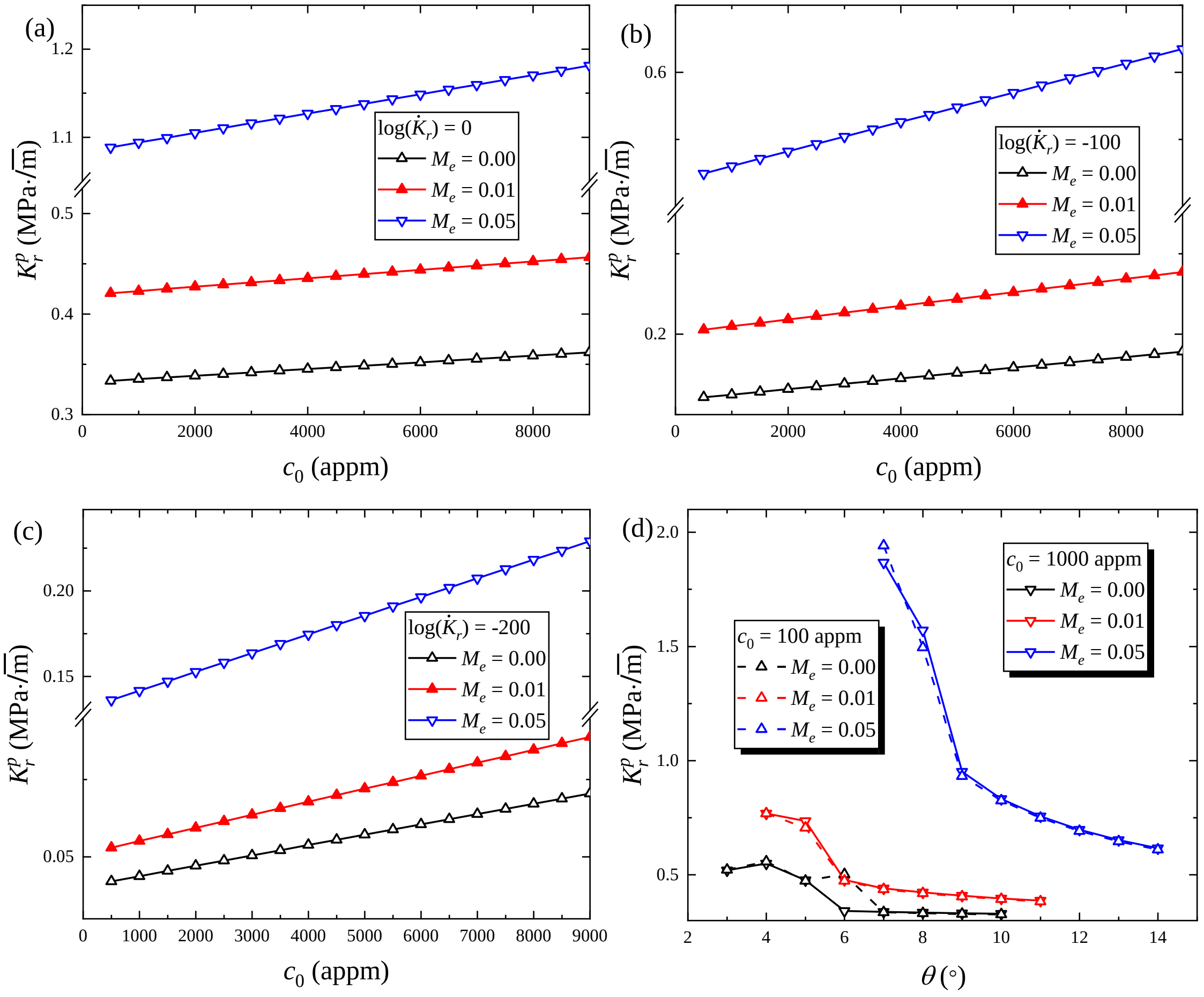

**Fig. 7**. The most probable (reduced) SIF $K_r^p$ as a function of the far-field hydrogen concentration $c_0$ for various loading rates under plane stress scenario: (a) $\log(\dot{K}_r) = 0$ MPa$\sqrt{\text{m}}$/s; (b) $\log(\dot{K}_r) = -100$ MPa$\sqrt{\text{m}}$/s; (c) $\log(\dot{K}_r) = -200$ MPa$\sqrt{\text{m}}$/s. The slip angle is set as $\theta = 8°$ here. (d) The most probable (reduced) SIF $K_r^p$ *v.s.* the slip angle $\theta$ under constant loading rate $\log(\dot{K}_r) = 0$ MPa$\sqrt{\text{m}}$/s.

As shown in **Fig. 7**, the most probable (reduced) SIF $K_r^p$ generally increases with the hydrogen concentration $c_H$ almost linearly. Let $K_{II}^p = M_e K_I^p$, and substitute it into **Eq**.(2.1.39), the most probable (applied) mode-I SIF $K_I^p$ is, $K_I^p = K_r^p/\sqrt{1+M_e^2}$. Thus, the most probable mode-I SIF $K_I^p$ required for dislocation emission is simply scaled with the reduced SIF $K_r^p$, which is indeed equivalent to the ***J***-integral ($J = (K_I^2 + K_{II}^2)/E'$, where, $E' = E$ and $E/(1-\upsilon^2)$ for plane stress and plane strain, respectively) according to its definition by **Eq**.(2.1.39).

Up to present, we are discussing the dislocation emission under plane stress sense. **Fig. A1** (see **Appendix A**) shows the results under plane strain scenario, which is almost the same as the plane stress counterpart only with slight (even ignorable) difference of $K_r^p$ values.

It is noted that all above results are evaluated by hypothesizing the ***USF*** energy $\gamma_{usf}$ is not influenced by the hydrogen charging (see **Table 1**). However, the atomistic simulations (see **Appendix C**) demonstrate that $\gamma_{usf}$ might be significantly influenced by the hydrogen concentration $c_H$ in metals. Assuming the linear correlation $\gamma_{usf} = \gamma_{usf}^0 + k \cdot c_H$ (where, $\gamma_{usf}^0$ is exactly the above-used ***USF*** energy of uncharged specimens in **Table 1**, and $k$ is the fitting parameter), and without loss of generality, we conduct parametric study of the slope $k$, as shown in **Fig. 8** for plane stress sense. It is found that for $M_e = 0.00$ scenario, the most probable SIF $K_r^p$ would increase with increasing ***USF*** energy slope $k$.

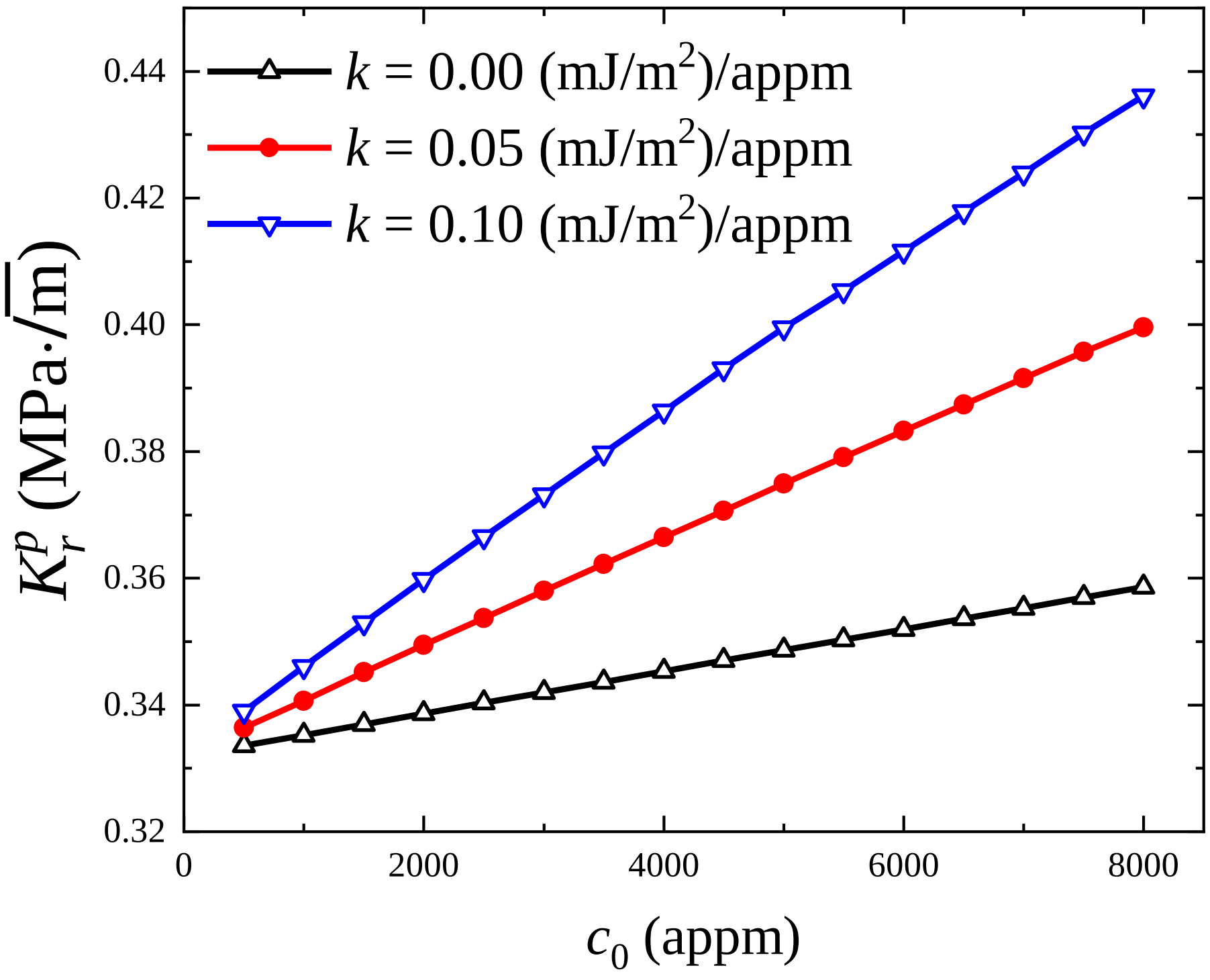

**Fig. 8**. The most probable (reduced) SIF $K_r^p$ as a function of the far-field hydrogen concentration $c_0$ for different ***USF*** energy slopes $k$ under plane stress scenario with $M_e = 0.00$.

### *3.2 Hydrogen transport within the transient DFZ in the vicinity of the crack tip*

Before computing the hydrogen transport equation within the transient elastic core (*i.e.*, the ***DFZ***), we need to evaluate its size by using the **Eq**.(2.2.1). Assuming the validation of the large-strain limit of the generalized Ramberg-Osgood (***RO***) constitutive law:

$$\frac{\tilde{\sigma}}{\sigma_0} = \left(\frac{E\tilde{\varepsilon}_p}{\beta\sigma_0}\right)^n \tag{4.2.1}$$

where, $\tilde{\sigma}$ is the effective stress ($= \sqrt{3s_{ij}s_{ij}/2}$, where $s_{ij}$ is the stress deviator), $\sigma_0$ is the uniaxial yield stress, $\tilde{\varepsilon}_p$ is the effective plastic strain ($= \sqrt{3\varepsilon_{ij}\varepsilon_{ij}/2}$), $E$ is the Young's modulus, $n$ is the work-hardening exponent, $\beta$ is an empirical prefactor of order unity (= 3/7 in the original ***RO*** formulation). The **Eq.**(2.2.1) can be renormalized as,

$$\frac{R_c}{b} = \xi\left(\frac{\alpha}{\lambda}\right)^2 \frac{E}{\sigma_0}\left(\frac{W_{ad}}{\sigma_0 b}\right)^{-1} \tag{4.2.2}$$

**Fig. 9** shows the normalized elastic core size $R_c/b$ as a function of the normalized adhesion work $W_{ad}/\sigma_0 b$, with the Young's modulus $E$ = 200 GPa, yield stress $\sigma_0$ = 300 MPa. The results demonstrate that the ***DFZ*** size of plane strain is larger than that of plane stress, while the ***DFZ*** size increases with increasing value of $\alpha$. Even for a relatively small value $\alpha = 0.5$, at the normalized work of adhesion $W_{ad}/\sigma_0 b = 0.5$, the normalized ***DFZ*** size ($R_c/b$) is about 1.1×$10^4$ and 0.49×$10^4$ for plane strain and plane stress, respectively. Thus, quoting **Eq.**(2.2.20), we can expect that the elastic core size of the intergranular fracture is larger than that of the transgranular cleavage, while the introduction of hydrogen would further enlarge the size according to **Eq.**(2.2.2).

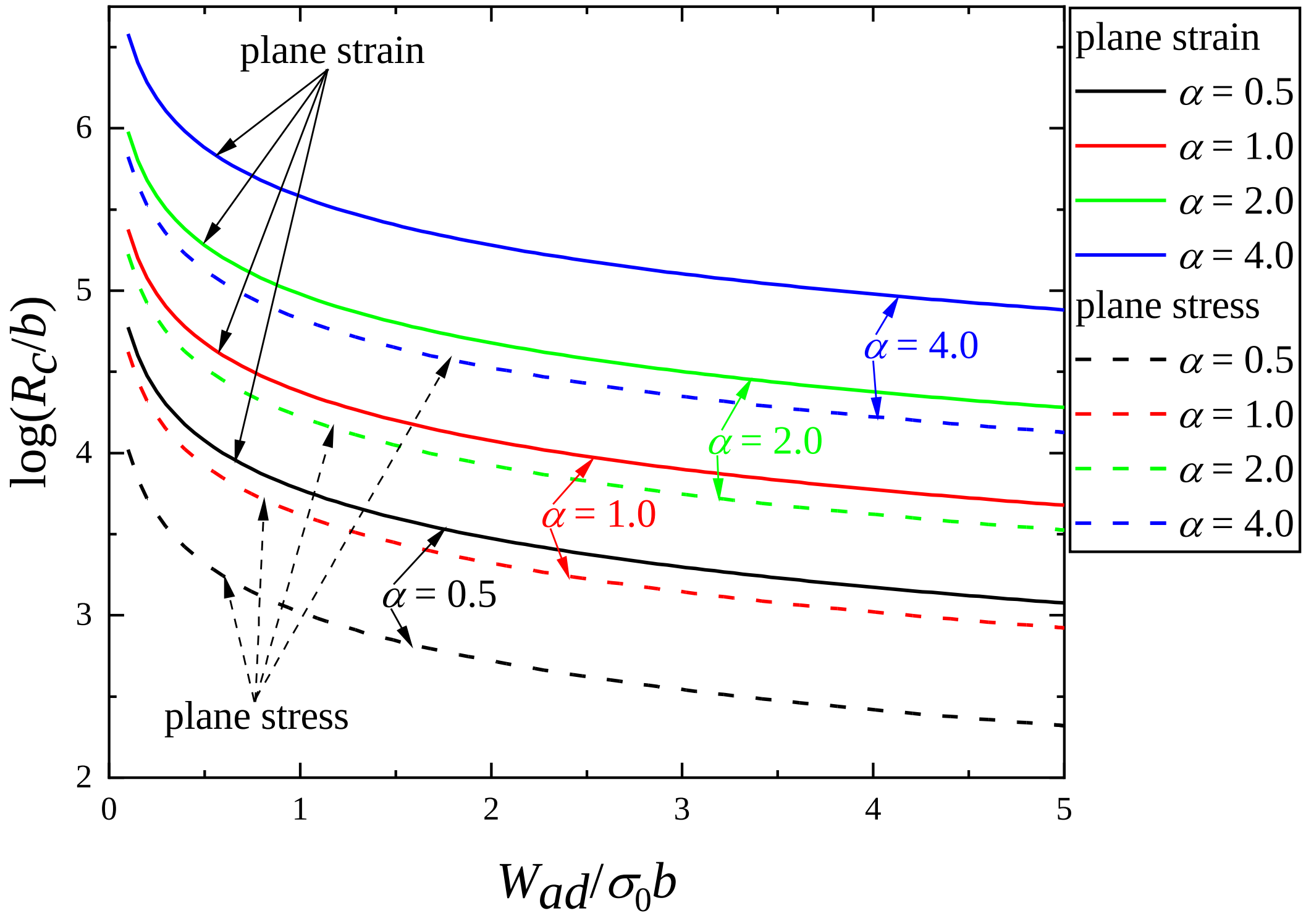

**Fig. 9**. The variation of the elastic core size with the work of adhesion. $\upsilon$ = 0.3 for plane strain conditions.

**Table 2**. Anisotropic ***LEFM*** parameters (unit: $10^{-3}$/GPa) for intergranular cleavage in bcc Fe loaded under the plane strain sense, with more details can be found in Ref.[104].

| index | orientations | $p_1$ | $p_2$ | $q_1$ | $q_2$ |
|---|---|---|---|---|---|
| #1 | $(112)/[1\bar{1}0]$ | 9.2859 | -0.3134 | 1.6121 | 7.5348 |
| #2 | $(111)/[1\bar{1}0]$ | 13.0878 | 0.8761 | -0.5444 | 5.0467 |
| #3 | $(114)/[1\bar{1}0]$ | 0.3880 | 0.3880 | 4.4901 | 4.4901 |
| #4 | $(221)/[1\bar{1}0]$ | 3.5750 | 3.5750 | 0.8813 | 0.8813 |
| #5 | $(113)/[1\bar{1}0]$ | 3.8058 | -0.3102 | 3.0627 | 6.5005 |
| #6 | $(332)/[1\bar{1}0]$ | 8.2325 | 2.1959 | -0.3929 | 2.7287 |
| #7 | $(223)/[1\bar{1}0]$ | 12.5168 | 0.1591 | 0.7090 | 6.8899 |
| #8 | $(334)/[1\bar{1}0]$ | 13.2862 | 0.3638 | 0.3067 | 6.4217 |
| #9 | $(116)/[1\bar{1}0]$ | -0.6764 | -0.6764 | 3.8875 | 3.8875 |
| #10 | $(552)/[1\bar{1}0]$ | 2.4688 | 2.4688 | 0.8504 | 0.8504 |
| #11 | $(118)/[1\bar{1}0]$ | -1.0218 | -1.0218 | 3.4829 | 3.4829 |
| #12 | $(335)/[1\bar{1}0]$ | 11.5017 | -0.0239 | 1.0557 | 7.2239 |

Considering the crack path might be reflected within the ***DFZ***, we compute the anisotropic ***LEFM*** parameters under the plane strain sense as shown in **Table 2**. For brevity, the plane stress counterpart would not be evaluated here, while the mathematical details can be found in Ref.s [191, 192]. Without loss of generality, we would still only consider the case of isotropic material here, and one can easily extend the present analysis to the anisotropic case by the **Eq**.s(2.2.13)-(2.2.16). According to Anderson's book [193], we evaluate the angular distribution of the *von Mises* equivalent stress (**Eq**.(2.2.17)) and the volumetric strain, as shown in **Fig. 10**. It is found that the normalized *von Mises* stress $\bar{\sigma}_{vm}$ of mode-II loading is generally larger than that of mode-I loading (see **Fig. 10**-(a)). **Fig. 10**-(b) shows that the volumetric strain would experience sharp transition near the crack tip for mode-II loading, especially for the plane strain scenario. The **Eq**.(2.2.30) is numerically solved in a 40×40 μm domain with the timestep as $\Delta t = 0.001$ s, and the mesh size as $\Delta l = 0.2$ μm, with an edge crack lying on the left center line and its tip located at the domain center. The material parameters are set as, the Young's modulus $E = 207$ GPa, Poisson's ratio $\upsilon = 0.3$, Burgers vector $b = 0.25$ nm, NILS diffusivity $D_L = 1.5\times10^{-8}$ m$^2$/s, while assuming the inclusion particles have the same radius $r = 1$ μm, environmental temperature $T = 300$ K.

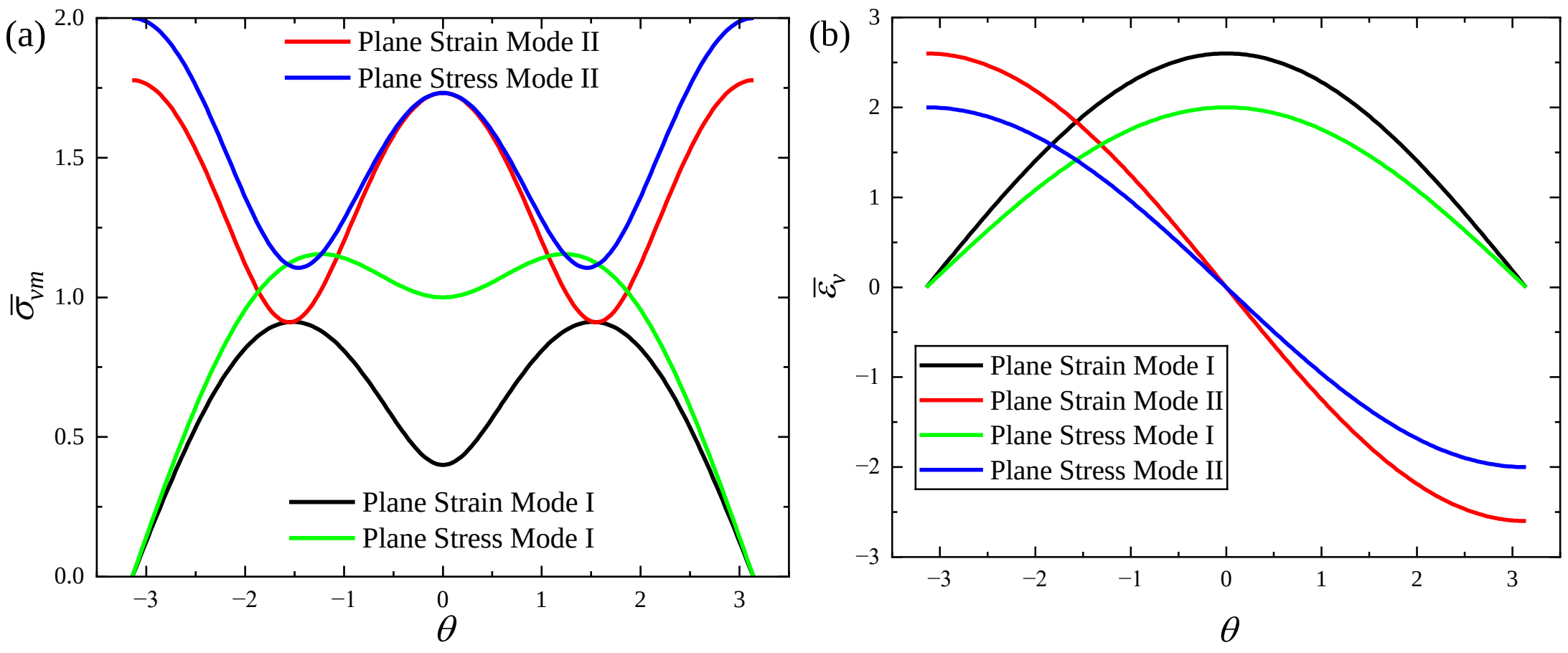

**Fig. 10**. Angular distribution of (a) the normalized *von Mises* stress $\bar{\sigma}_{vm}$ (= $\sigma_{vm}\sqrt{2\pi r}/K_I$, or $\sigma_{vm}\sqrt{2\pi r}/K_{II}$ for mode-I or mode-II loading, respectively), (b) the normalized volumetric strain $\bar{\varepsilon}_v$ (= $\varepsilon_v E\sqrt{2\pi r}/(1-2\upsilon)K_I$, or $\varepsilon_v E\sqrt{2\pi r}/(1-2\upsilon)K_{II}$ for mode-I or mode-II loading, respectively).

**Fig. 11** shows the distributions of stress components and hydrogen concentration near the crack tip under mode-I loading with a constant loading rate $\dot{K}_I$ for plane strain scenario. The results agree well with Sofronis and McMeeking [151] and Krom *et al*. [149], where the NILS hydrogen concentration $C_L$ on the axis of symmetry ($\theta$ = 0°, *i.e.*, the ligament) increases initially, and attains its maxima at some distance from the crack tip.

**Fig. 12** shows the results under mode-II loading for plane strain scenario. Unlike the mode-I counterpart, the hydrogen concentration distribution does not follow a symmetric fashion in the mode-II case.

**Fig. 13** and **Fig. 14** show the plane stress results under mode-I and mode-II loading, respectively. The hydrogen concentrations share the similar feature as their plane strain counterparts, but differ in the values.

It is noted that up to the applied loading of $K_{I,II}$ = 2.5 MPa$\sqrt{\text{m}}$, the negative value of the trapped hydrogen concentration $C_T$ would appear in the plane strain mode-II scenario, since the extremely large (compressive) strain appears at one side of the sharp crack (see **Fig. 10**-(b)). The negative value of $C_T$ indicates strong repulsion of hydrogen atoms from the compressive regime. However, this phenomenon would not occur if real microstructure-based FE models were created from electron microscopy tomography and 3D electron back scattered diffraction experiments [194], where the (mathematically) sharp crack would not appear since the concentrated strain at the crack tip would be released via plastic deformation.

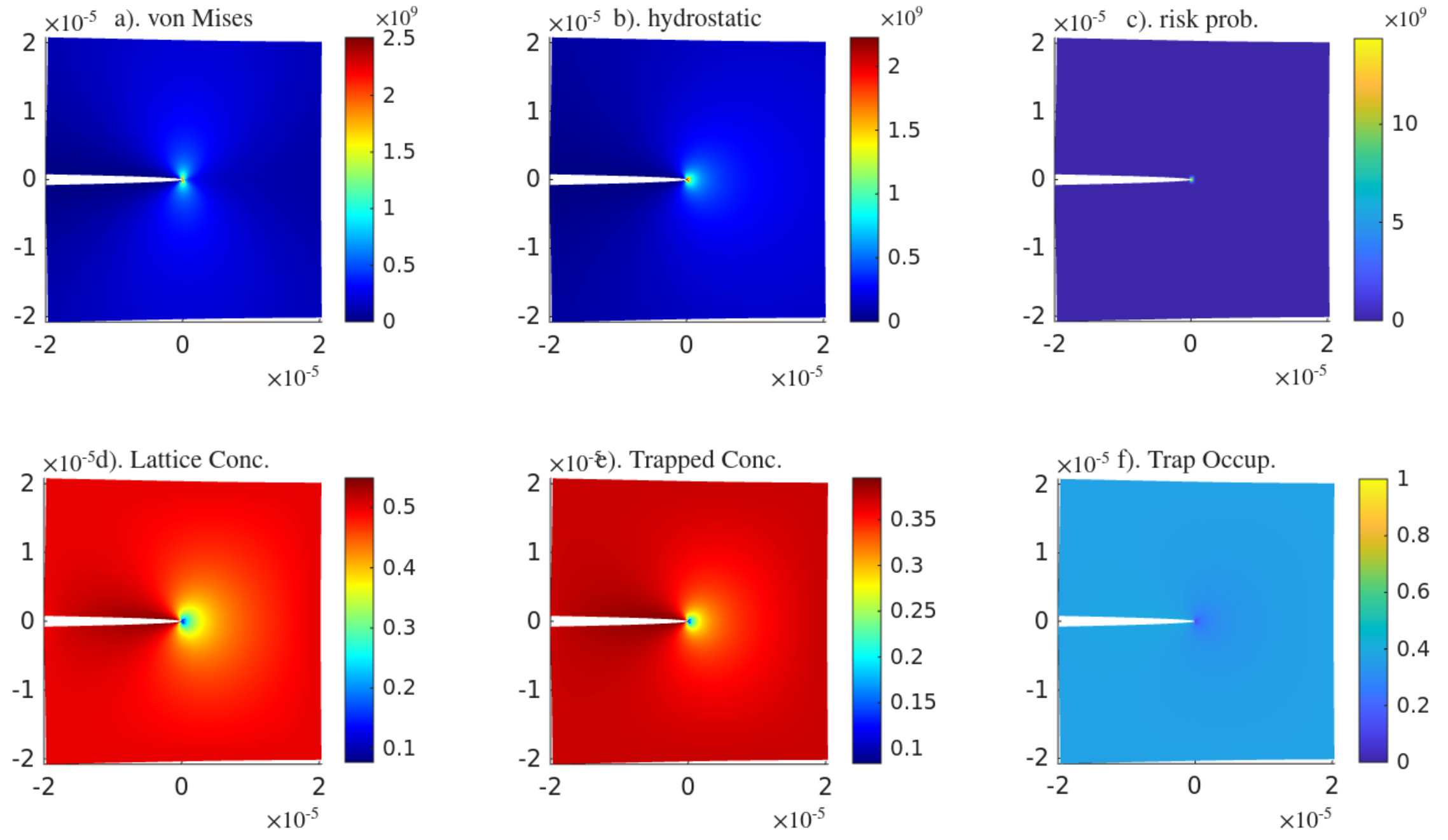

**Fig. 11**. Coupled crack tip fields and hydrogen transport for *plane strain* mode-I loading at $t$ = 5 s under the loading rate of $\dot{K}_I$ = 0.5 MPa$\sqrt{\text{m}}$/s. (a) *von Mises* stress $\sigma_{vm}$; (b) hydrostatic stress $\sigma_H$; (c) risk probability d$\phi$; (d) NILS hydrogen concentration $c_L$; (e) trapped hydrogen concentration $c_T$; (f) hydrogen occupancy of trapped sites $\theta_T$. The international system of units (SI) is used in this figure.

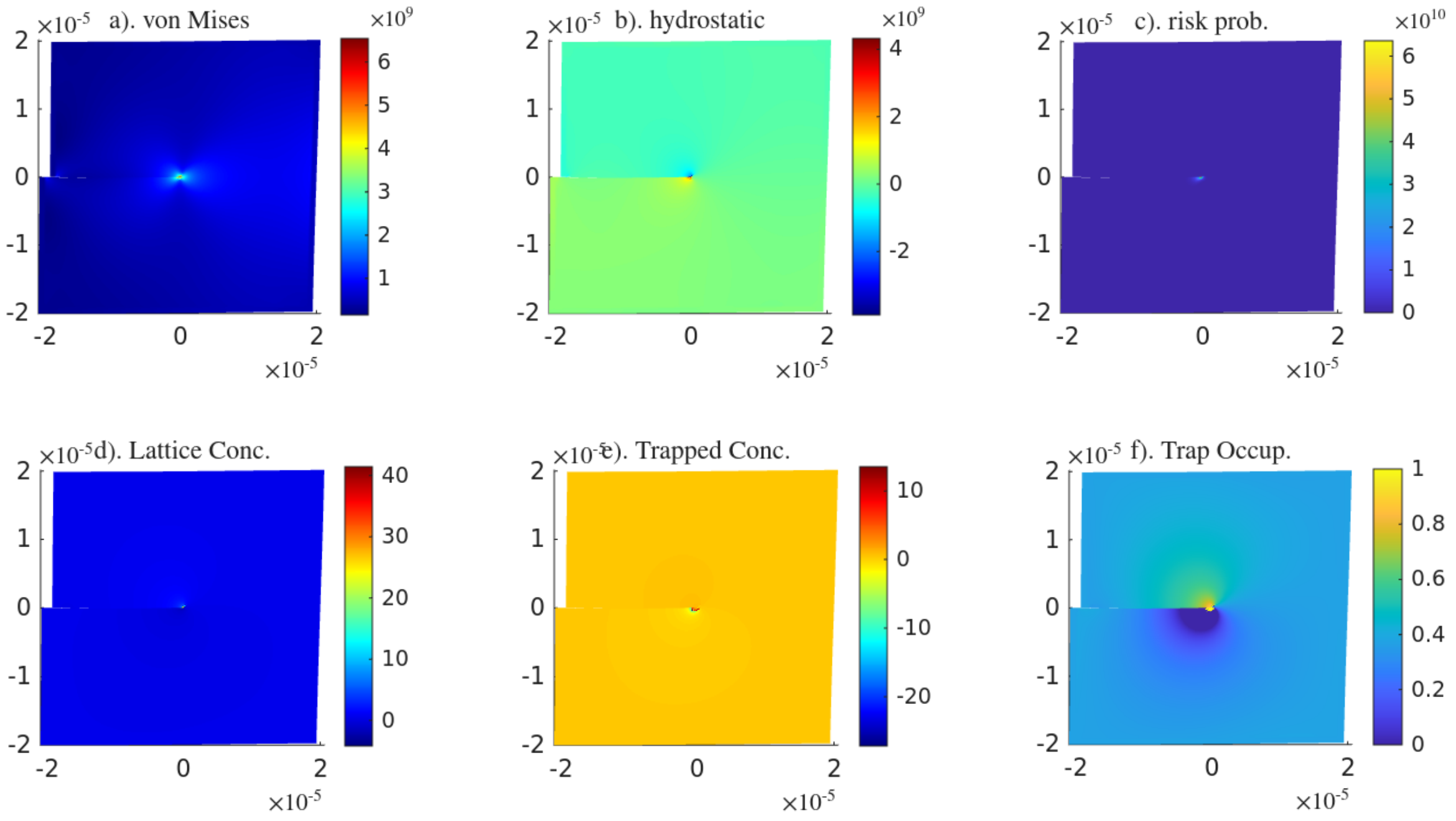

**Fig. 12**. Coupled crack tip fields and hydrogen transport for *plane strain* mode-II loading at $t$ = 5 s under the loading rate of $\dot{K}_{II}$ = 0.5 MPa$\sqrt{\text{m}}$/s. (a) *von Mises* stress $\sigma_{vm}$; (b) hydrostatic stress $\sigma_H$; (c) risk probability d$\phi$; (d) NILS hydrogen concentration $c_L$; (e) trapped hydrogen concentration $c_T$; (f) hydrogen occupancy of trapped sites $\theta_T$. The international system of units (SI) is used in this figure.

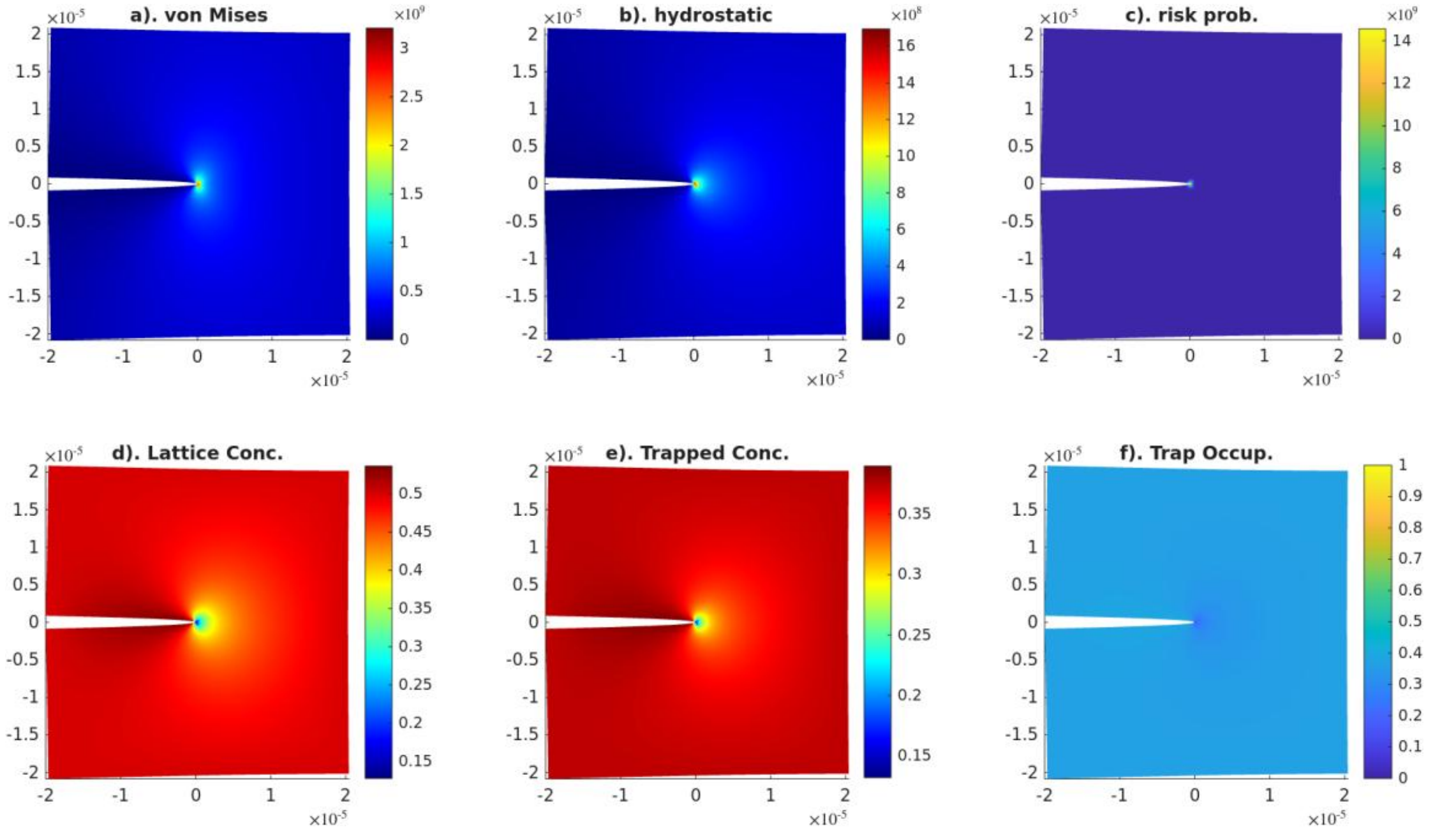

**Fig. 13**. Coupled crack tip fields and hydrogen transport for *plane stress* mode-I loading at $t$ = 5 s under the loading rate of $\dot{K}_I$ = 0.5 MPa$\sqrt{\text{m}}$/s. (a) *von Mises* stress $\sigma_{vm}$; (b) hydrostatic stress $\sigma_H$; (c) risk probability d$\phi$; (d) NILS hydrogen concentration $c_L$; (e) trapped hydrogen concentration $c_T$; (f) hydrogen occupancy of trapped sites $\theta_T$. The international system of units (SI) is used in this figure.

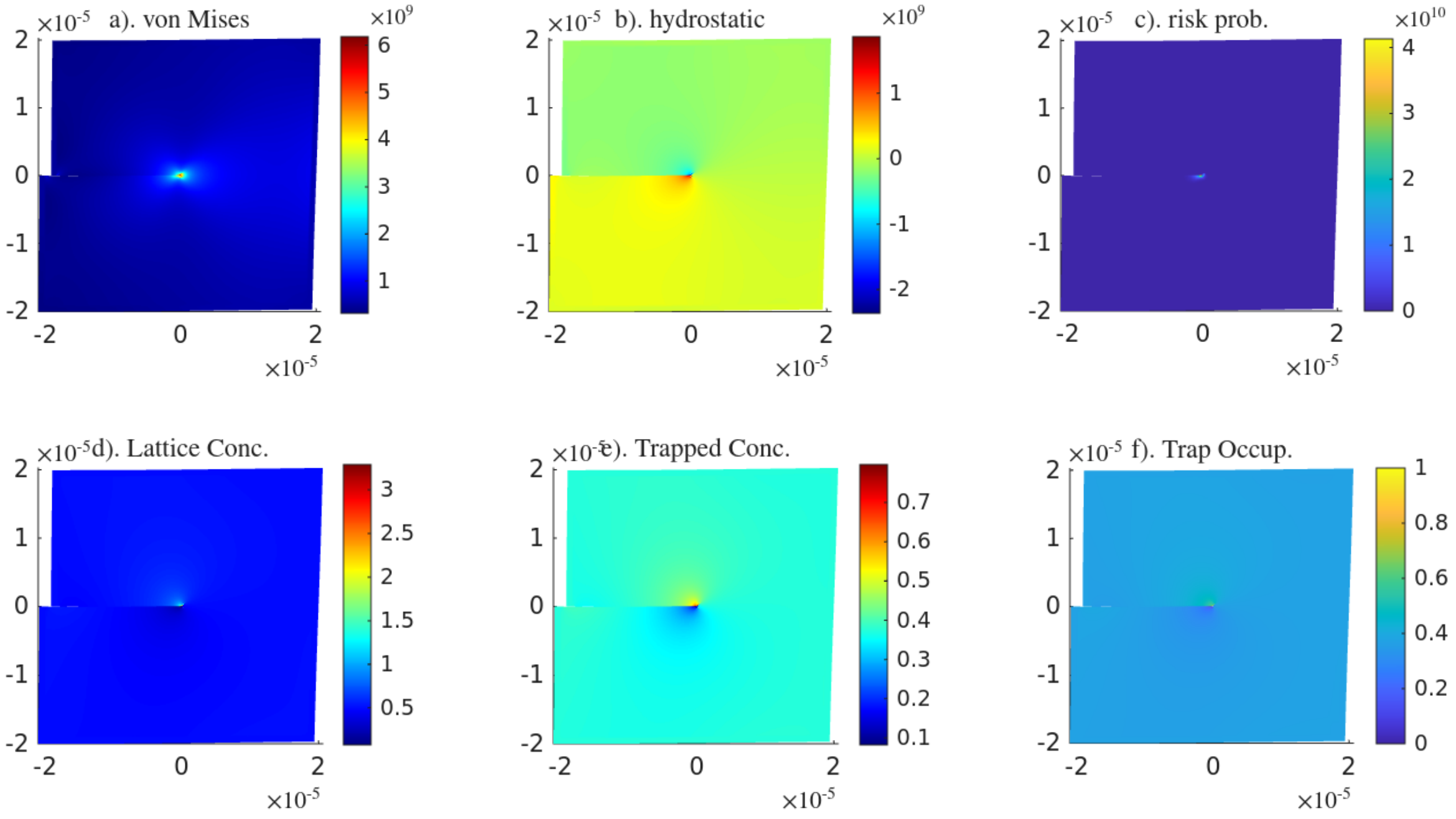

**Fig. 14**. Coupled crack tip fields and hydrogen transport for *plane stress* mode-II loading at $t$ = 5 s under the loading rate of $\dot{K}_{II}$ = 0.5 MPa$\sqrt{\text{m}}$/s. (a) *von Mises* stress $\sigma_{vm}$; (b) hydrostatic stress $\sigma_H$; (c) risk probability d$\phi$; (d) NILS hydrogen concentration $c_L$; (e) trapped hydrogen concentration $c_T$; (f) hydrogen occupancy of trapped sites $\theta_T$. The international system of units (SI) is used in this figure.

It is found in the above figures although the high local stress concentrations at the crack tip enable localized plasticity, *i.e.*, the dislocation emission in the Phase-I, the widespread bulk yielding (the yield strength of bcc Fe is ~ 10 GPa) is not required [103, 104, 195]. As shown in **Fig. 15**-(a), the NILS hydrogen concentration $C_L$ attains its minimum near the crack tip. **Fig. 15**-(b) shows that the total failure probability $\Phi$ grows faster under the mode-II loading than under mode-I loading, meanwhile it is larger under the plane strain than plane stress sense.

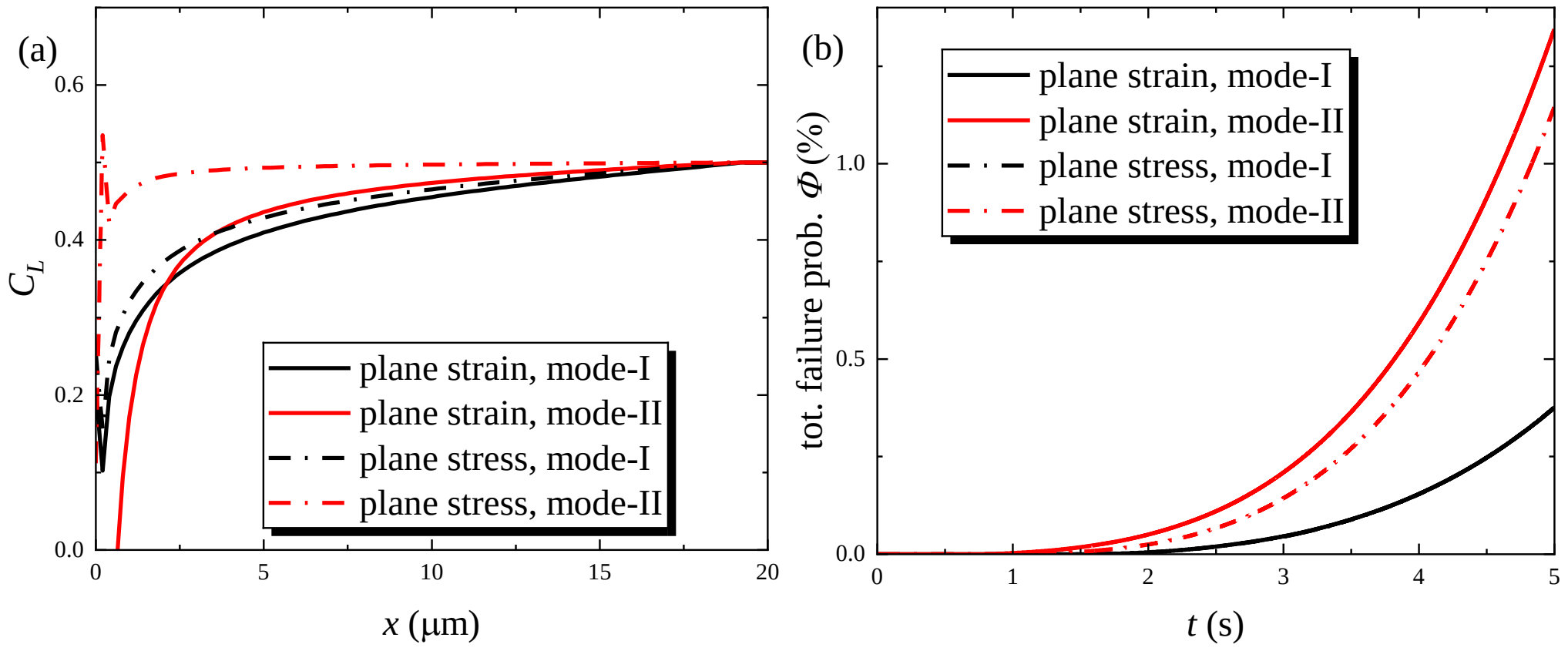

**Fig. 15**. (a) The distribution of the NILS hydrogen concentration $C_L$ along the ligament; (b) the total failure probability $\Phi$ as a function of the loading time $t$, with the loading rate $\dot{K}_{I,II} = 0.5\ \text{MPa}\sqrt{\text{m}}/\text{s}$. It is noted that the $\Phi$ functions are almost the same under the mode-I loading for either plane strain or plane stress scenarios.

## 3.3 Evolution of the stochastic void dynamics

**Fig. 16** shows the ***HRR*** solution under plane strain condition for the mode-II crack tip field for $n$ = 3 and 6, and agrees well with the results of Dai *et al*. [196], thus validates the correctness and accuracy of the analytical procedure formulated in **Section 2.3**.

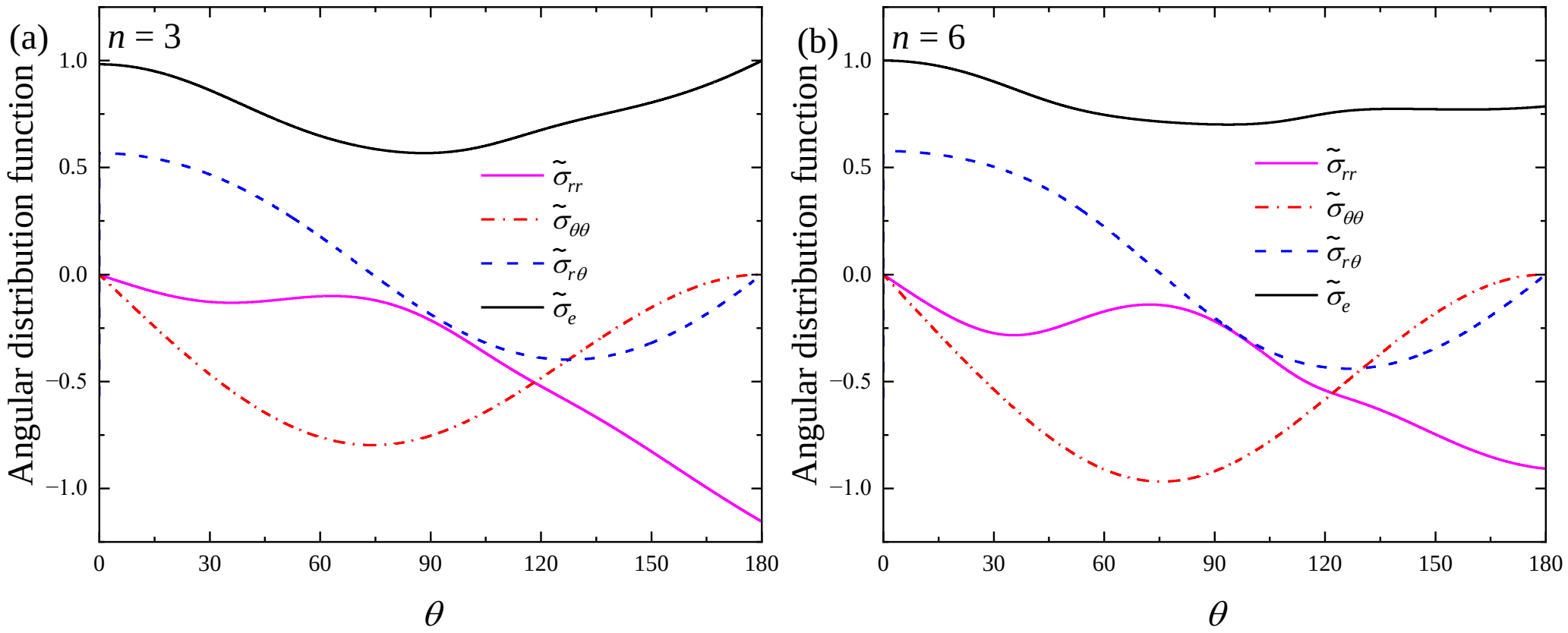

**Fig. 16**. ***HRR*** solution of the angular distribution functions of stress components under plane strain condition for the mode-II crack tip field of (a) $n$ = 3; (b) $n$ = 6 ***RO*** materials, respectively.

**Fig. 17** and **Fig. 18** show the normalized components of the plastic strain gradient tensor $\tilde{\eta}^{p}_{ijk} = \eta^{p}_{ijk}/(\beta_0 r^{m-1})$ for $n$ = 3 and 6 ***RO*** materials, respectively.

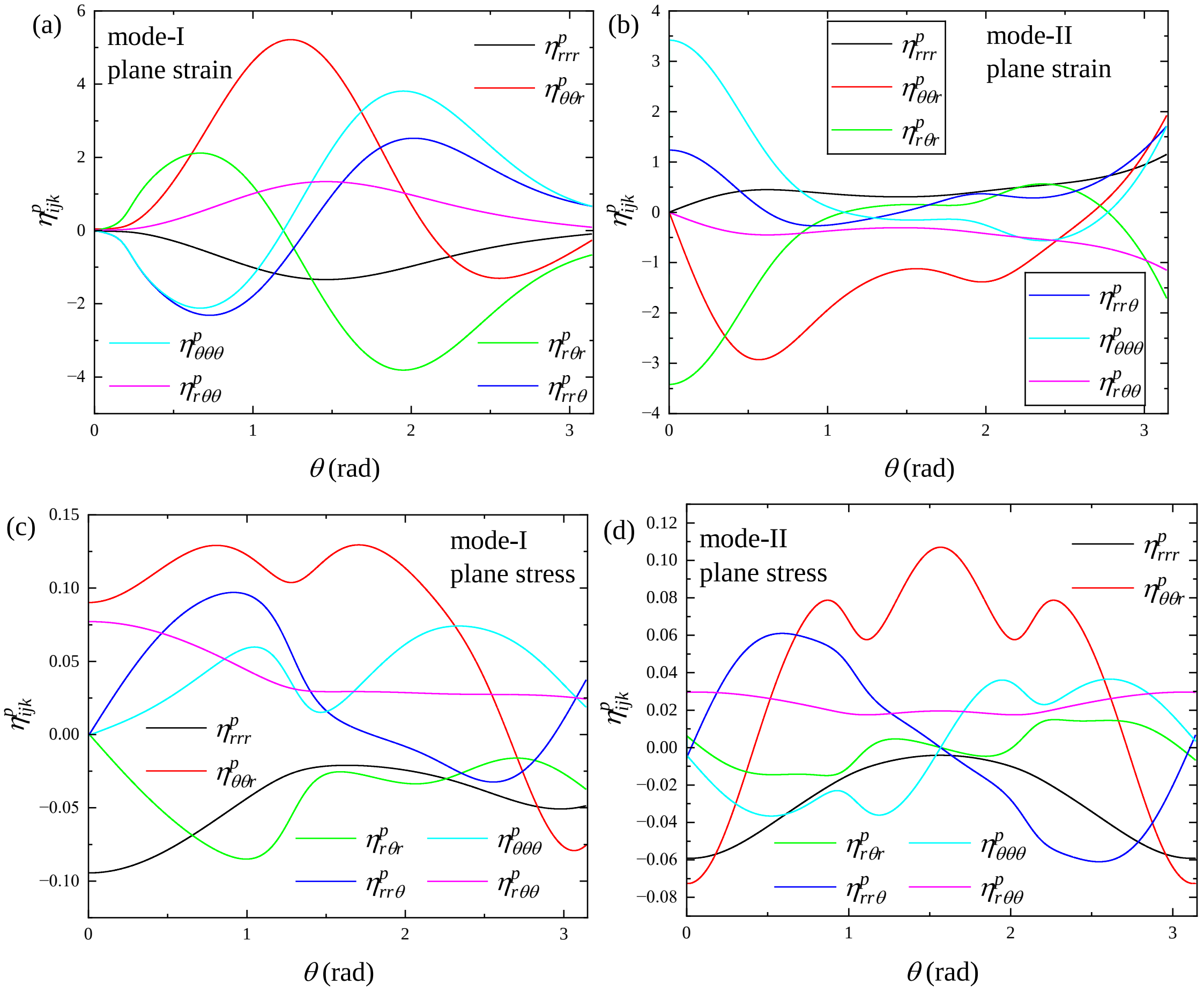

**Fig. 17**. Normalized components of the plastic strain gradient tensor $\tilde{\eta}^{p}_{ijk} \equiv \eta^{p}_{ijk}/(\beta_0 r^{m-1}) \rightarrow \eta^{p}_{ijk}$ for $n$ = 3 ***RO*** material.

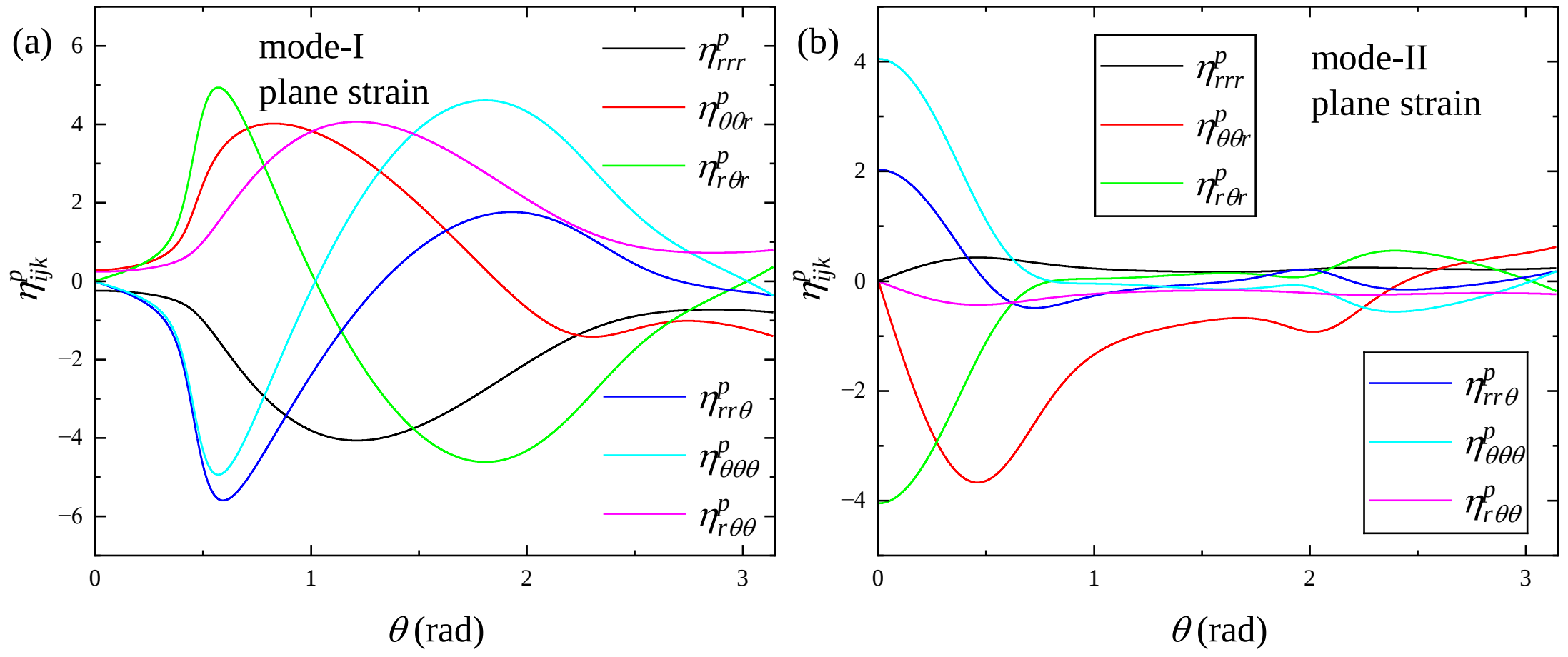

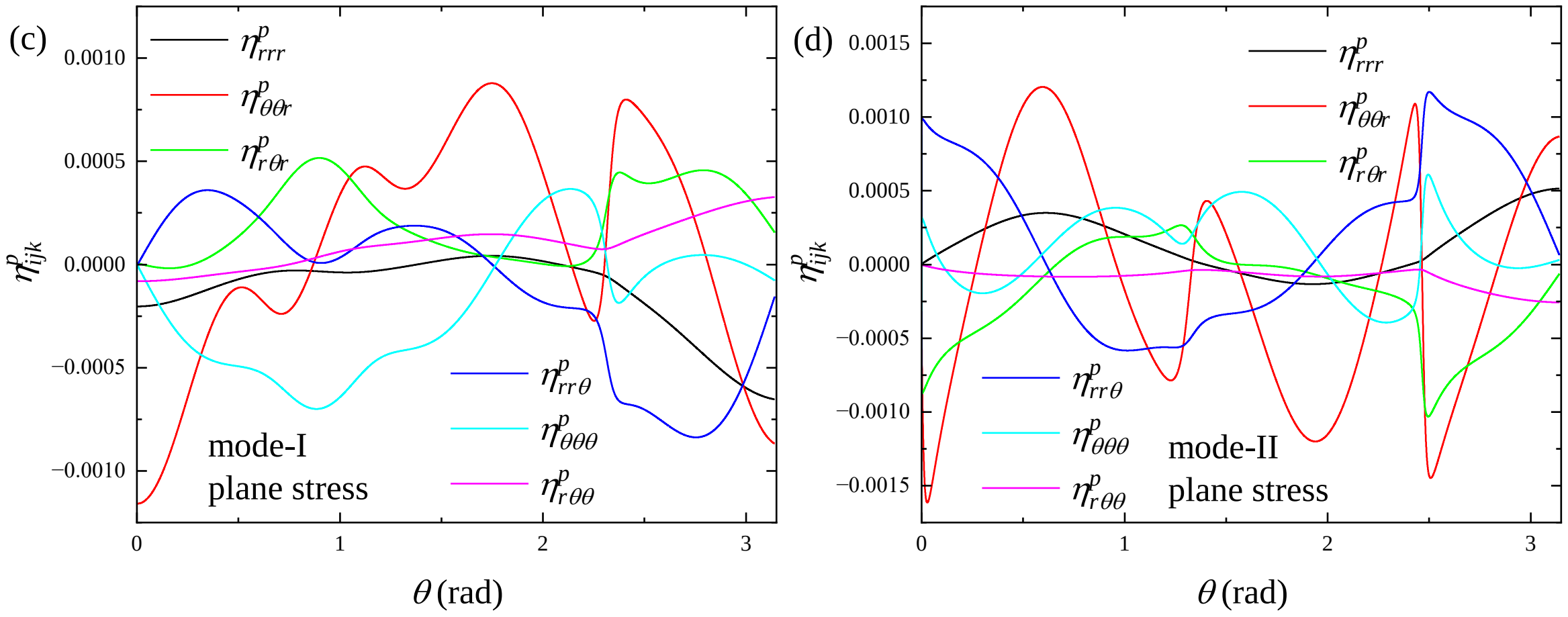

**Fig. 18**. Normalized components of the plastic strain gradient tensor $\tilde{\eta}^p_{ijk} \equiv \eta^p_{ijk}/(\beta_0 r^{m-1}) \to \eta^p_{ijk}$ for $n$ = 6 ***RO*** material.

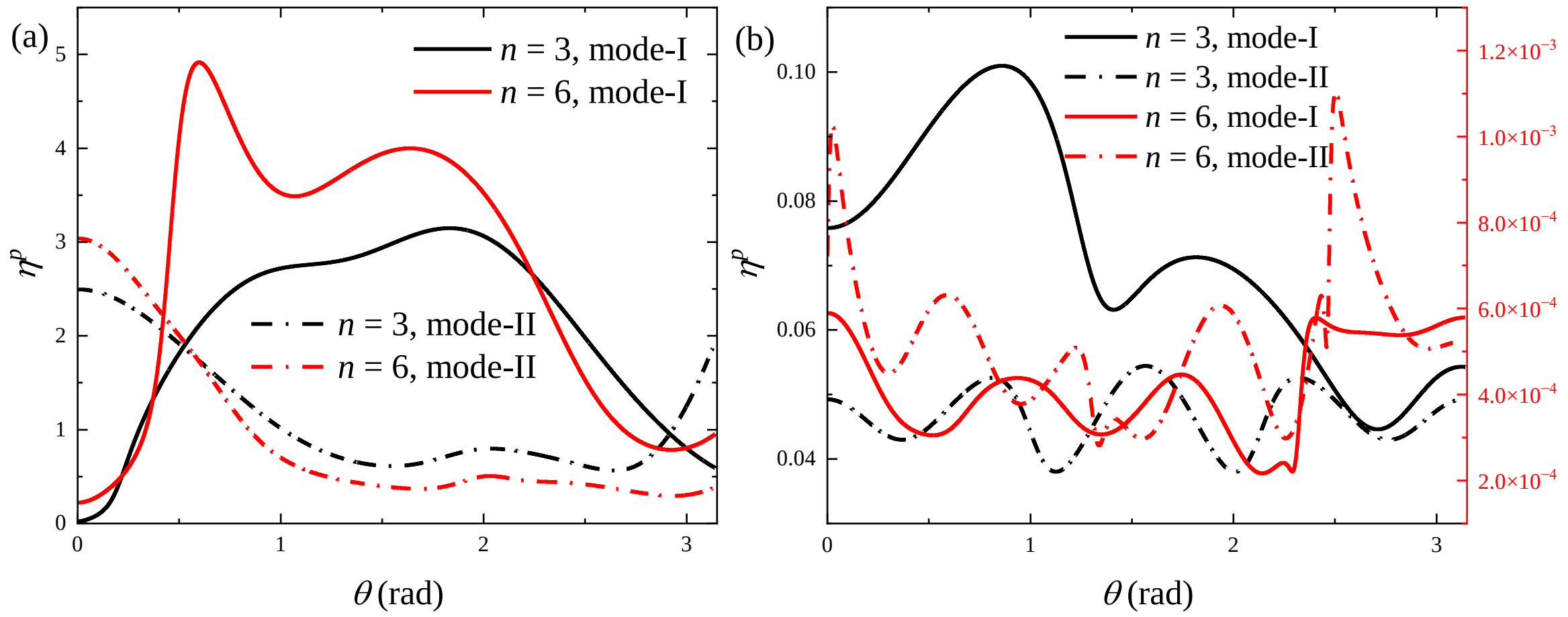

**Fig. 19**. Normalized effective plastic strain gradient $\tilde{\eta}^p \equiv \eta^p/(\beta_0 r^{m-1}) \to \eta^p$ for different ***RO*** materials under (a) plane strain; (b) plane stress loading, respectively.

**Fig. 19** compares the effective plastic strain gradient $\eta^p$ for $n$ = 3 and $n$ = 6 ***RO*** materials. It is found that the variance of hardening parameter $n$ would significantly change $\eta^p$, thus the density of ***GND***s. Previous experiments [197] have indicated that the ***RO*** hardening parameter $n$ would change upon the hydrogen charging. While the effective plastic strain gradient $\eta^p$ for $n$ = 3 and $n$ = 6 are still on the same order of magnitude under plane strain sense, they would differ by 2 ~ 3 orders of magnitude under plane stress sense. Thus, the hydrogen-induced variation of ***GND***s' density would be also strongly dependent on the loading patterns.

By using the Heun's method, here we solve the ***Langevin equation*** **Eq**.(2.3.81) with the timestep as $\Delta t = 0.001$. **Fig. 20** shows 10,000 trajectories of the stochastic void growth with $\beta = 1$ and $\Sigma = 1$. **Fig. 21** shows the mean void radius $\langle u \rangle$ and dispersion $\langle (\delta u)^2 \rangle$ with different values of $\beta \in \{1,2\}$ and $\Sigma \in \{1,2\}$, and agrees well with the results of Kharchenko *et al*. [184].

**Fig. 22** shows the spatial evolution of the normalized vacancy and hydrogen interstitial concentrations with $\theta = 0.01$, $P = 0.25$, $\epsilon = 1000$, $\kappa = 0.01$, $\varepsilon' = 0.01$, and $\eta_v' = \eta_i' = 0.001$ in **Eq**.(2.3.91). By tuning the void sinks intensity $\theta$, we further evaluate the mean vacancy concentration $\langle \tilde{c}_v \rangle$ and its dispersion $\langle (\delta \tilde{c}_v)^2 \rangle$ in **Fig. 23**. **Fig. 24** shows that the mean void radius $\langle R \rangle$ behaves in an algebraic manner $\langle R \rangle \propto t^z$, where the growth exponent $z$ decreases with the increasing $\theta$. As argued by Kharchenko *et al*. [184], at large contribution of dislocations as sinks of point defects (*i.e.*, $\theta \to 0$), the time dependence of $\langle R \rangle$ would decay into the Lifshitz-Allen-Cahn law with $z = 1/2$. With the growth of (preexisting) voids contribution (*i.e.*, $\theta \uparrow$), the exponent $z$ decreases and would arrive to a lower value $z = 1/3$ related to the Lifshitz-Slyozov-Wagner theory [184]. We thus could arrive at that, hydrogen → ***RO*** constitution ($n$) → dislocation density ($\rho_D$) → void dynamics ($z$).

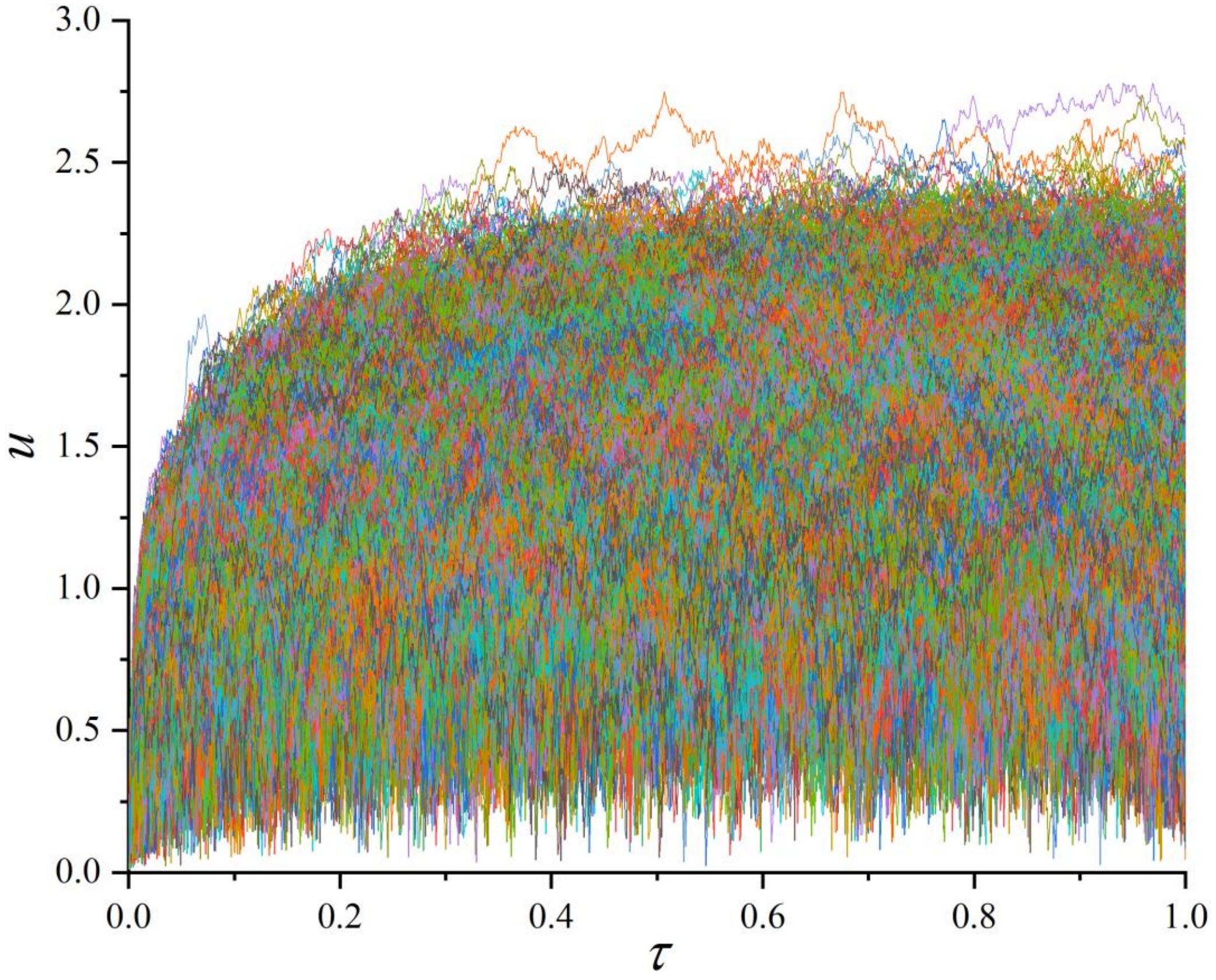

**Fig. 20**. 10,000 stochastic trajectories of the Langevin dynamics defined by **Eq**.(2.3.81) with $\beta = 1$, $\Sigma = 1$.

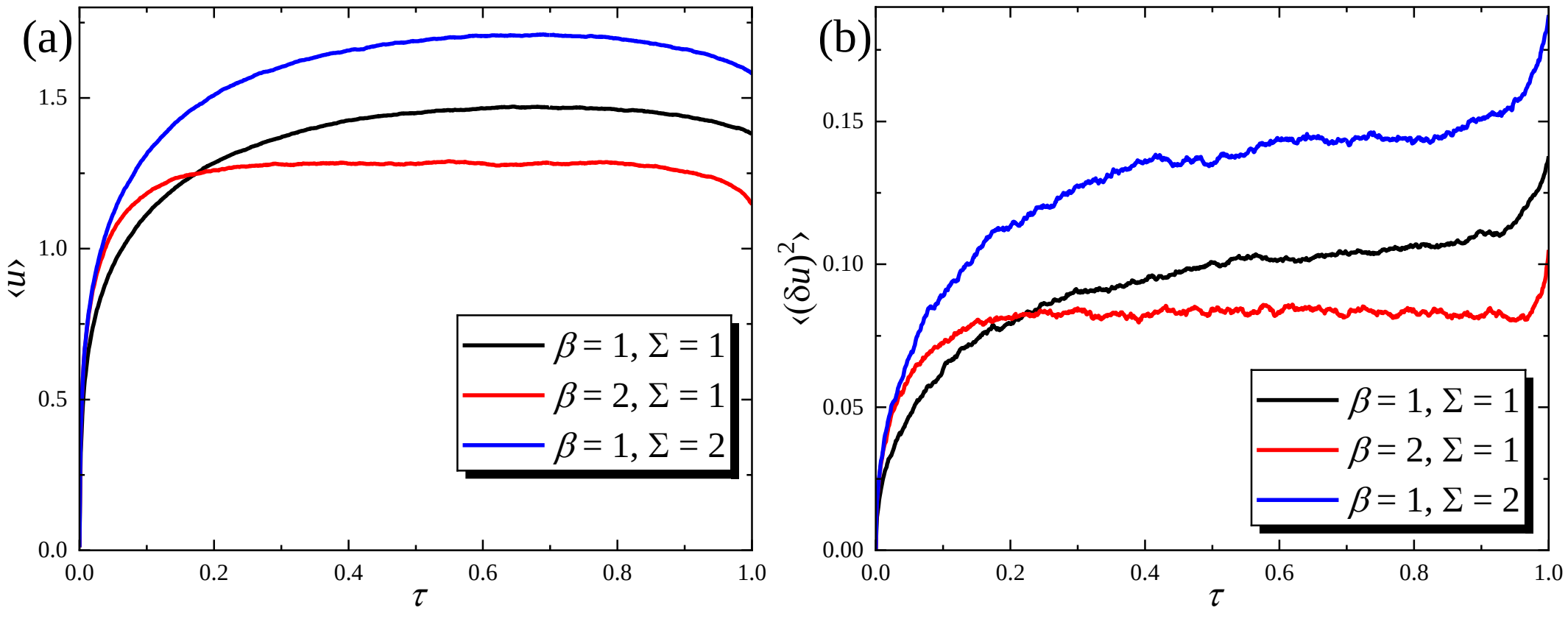

**Fig. 21**. Protocols for (a) mean void radius $\langle u \rangle$, and (b) dispersion $\langle (\delta u)^2 \rangle$ at different $\beta$ and $\Sigma$.

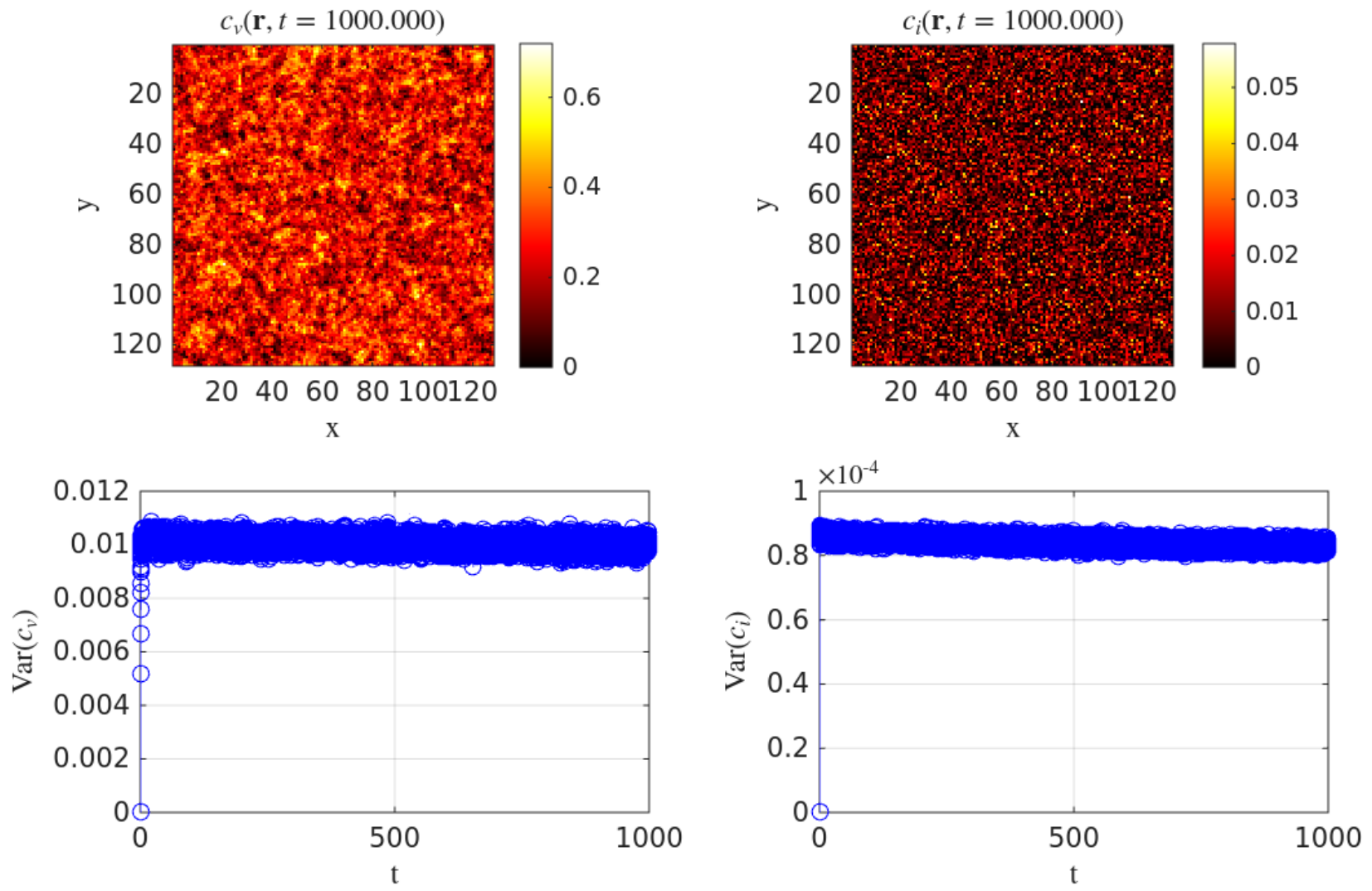

**Fig. 22**. Snapshots of the vacancy and (hydrogen) interstitial concentration inside the matrix phase with $\theta = 0.01$, $P = 0.25$, $\epsilon = 1000$, $\kappa = 0.01$, $\varepsilon' = 0.01$, $\eta_v' = \eta_i' = 0.001$ in **Eq.**(2.3.91).

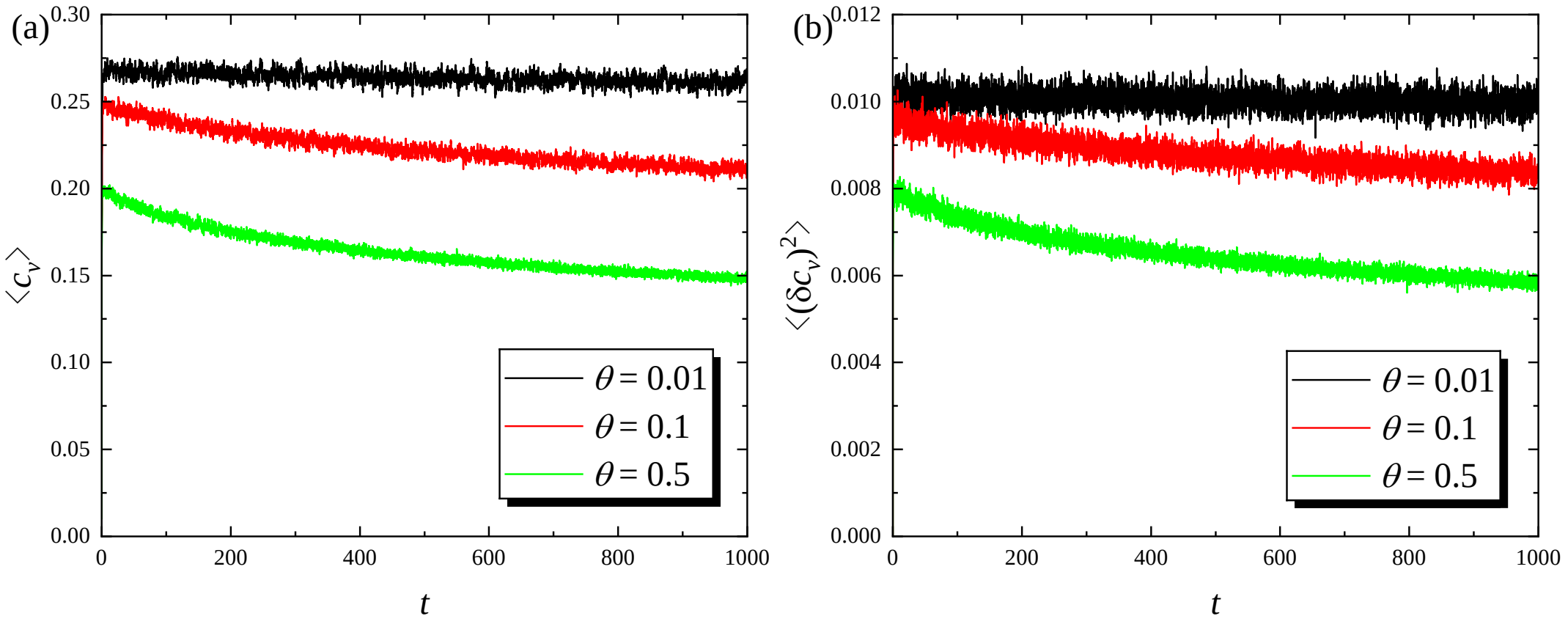

**Fig. 23**. Time dependencies of (a) the averaged vacancy concentration $\langle \tilde{c}_v \rangle$, and (b) its dispersion $\langle (\delta \tilde{c}_v)^2 \rangle$.

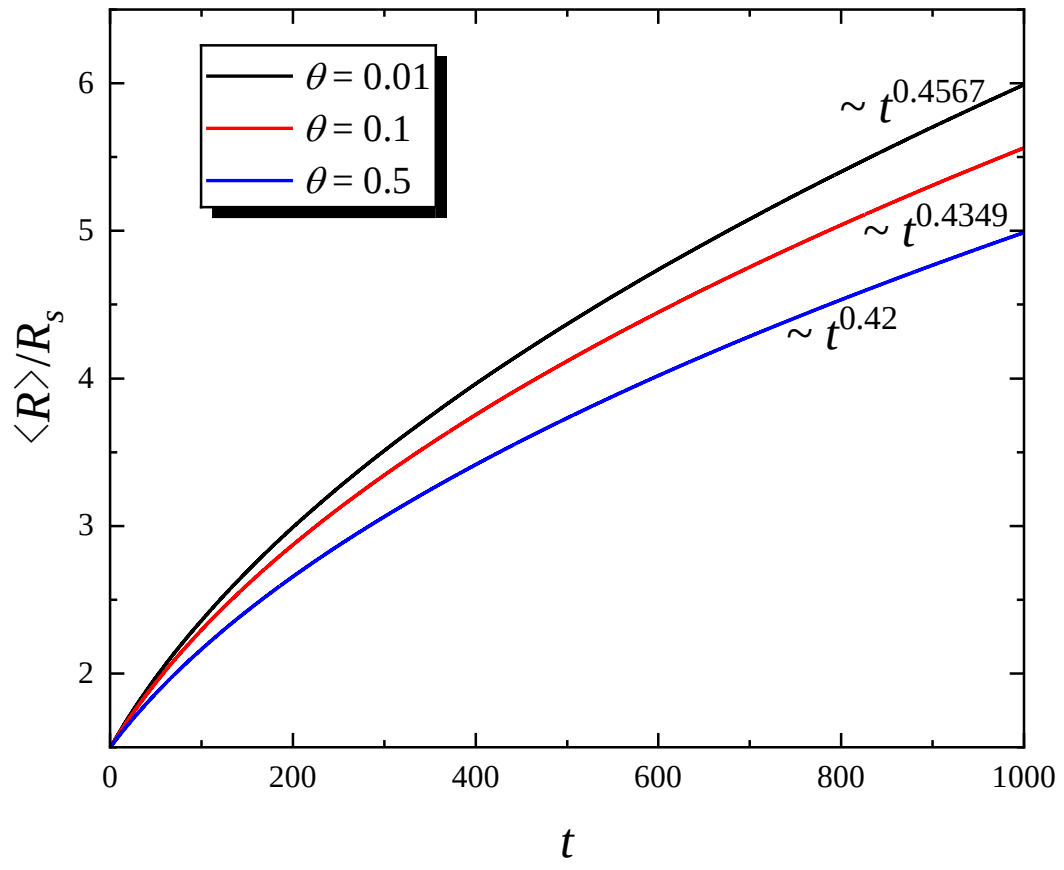

**Fig. 24**. The normalized mean radius of void $\langle R\rangle/R_s$ *v.s.* time $t$. The fitted slopes are obtained by fitting the data in the regime $t \in [500{,}1000]$.

# 4. Discussion

While the numerical results of the ***HERB*** framework have been presented in **Section 3**, we would discuss the underlying physics here.

## *4.1 Effects of hydrogen atmosphere on the core structure of dislocation*

Although the hydrogen effect is not linearly involved in **Eq**.(2.1.19) and **Eq**.(2.1.21), the model predicted most probable SIF $K_r^p$ almost linearly increases with the far-field hydrogen concentration $c_0$ in **Section 3.1**. Compared with our previous study (where only the pure mode-I loading was considered) [103], which also predicts the monotonic dependence of the most probable SIF on the hydrogen concentration, the introduction of the minimization of the strain energy density in present study correctly capture crack-tip field under mixed mode-I+II loading. Another noticeable modification in present model, is exactly the consideration of hydrogen distribution, which was approximately treated as the rate-dependent distribution proposed by Song and Curtin [25] in our previous model [103]. It is noted that the application of **Eq**.(2.1.19) and **Eq**.(2.1.21) actually indicates that the thermodynamic equilibrium is hypothesized in the Phase-I of the present model, while the kinematic transportation of hydrogen atoms is not considered. However, for the atomically transient fracture within the characteristic time period of lattice vibration, it is adequate to assume the hydrogen is already distributed in the equilibrium status in the immediate vicinity of the crack tip.

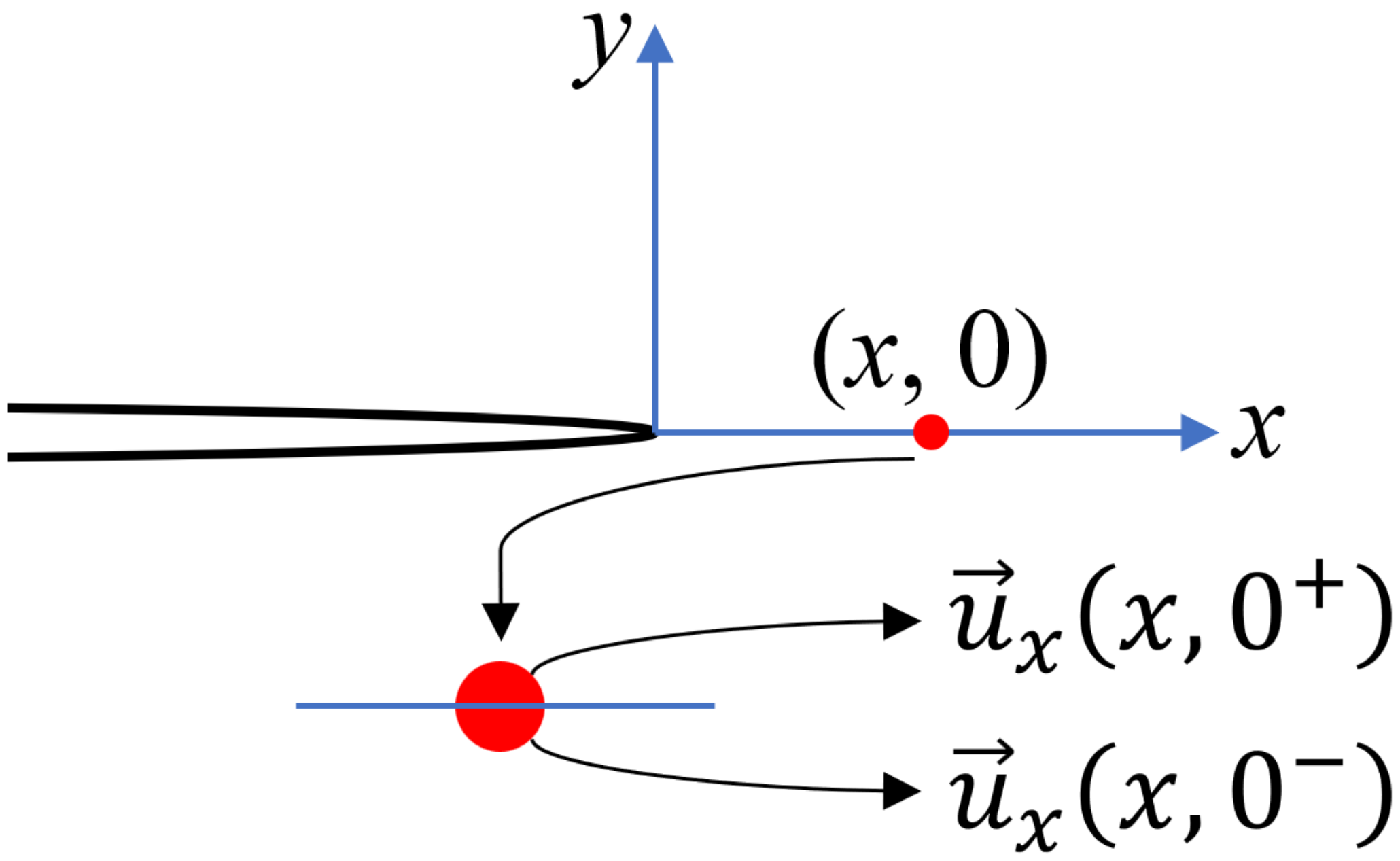

**Fig. 25**. Schematic configuration considered in Rice's theory. The red sphere represents a material point ($x$, 0) on the $x$-axis.

Nonetheless, it would be not enough to employ an equilibrium hydrogen distribution, considering the strong singularity within the dislocation core region, and the effect of hydrogen atmosphere on the core structure [47, 48, 198]. Let us revisit the seminal work by Rice and coworkers [92, 93, 199], and consider a simi-infinite crack lying on the plane $y = 0$ for $x < 0$ (see **Fig. 25**), and we now analyze the formation of slip (dislocation emission) on the plane $y = 0$ for $x > 0$, with defining the slip discontinuity distribution $\vec{\zeta}(x) = \vec{u}_x(x, 0^+) - \vec{u}_x(x, 0^-)$, lattice misfit potential energy $\Phi(\vec{\zeta})$, and lattice restoring stress $\tau[\zeta(x)] \equiv \Phi'[\zeta(x)] = \frac{\mathrm{d}\Phi}{\mathrm{d}\zeta}$. The stress $\tau_{xy}$ at any point $x$ on the slip plane arises from two sources, the external load $K$ and the internal dislocation distribution $b(x)$. Thus, we have, a) the external load stress: $\sigma_{load}(x) = \frac{K_{II}}{\sqrt{2\pi x}}$, is the stress at distance $x$ ahead of the tip for a mode-II crack loaded by the SIF $K_{II}$; b) the elastic back-stress: $\sigma_{self}(x; \zeta) = \frac{\mu}{2\pi(1-\upsilon)} \int_0^\infty \sqrt{\frac{x'}{x}} \frac{\mathrm{d}\zeta(x')/\mathrm{d}x'}{x-x'} \mathrm{d}x'$. In an infinite medium, the stress at $x$ caused by a dislocation at $x'$, is ~ $1/(x - x')$. However, near a crack, the term $\sqrt{x'/x}$ has to be introduced to account for the traction-free boundary condition on the crack surfaces ($x < 0$).

Considering the hydrogen (solute atoms) distribution as the decoration of the misfit potential similar as **Eq**.(2.1.22), the total energy functional equal to the sum of the misfit energy and the elastic potential energy minus the work of the load can be recast as,

$$E[\vec{\zeta}] = E_0 + \int_0^{\infty}\{\Phi[\vec{\zeta}(x)] + V_{sol}[\vec{\zeta}(x); c_H, \cdots]\}\mathrm{d}x + E_{elastic}[\vec{\zeta}] - W_{load}[\vec{\zeta}] \tag{4.1.1}$$

where, the first term $E_0$ (equivalent to the first term $U_0$ in **Eq.**(2.1.22)) is the energy of the loaded elastic solid in which $\zeta$ is constrained to be 0 along the slit and shear stress ahead of the crack is $\tau_0 = K_{II}/\sqrt{2\pi x}$.

The second term is the solute-decorated misfit energy, in which the original ***PN*** potential $\Phi$ is related but not equal, to the generalized stacking fault (***GSF***) interplanar potential $\Psi$, where $\Phi(\zeta) = \Psi(\Delta) - h\tau^2/2\mu$ with $\zeta$ is the magnitude of $\vec{\zeta}(x)$, $h$ is the interplanar spacing along the direction perpendicular to the slip plane, and $\tau$ $(= \mathrm{d}\Phi/\mathrm{d}\zeta = \mathrm{d}\Psi/\mathrm{d}\Delta)$ is the shear stress (*i.e.*, the lattice restoring stress) on the slip plane in the direction of $\vec{\zeta}$. Inspired by recent studies [50, 200], where the so-called ***sPN*** model was proposed, the classical ***PN*** model is modified by scaling the interlayer potential $\Phi$ with a random variable $\omega$, to treat the stochastic occupancy of solute atoms within the core width,

$$\begin{aligned} E_{misfit} &= \int_0^{\infty}\{\Phi[\vec{\zeta}(x)] + V_{sol}[\vec{\zeta}(x); c_H, \cdots]\}\mathrm{d}x \\ &= \int_0^{\infty} \omega(x)\Phi[\vec{\zeta}(x)]\mathrm{d}x \\ &\approx \omega \int_0^{\infty} \Phi[\vec{\zeta}(x)]\mathrm{d}x \end{aligned} \tag{4.1.2}$$

With the random-amplitude interlayer potential, the equilibrium disregistry function is,

$$\zeta(x) = \frac{b}{\pi}\arctan\left(\frac{x}{\lambda}\right) + \frac{b}{2} \tag{4.1.3}$$

where, the equilibrium dislocation core width $\lambda$ is,

$$\lambda = \frac{h}{2(1-\upsilon)\omega} = \frac{\lambda_0}{\omega} \tag{4.1.4}$$

where, $\lambda_0 = h/2(1-\upsilon)$ is the core width in the classical ***PN*** model.

The third and fourth terms are also equivalent to their counterparts in **Eq.**(2.1.22), but only decay into the presently simple case with the slip is limited on the $y = 0$ plane, where the third term $E_{elastic}$ is the non-local elastic self-energy of the dislocation pile-up,

$$\begin{aligned} E_{elastic} &= \frac{1}{2}\int_0^{\infty} \zeta(x)\sigma_{self}(x;\zeta)\mathrm{d}x \\ &= \frac{\mu}{4\pi(1-\upsilon)}\int_0^{\infty} \zeta(x)\left[P.V.\int_0^{\infty}\sqrt{\frac{x'}{x}}\frac{1}{x-x'}\frac{\mathrm{d}\zeta(x')}{\mathrm{d}x'}\mathrm{d}x'\right]\mathrm{d}x \end{aligned} \tag{4.1.5}$$

And the fourth term $W_{load}$ is the work done by the external $K$-field,

$$W_{load} = \int_0^{\infty} \frac{K_{II}}{\sqrt{2\pi x}}\zeta(x)\mathrm{d}x \tag{4.1.6}$$

It is noted that instead of using the original Frenkel sinusoidal function [92], Warner and Curtin [121] proposed a newly analytic form of the ***GSF*** interplanar potential $\Psi(\Delta)$, which could be evaluated from the atomistic simulations to involve the effects of field variables, *e.g*. the hydrogen concentration $c_H$, temperature $T$, *i.e*., $\Psi(\Delta) \rightarrow \Psi(\Delta; c_H, T, \cdots)$ with the four extrema ($\gamma_{ssf}$, $\gamma_{usf}$, $\gamma_{stf}$, $\gamma_{utf}$) of the generalized stacking fault energy (***GSFE***) curve as functions of these field variables. Upon now, we could arrive at a newly stochastic Peierls-Rice-Beltz (***sPRB***) model, to deal with the dislocation emission from a crack tip under environment of solute atoms, considering the short-range fluctuation of solute concentration within the dislocation core region. However, we would not extend further discussions of the ***sPRB*** model here, and its details will be presented in another up-coming paper.

It is also noted that the present model for dislocation emission is still built upon the ***HELP*** mechanism, while recent experiments [65] and simulations [66] support the dual role of hydrogen on dislocation mobility. Thus, the last term $f_H$ (accounting for the hydrogen effect) in **Eq**.(2.1.1) should be dependent on the hydrogen concentration non-linearly in future studies. Besides, while only the dislocation gliding is considered in present framework, the twinning and phase transition [201] should also be involved in modified version.

### *4.2 Dynamic variation of the hydrogen trapping capacity under external loading*

As demonstrated by **Eq**.(2.2.26), the trap binding energy would change with the elastic strain even with the absence of global plasticity. While Gornostyrev and Katsnelson [202] have discussed the feasibility of the transition from coherent to incoherent state of the precipitate/matrix interface, by expressing the total elastic energy in terms of misfit dislocations and virtual dislocations responsible for the phase transformation, it is also anticipated that such a crossover between coherent and incoherent states might occur due to the variation of the atomic spacing (within the ***DFZ***) under external (elastic) loading.

Although we do not consider the trapping sites within the plastic zone temporally in present study, the precipitate/matrix interface might also undergo a transition from coherent to semi-coherent through dislocation nucleation or interactions with pre-existing dislocations [203]. Recent numerical simulations [204, 205] also reveal the strain energy relaxation by interactions between gliding dislocations and misfitting particles. Thus, the distribution of the trap binding energy ($\varphi(\Delta E_T)$) would also change dynamically upon loading, *i.e.*, $\varphi(\Delta E_T) \rightarrow \varphi(\Delta E_T; t)$, causing the first approximation (of this distribution $\varphi(\Delta E_T; t)$ collapsing into a single-valued function $\Delta E_T$ which changes with time $t$ ($\Delta E_T \rightarrow \Delta E_T(t)$), see **Eq**.(2.2.36)) to be not enough in practical engineering analysis. Specifically, as shown in **Fig. 26**, assuming we have in total 10000 particles with their trapping energy following the Gaussian distribution (mean energy $\vartheta_m$ = 1 fJ, standard deviation $\vartheta_{std}$ = 0.2 fJ) initially, it is found that according to **Eq.**(2.2.33), the energy distribution histogram would deviate from the Gaussian (normal) distribution upon elastic loading. The improvement can be made (in future studies) for the mean-field theory formulated in **Section 2.2**, by introducing the stochastic analysis, *i.e.*, $\Delta E_T(t) = \Delta E_T^0 + \xi(t)$, where $\Delta E_T^0$ is the statistical average of the initial distribution $\varphi(\Delta E_T; t = 0)$ before loading, and $\xi(t)$ is the white noise with statistical properties: $\langle \xi(t) \rangle = 0$, $\langle \xi(t)\xi(t') \rangle = 2\Delta E_T^0 \Sigma \delta(t - t')$, where $\Sigma$ is the noise intensity, and $\delta(t - t')$ is the Dirac delta function, similar as that we have defined in **Section 2.3**. In order to apply the present theory, engineers can utilize the numerical techniques, such as, phase field method, etc., well implemented in commercial FE software packages, to capture the energy fluctuation at the precipitate/matrix interface of each inclusion particle, by creating randomly distributed spherical inclusions (can be inferred from experiments) within the 3D computational domain.

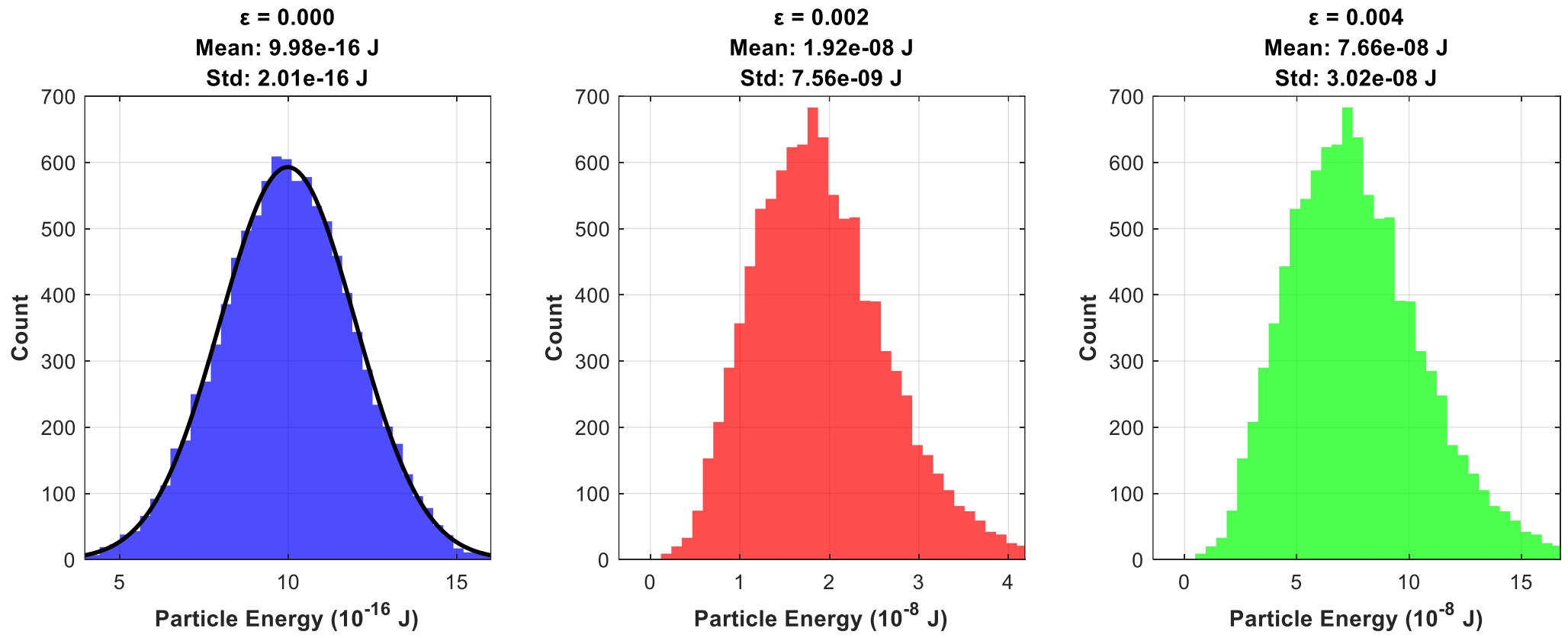

**Fig. 26**. Histograms of the particle energy under different elastic strains, *i.e.*, $\varepsilon$ = 0.000, 0.002, and 0.004, respectively, assuming 10000 particles with their radius following the Gaussian distribution (mean radius $r_m$ = 50 nm, standard deviation $r_{std}$ = 10 nm) are embedded in the matrix. The material parameters are set as, shear modulus *G* = 80 GPa, bulk modulus *K* = 160 GPa, Poisson's ration $\upsilon$ = 0.33, Burgers vector *b* = 2.5 Å. The reference dilatation strain is set as $\varepsilon_0 = 0$ here.

### *4.3 Quasi-cleavage fracture via void nucleation*

It is noted that the distributions of ***GND***s and ***SSD***s densities have been evaluated in previous studies [206-208] by implementing the constitutive relation of conventional ***MSG*** in commercial FE software packages. Specifically, Martínez-Pañeda and Fleck [209] have revealed that higher-order ***SGP*** models predicted an elastic stress state very close to the crack tip, akin to a ***DFZ***, thus supported the division of elastic and plastic zones in present study. While it is already evidenced that the conventional ***MSG*** and classical plasticity (*i.e.*, the ***HRR*** field) give identical stress distributions at a distance ~ 1 μm away from the crack tip [208, 210], it further validates the feasibility of our model that cooperating ***HRR*** with ***MSG*** to obtain the semi-analytic solution of dislocation densities in **Section 2.3**. As shown in experiments by Radu *et al*. [197], the ***RO*** hardening exponent ($n$) could change upon the variation of hydrogen concentration (while the material parameter ($\alpha$) remains almost constant), leading to the variation of the dislocation density field. Similar findings of hydrogen effect on the ***RO*** constitution were also reported by Nazar *et al*. recently [211].

While the incipient stage of void formation under hydrogen environment has been evaluated in present study, the knowledge about void growth is required to bridge the ***HERB*** framework with continuum models, *e.g*. the Gurson model [212-216]. Combining large-scale MD simulations and thermodynamic modelling, Zhao *et al*. [217] further demonstrated that the plasticity has to be initiated before homogeneous nucleation of voids, while the insertion of hydrogen atoms does not contribute to the void nucleation significantly. In previous studies [123, 218-220], it is revealed that the void growth can be accomplished by the emission of shear loops, even taking place at voids as small as a tri-vacancy [221, 222]. It is thus anticipated that hydrogen might affect the void growth by interacting with emitted loops, which would contribute new terms in the right side of **Eq**.(2.3.70) and discussed in future studies.

It is also noted that previous experiments [223, 224] have revealed a mechanism of quasi-cleavage fracture, proceeding by the formation of a nanocrack in the ***DFZ*** of large elastic distortion and then by linking the nanocrack with the main crack. While a mechanism driven by dislocation pileups against the ***DFZ*** was proposed by Zhu *et al*. [105] to explore the quasi-cleavage process, they did not consider the void nucleation might occur outside the ***DFZ*** concurrently. Informed by the quasi-cleavage enhanced by void nucleation in ***HE*** experiments [111], it is thus anticipated that the hydrogen-induced catastrophic failure should be a synergistic effect of multiple ***HE*** mechanisms [225]. It is emphasized that either the quasi-cleavage or void-informed ductile fracture are both fundamental fracture modes of materials, while the hydrogen plays a role as catalyst. Therefore, the present ***HERB*** framework might be conceptualized as a hydrogen-catalytic fracture mechanism, by treating the fracture process as a high-dimensional reaction space. In such a complex system far from equilibrium, involving plastic deformation, mass transport, heat dissipation, etc., the maximum entropy production principle (***MEPP***, also known as Ziegler's maximum dissipation principle (***MDP***) [226, 227]), perhaps provides the "easiest and most accessible" evolution [228], while a hydrogen-enhanced entropy (***HEENT***) mechanism [229, 230] was proposed recently. And indeed, Zajkani and Khonsari [231] have developed a thermomechanically-consistent framework based on Ziegler's ***MDP***, to establish the correlation between the mechanical degradation rate and the progression of the fatigue cycle. Gaede *et al*. [232] have used Ziegler's ***MDP*** to relate the dissipation potential to the damage threshold and damage evolution law. Whaley [233] has revealed that the ***MEPP*** (or Ziegler's ***MDP***) is equivalent to the stability criterion for isothermal, quasi-static fracture, *i.e*., $\left.\frac{\partial W}{\partial \varepsilon}\right|_t = 0$ and $\left.\frac{\partial^2 W}{\partial \varepsilon^2}\right|_t < 0$, where $W$ is the crack

tip strain energy density. Thus, the competition among various ***HE*** mechanisms might be constrained by the ***MDP*** in future studies, since the increased entropy associated with hydrogen trapping increases the overall thermodynamic stability of the system [229]. Mathematical details of the thermodynamic framework would be presented in another upcoming paper.

## 5. Conclusions

Stemming from the classical ***Rice-Beltz*** model, I construct a multiscale framework to describe the crack-tip dislocation emission, and subsequently both the near- and far-field mechanical responses under hydrogen environment.

1) A theoretical model for predicting the critical SIF required for dislocation emission under mixed mode-I+II loading is established. Compared with our previous studies, the novelty is the introduction of the minimization of the strain energy density to determine the renormalized $K_I^*$ and $K_{II}^*$. Considering the stochastic feature of the hydrogen distribution within the core width, I also discussed the feasibility to build the ***sPRB*** model in future studies.

2) By considering the variation of the trapping energy induced by the changing of the lattice misfit under dynamic loading, I rewrite the classical ***SMKKB*** hydrogen transport equation in terms of the trapping energy and (elastic) volumetric strain. Acutely aware of the fact that the trapping capability of inclusions (*e.g.* carbides in micro-alloyed steels) is not immutable, our present model can explain the ***HE*** phenomenon experimentally observed without the presence of global plasticity. While the mean-field approximation of the trapping energy distribution is employed in present study for brevity, the modification via introducing stochastic noise can be realized in future studies.

3) By incorporating the ***HRR*** solution into the ***MSG*** plasticity theory, I obtain the semi-analytical expression of the ***GND***s' density field for plane problems.

4) The ***Langevin equation*** is employed to depict the stochastic nature of the void growth dynamics via the condensation of vacancies under hydrogen environment.

An essential ingredient of the present ***HERB*** framework is the stochastic analysis, which captures the intrinsic fluctuation of the plasticity-involved process, but not considered previously in the ***HE*** community to the author's knowledge. The efforts made in present study would also shed a light on establishing a thermomechanically-consistent framework to constraint the competition among various ***HE*** mechanisms.

### CRediT authorship contribution statement

**K. Zhao**: Conceptualization, Writing – original draft, Writing – review & editing, Project administration, Funding acquisition.

### Declaration of competing interest

The authors declare that they have no known competing financial interests or personal relationships that could have appeared to influence the work reported in this paper.

### Acknowledgements

Financial supports provided by the National Natural Science Foundation of China (Grant No. 12102145), Natural Science Foundation of Jiangsu Province (Grant No. BK20210444) are acknowledged.

## Appendix A: Minimization of the strain energy density under mixed-mode loading

Revisit the **Eq**.(2.1.27) for the constraint from the shear stress equivalence, we can define $\eta = \rho/r$ for compactness, and let $s = \sin(\theta/2)$, $c = \cos(\theta/2)$. Using trigonometric identities, one can have,

$$\begin{aligned} \sin\theta &= 2sc \\ \cos\theta &= c^2 - s^2 \end{aligned} \tag{A.1}$$

The coefficients defined in **Eq**.(2.1.34) thus simplify to,

$$\begin{aligned} B &= s(1 + c^2 - s^2 + \eta) \\ C &= c(-1 + 3(c^2 - s^2) - \eta) \end{aligned} \tag{A.2}$$

The shear stress equivalence constraint **Eq**.(2.1.27) rearranges to,

$$BK_I^* + CK_{II}^* = D, \text{with}, D = 2\sqrt{2\pi r}f/b \tag{A.3}$$

The reduced stress fields are recast as,

$$\begin{cases} \sigma_{xx}^* = \frac{1}{\sqrt{2\pi r}}(K_I^* f_1 + K_{II}^* g_1) \\ \sigma_{yy}^* = \frac{1}{\sqrt{2\pi r}}(K_I^* f_2 + K_{II}^* g_2) \\ \tau_{xy}^* = \frac{1}{\sqrt{2\pi r}}(K_I^* f_3 + K_{II}^* g_3) \end{cases} \tag{A.4}$$

with using identities for multiple angles,

$$\begin{aligned} \sin(3\theta/2) &= 3s - 4s^3 \\ \cos(3\theta/2) &= 4c^3 - 3c \end{aligned} \tag{A.5}$$

The angular functions simplify to,

$$\begin{cases} f_1 = c - sc(3s - 4s^3) - \frac{\eta}{2}(4c^3 - 3c) = c\left(1 - 3s^2 + 4s^4 + \frac{3\eta}{2}\right) - 2\eta c^3 \\ f_2 = c + sc(3s - 4s^3) + \frac{\eta}{2}(4c^3 - 3c) = c\left(1 + 3s^2 - 4s^4 - \frac{3\eta}{2}\right) + 2\eta c^3 \\ f_3 = sc(4c^3 - 3c) - \frac{\eta}{2}(3s - 4s^3) = sc^2(4c^2 - 3) - \frac{\eta s}{2}(3 - 4s^2) \\ g_1 = -2s - sc(4c^3 - 3c) + \frac{\eta}{2}(3s - 4s^3) = -2s - sc^2(4c^2 - 3) + \frac{\eta s}{2}(3 - 4s^2) \\ g_2 = f_3 \\ g_3 = f_1 \end{cases} \tag{A.6}$$

These simplifications reduce higher powers and facilitate squaring in the energy expression.

The strain energy density is,

$$V_\varepsilon = \frac{1}{2E}\left[(\sigma_{xx}^*)^2 + \left(\sigma_{yy}^*\right)^2 - 2\upsilon\sigma_{xx}^*\sigma_{yy}^* + 2(1+\upsilon)\left(\tau_{xy}^*\right)^2\right] \tag{A.7}$$

for plane stress, and,

$$V_\varepsilon = \frac{1+\upsilon}{2E}\left[(1-\upsilon)(\sigma_{xx}^*)^2 + (1-\upsilon)\left(\sigma_{yy}^*\right)^2 - 2\upsilon\sigma_{xx}^*\sigma_{yy}^* + 2\left(\tau_{xy}^*\right)^2\right] \tag{A.8}$$

for plane strain.

Let $k = 1/\sqrt{2\pi r}$, substituting the stresses yields a quadratic form:

$$V_\varepsilon = \frac{k^2}{2E}[A_{11}(K_I^*)^2 + 2A_{12}K_I^*K_{II}^* + A_{22}(K_{II}^*)^2] \tag{A.9}$$

for plane stress, and,

$$V_\varepsilon = \frac{(1+\upsilon)k^2}{2E}[A'_{11}(K_I^*)^2 + 2A'_{12}K_I^*K_{II}^* + A'_{22}(K_{II}^*)^2] \tag{A.10}$$

for plane strain, where the coefficients, incorporating the simplified angular functions, are,

$$\begin{aligned} A_{11} &= f_1^2 + f_2^2 - 2\upsilon f_1 f_2 + 2(1+\upsilon)f_3^2 \\ A_{12} &= f_1 g_1 + f_2 g_2 - \upsilon(f_1 g_2 + f_2 g_1) + 2(1+\upsilon)f_3 g_3 \\ A_{22} &= g_1^2 + g_2^2 - 2\upsilon g_1 g_2 + 2(1+\upsilon)g_3^2 \end{aligned} \tag{A.11}$$

for plane stress, and,

$$
\begin{aligned}
A'_{11} &= (1-v)(f_1^2+f_2^2) - 2vf_1f_2 + 2f_3^2 \\
A'_{12} &= (1-v)(f_1g_1+f_2g_2) - v(f_1g_2+f_2g_1) + 2f_3g_3 \\
A'_{22} &= (1-v)(g_1^2+g_2^2) - 2vg_1g_2 + 2g_3^2
\end{aligned}
\tag{A.12}
$$

for plane strain.

To minimize the quadratic $V_\varepsilon$ subject to the linear constraint, one can use the method of Lagrange multipliers or direct substitution. By defining the vector $\mathbf{K} = [K_I^*, K_{II}^*]^T$, quadratic matrix $\boldsymbol{Q} = \begin{bmatrix} A_{11} & A_{12} \\ A_{21} & A_{22} \end{bmatrix}$ ($A_{12} = A_{21}$), constraint vector $\boldsymbol{a} = [B, C]^T$, and scalar $D$, the objective is: $\min \frac{1}{2}\mathbf{K}^T\boldsymbol{Q}\mathbf{K}$ s.t. $\boldsymbol{a}^T\mathbf{K} = D$, and the closed-form solution is,

$$
\mathbf{K} = \boldsymbol{Q}^{-1}\boldsymbol{a}\left(\frac{D}{\boldsymbol{a}^T\boldsymbol{Q}^{-1}\boldsymbol{a}}\right) \tag{A.13}
$$

where, the inverse matrix is,

$$
\boldsymbol{Q}^{-1} = \frac{1}{\det \boldsymbol{Q}}\begin{bmatrix} A_{22} & -A_{12} \\ -A_{21} & A_{11} \end{bmatrix} \tag{A.14}
$$

with the determinant is,

$$
\det \boldsymbol{Q} = A_{11}A_{22} - A_{12}^2 \tag{A.15}
$$

Then,

$$
\boldsymbol{a}^T\boldsymbol{Q}^{-1}\boldsymbol{a} = \frac{A_{11}C^2 - 2A_{12}BC + A_{22}B^2}{\det Q} \tag{A.16}
$$

It is also noted that $V_\varepsilon$ is a positive definite quadratic form in the stress intensity factors, meaning that the Hessian (second order derivative matrix) is positive definite, guaranteeing a global minimum rather than a maximum or saddle point.

Alternatively, via substitution of $K_{II}^* = (D - BK_I^*)/C$ (or $K_I^* = (D - CK_{II}^*)/B$) into $V_\varepsilon$, and let $\mathrm{d}V_\varepsilon/\mathrm{d}K_I^* = 0$ (or $\mathrm{d}V_\varepsilon/\mathrm{d}K_{II}^* = 0$), one can have,

$$
\begin{aligned}
&A_{11}C^2K_I^* + A_{12}C(D - 2BK_I^*) + A_{22}B(-D + BK_I^*) = 0 \\
&\Rightarrow K_I^* = \frac{D(A_{22}B - A_{12}C)}{A_{11}C^2 - 2A_{12}BC + A_{22}B^2}
\end{aligned}
\tag{A.17}
$$

or,

$$
\begin{aligned}
&A_{11}C(-D + CK_{II}^*) + A_{12}B(D - 2CK_{II}^*) + A_{22}B^2K_{II}^* = 0 \\
&\Rightarrow K_{II}^* = \frac{D(A_{11}C - A_{12}B)}{A_{11}C^2 - 2A_{12}BC + A_{22}B^2}
\end{aligned}
\tag{A.18}
$$

Actually, the denominator in **Eq.**(A.17) & **Eq.**(A.18) $\Delta = A_{11}C^2 - 2A_{12}BC + A_{22}B^2$ represents $\boldsymbol{a}^T\boldsymbol{Q}^{-1}\boldsymbol{a} \cdot \det \boldsymbol{Q}$, ensuring the convexity (assuming $\boldsymbol{Q} > 0$).

Assuming the ***USF*** energy is independent of the local hydrogen concentration $c_H$, **Fig. A1** shows the most probable (reduced) SIF $K_r^p$ as a function of the environmental hydrogen concentration $c_0$ under plane strain, is almost the same as its plane stress counterpart shown in **Fig. 7**.

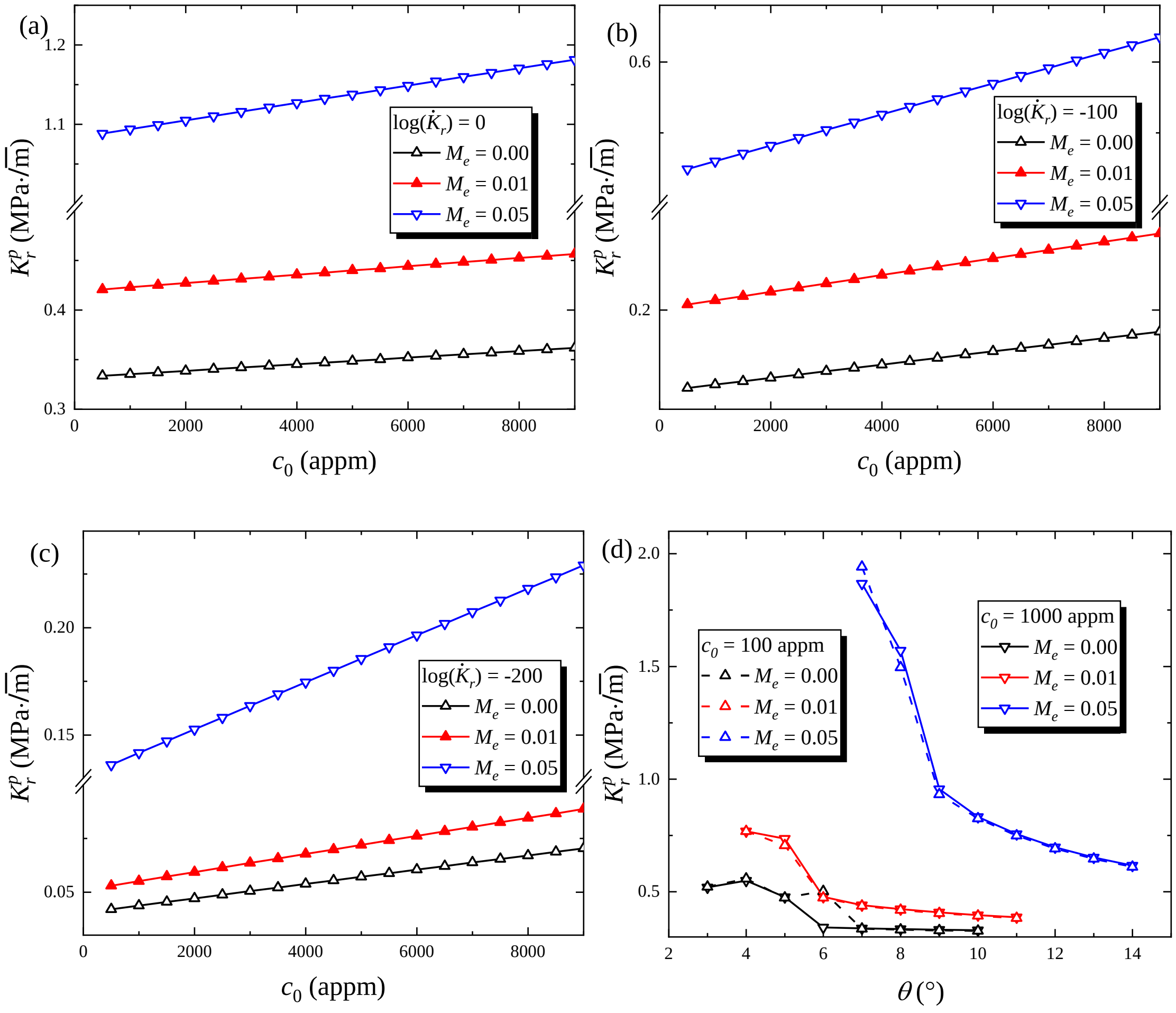

**Fig. A1**. The most probable (reduced) SIF $K_r^p$ as a function of the hydrogen concentration $c_H$ for various loading rates under plane strain scenario: (a) $\log(\dot{K}_r)$ = 0 MPa$\sqrt{\text{m}}$/s; (b) $\log(\dot{K}_r)$ = -100 MPa$\sqrt{\text{m}}$/s; (c) $\log(\dot{K}_r)$ = -200 MPa$\sqrt{\text{m}}$/s. The slip angle is set as $\theta$ = 8° here. (d) The most probable (reduced) SIF $K_r^p$ *v.s.* the slip angle $\theta$ under constant loading rate $\log(\dot{K}_r)$ = 0 MPa$\sqrt{\text{m}}$/s.

# Appendix B: Derivation of the plastic strain gradient in cylindrical coordinates

Generally, the gradient of a second-order tensor field $\boldsymbol{T}$ is defined as the following third-order tensor,

$$\text{grad}\,\boldsymbol{T} = \frac{\partial \boldsymbol{T}}{\partial x_k} \otimes \boldsymbol{e}_k = \frac{\partial T_{ij}}{\partial x_k} \boldsymbol{e}_i \otimes \boldsymbol{e}_j \otimes \boldsymbol{e}_k \tag{B.1}$$

Alternatively, we can also write it as,

$$\boldsymbol{\nabla T} = \frac{\partial (T_{jk} \boldsymbol{e}_j \otimes \boldsymbol{e}_k)}{\partial x_i} \otimes \boldsymbol{e}_i = \frac{\partial T_{jk}}{\partial x_i} e_j \otimes e_k \otimes e_i = T_{jk,i} e_j \otimes e_k \otimes e_i \tag{B.2}$$

However, it should be noted that this gradient differs from the quantity,

$$\nabla \otimes \boldsymbol{T} = \boldsymbol{e}_i \frac{\partial}{\partial x_i} \otimes \left( T_{jk} \boldsymbol{e}_j \otimes \boldsymbol{e}_k \right) = \frac{\partial T_{jk}}{\partial x_i} e_i \otimes e_j \otimes e_k \tag{B.3}$$

Assuming $\boldsymbol{T}(r, \theta, z)$ to be a tensor field with components $T_{\alpha\beta}(r, \theta, z)$ (where, $\alpha, \beta \in \{r, \theta, z\}$) in cylindrical coordinate systems, then, we can use the tensor product representation of the tensor field to find the derivatives of $\boldsymbol{T}$ with respect to $r$, $\theta$ and $z$,

$$\begin{aligned} \frac{\partial \boldsymbol{T}(r,\theta,z)}{\partial r} = & \frac{\partial T_{rr}}{\partial r} e_r \otimes e_r + \frac{\partial T_{r\theta}}{\partial r} e_r \otimes e_\theta + \frac{\partial T_{rz}}{\partial r} e_r \otimes e_z \\ & + \frac{\partial T_{\theta r}}{\partial r} e_\theta \otimes e_r + \frac{\partial T_{\theta\theta}}{\partial r} e_\theta \otimes e_\theta + \frac{\partial T_{\theta z}}{\partial r} e_\theta \otimes e_z \\ & + \frac{\partial T_{zr}}{\partial r} e_z \otimes e_r + \frac{\partial T_{z\theta}}{\partial r} e_z \otimes e_\theta + \frac{\partial T_{zz}}{\partial r} e_z \otimes e_z \end{aligned} \tag{B.4}$$

$$\begin{aligned} \frac{\partial \boldsymbol{T}(r,\theta,z)}{\partial \theta} = & \frac{\partial T_{rr}}{\partial \theta} e_r \otimes e_r + \frac{\partial T_{r\theta}}{\partial \theta} e_r \otimes e_\theta + \frac{\partial T_{rz}}{\partial \theta} e_r \otimes e_z \\ & + \frac{\partial T_{\theta r}}{\partial \theta} e_\theta \otimes e_r + \frac{\partial T_{\theta\theta}}{\partial \theta} e_\theta \otimes e_\theta + \frac{\partial T_{\theta z}}{\partial \theta} e_\theta \otimes e_z \\ & + \frac{\partial T_{zr}}{\partial \theta} e_z \otimes e_r + \frac{\partial T_{z\theta}}{\partial \theta} e_z \otimes e_\theta + \frac{\partial T_{zz}}{\partial \theta} e_z \otimes e_z \\ & + T_{rr} e_r \otimes e_\theta + T_{rr} e_\theta \otimes e_r - T_{r\theta} e_r \otimes e_r + T_{r\theta} e_\theta \otimes e_\theta + T_{rz} e_\theta \otimes e_z \\ & + T_{\theta r} e_\theta \otimes e_\theta - T_{\theta r} e_r \otimes e_r - T_{\theta\theta} e_\theta \otimes e_r - T_{\theta\theta} e_r \otimes e_\theta - T_{\theta z} e_r \otimes e_z \\ & + T_{zr} e_z \otimes e_\theta - T_{z\theta} e_z \otimes e_r \end{aligned} \tag{B.5}$$

$$\begin{aligned} \frac{\partial \boldsymbol{T}(r,\theta,z)}{\partial z} = & \frac{\partial T_{rr}}{\partial z} e_r \otimes e_r + \frac{\partial T_{r\theta}}{\partial z} e_r \otimes e_\theta + \frac{\partial T_{rz}}{\partial z} e_r \otimes e_z \\ & + \frac{\partial T_{\theta r}}{\partial z} e_\theta \otimes e_r + \frac{\partial T_{\theta\theta}}{\partial z} e_\theta \otimes e_\theta + \frac{\partial T_{\theta z}}{\partial z} e_\theta \otimes e_z \\ & + \frac{\partial T_{zr}}{\partial z} e_z \otimes e_r + \frac{\partial T_{z\theta}}{\partial z} e_z \otimes e_\theta + \frac{\partial T_{zz}}{\partial z} e_z \otimes e_z \end{aligned} \tag{B.6}$$

Then, the gradient of a tensor field $\boldsymbol{T}$ denoted as $\nabla\boldsymbol{T}$ (or, grad $\boldsymbol{T}$) can be obtained with components to be $(\nabla T)_{\alpha\beta\lambda}$ (where, $\alpha, \beta, \lambda \in \{r, \theta, z\}$):

when $\lambda = r$ or $\lambda = z$,

$$(\nabla T)_{\alpha\beta\lambda} = \left(\frac{\partial T}{\partial \lambda} e_\beta\right) \cdot e_\alpha \tag{B.7}$$

*i.e.*,

$$(\nabla T)_{\alpha\beta r} = \frac{\partial T_{\alpha\beta}}{\partial r} \tag{B.8}$$

$$(\nabla T)_{\alpha\beta z} = \frac{\partial T_{\alpha\beta}}{\partial z} \tag{B.9}$$

However, when $\lambda = \theta$, the components have the following form,

$$\begin{aligned}
(\nabla T)_{rr\theta} &= \frac{\partial T_{rr}}{r\partial\theta} - \frac{T_{r\theta}}{r} - \frac{T_{\theta r}}{r} \\
(\nabla T)_{r\theta\theta} &= \frac{\partial T_{r\theta}}{r\partial\theta} + \frac{T_{rr}}{r} - \frac{T_{\theta\theta}}{r} \\
(\nabla T)_{rz\theta} &= \frac{\partial T_{rz}}{r\partial\theta} - \frac{T_{\theta z}}{r} \\
(\nabla T)_{\theta r\theta} &= \frac{\partial T_{\theta r}}{r\partial\theta} + \frac{T_{rr}}{r} - \frac{T_{\theta\theta}}{r} \\
(\nabla T)_{\theta\theta\theta} &= \frac{\partial T_{\theta\theta}}{r\partial\theta} + \frac{T_{r\theta}}{r} + \frac{T_{\theta r}}{r} \\
(\nabla T)_{\theta z\theta} &= \frac{\partial T_{\theta z}}{r\partial\theta} + \frac{T_{rz}}{r} \\
(\nabla T)_{zr\theta} &= \frac{\partial T_{zr}}{r\partial\theta} - \frac{T_{z\theta}}{r} \\
(\nabla T)_{z\theta\theta} &= \frac{\partial T_{z\theta}}{r\partial\theta} + \frac{T_{zr}}{r} \\
(\nabla T)_{zz\theta} &= \frac{\partial T_{zz}}{r\partial\theta}
\end{aligned} \tag{B.10}$$

Alternatively, Zhao and Pedroso [234] also give the 27 physical components of strain gradient directly derived from the displacement fields in cylindrical coordinates, considering the ***HRR*** solution of the dimensionless displacement fields have been obtained as $\tilde{u}_r(\theta)$ and $\tilde{u}_\theta(\theta)$ in **Section 2.3**. However, Swaddiwudhipong *et al*. [235] have argued that the strain gradient tensor, $\boldsymbol{\eta} = \nabla\nabla\boldsymbol{u}$, derived based on the total strain and deformation is applicable for expressions used in the effective plastic strain gradient ($\eta^p$) and presented the corresponding formulation in cylindrical coordinate system, since the elastic component is usually small and can be ignored [169].

For brevity, we further denote the covariant derivative for the components of the plastic strain field $\varepsilon^p = \varepsilon^p(r,\theta,z)$ as $\nabla\varepsilon^p_{ij,k}$, which is also equivalent to the representation $(\nabla\varepsilon^p)_{ijk}$, with $i,j,k \in \{r,\theta,z\}$. Then, the components of the plastic strain gradient tensor field are obtained via the cancellation of the Christoffel-symbols-related terms, meanwhile considering the symmetry of the strain field [234],

$$\eta^p_{rrr} = \nabla\varepsilon^p_{rr,r} + \nabla\varepsilon^p_{rr,r} - \nabla\varepsilon^p_{rr,r} = \nabla\varepsilon^p_{rr,r} \tag{B.11-1}$$

$$\eta^p_{r\theta r} = \nabla\varepsilon^p_{rr,\theta} + \nabla\varepsilon^p_{\theta r,r} - \nabla\varepsilon^p_{r\theta,r} = \nabla\varepsilon^p_{rr,\theta} \tag{B.11-2}$$

$$\eta^p_{rzr} = \nabla\varepsilon^p_{rr,z} + \nabla\varepsilon^p_{zr,r} - \nabla\varepsilon^p_{rz,r} = \nabla\varepsilon^p_{rr,z} \tag{B.11-3}$$

$$\eta^p_{\theta rr} = \nabla\varepsilon^p_{\theta r,r} + \nabla\varepsilon^p_{rr,\theta} - \nabla\varepsilon^p_{\theta r,r} = \nabla\varepsilon^p_{rr,\theta} \tag{B.11-4}$$

$$\eta^p_{\theta\theta r} = \nabla\varepsilon^p_{\theta r,\theta} + \nabla\varepsilon^p_{\theta r,\theta} - \nabla\varepsilon^p_{\theta\theta,r} = 2\nabla\varepsilon^p_{\theta r,\theta} - \nabla\varepsilon^p_{\theta\theta,r} \tag{B.11-5}$$

$$\eta^p_{\theta zr} = \nabla\varepsilon^p_{\theta r,z} + \nabla\varepsilon^p_{zr,\theta} - \nabla\varepsilon^p_{\theta z,r} \tag{B.11-6}$$

$$\eta^p_{zrr} = \nabla\varepsilon^p_{zr,r} + \nabla\varepsilon^p_{rr,z} - \nabla\varepsilon^p_{zr,r} = \nabla\varepsilon^p_{rr,z} \tag{B.11-7}$$

$$\eta^p_{z\theta r} = \nabla\varepsilon^p_{zr,\theta} + \nabla\varepsilon^p_{\theta r,z} - \nabla\varepsilon^p_{z\theta,r} \tag{B.11-8}$$

$$\eta^p_{zzr} = \nabla\varepsilon^p_{zr,z} + \nabla\varepsilon^p_{zr,z} - \nabla\varepsilon^p_{zz,r} = 2\nabla\varepsilon^p_{zr,z} - \nabla\varepsilon^p_{zz,r} \tag{B.11-9}$$

$$\eta^p_{rr\theta} = \nabla\varepsilon^p_{r\theta,r} + \nabla\varepsilon^p_{r\theta,r} - \nabla\varepsilon^p_{rr,\theta} = 2\nabla\varepsilon^p_{r\theta,r} - \nabla\varepsilon^p_{rr,\theta} \tag{B.11-10}$$

$$\eta^p_{r\theta\theta} = \nabla\varepsilon^p_{r\theta,\theta} + \nabla\varepsilon^p_{\theta\theta,r} - \nabla\varepsilon^p_{r\theta,\theta} = \nabla\varepsilon^p_{\theta\theta,r} \tag{B.11-11}$$

$$\eta^p_{rz\theta} = \nabla\varepsilon^p_{r\theta,z} + \nabla\varepsilon^p_{z\theta,r} - \nabla\varepsilon^p_{rz,\theta} \tag{B.11-12}$$

$$\eta^p_{\theta r\theta} = \nabla\varepsilon^p_{\theta\theta,r} + \nabla\varepsilon^p_{r\theta,\theta} - \nabla\varepsilon^p_{\theta r,\theta} = \nabla\varepsilon^p_{\theta\theta,r} \tag{B.11-13}$$

$$\eta^p_{\theta\theta\theta} = \nabla\varepsilon^p_{\theta\theta,\theta} + \nabla\varepsilon^p_{\theta\theta,\theta} - \nabla\varepsilon^p_{\theta\theta,\theta} = \nabla\varepsilon^p_{\theta\theta,\theta} \tag{B.11-14}$$

$$\eta^p_{\theta z\theta} = \nabla\varepsilon^p_{\theta\theta,z} + \nabla\varepsilon^p_{z\theta,\theta} - \nabla\varepsilon^p_{\theta z,\theta} = \nabla\varepsilon^p_{\theta\theta,z} \tag{B.11-15}$$

$$\eta^p_{zr\theta} = \nabla\varepsilon^p_{z\theta,r} + \nabla\varepsilon^p_{r\theta,z} - \nabla\varepsilon^p_{zr,\theta} \tag{B.11-16}$$

$$\eta^p_{z\theta\theta} = \nabla\varepsilon^p_{z\theta,\theta} + \nabla\varepsilon^p_{\theta\theta,z} - \nabla\varepsilon^p_{z\theta,\theta} = \nabla\varepsilon^p_{\theta\theta,z} \tag{B.11-17}$$

$$\eta^p_{zz\theta} = \nabla\varepsilon^p_{z\theta,z} + \nabla\varepsilon^p_{z\theta,z} - \nabla\varepsilon^p_{zz,\theta} = 2\nabla\varepsilon^p_{z\theta,z} - \nabla\varepsilon^p_{zz,\theta} \tag{B.11-18}$$

$$\eta^p_{rrz} = \nabla\varepsilon^p_{rz,r} + \nabla\varepsilon^p_{rz,r} - \nabla\varepsilon^p_{rr,z} = 2\nabla\varepsilon^p_{rz,r} - \nabla\varepsilon^p_{rr,z} \tag{B.11-19}$$

$$\eta^p_{r\theta z} = \nabla\varepsilon^p_{rz,\theta} + \nabla\varepsilon^p_{\theta z,r} - \nabla\varepsilon^p_{r\theta,z} \tag{B.11-20}$$

$$\eta^p_{rzz} = \nabla\varepsilon^p_{rz,z} + \nabla\varepsilon^p_{zz,r} - \nabla\varepsilon^p_{rz,z} = \nabla\varepsilon^p_{zz,r} \tag{B.11-21}$$

$$\eta^p_{\theta rz} = \nabla\varepsilon^p_{\theta z,r} + \nabla\varepsilon^p_{rz,\theta} - \nabla\varepsilon^p_{\theta r,z} \tag{B.11-22}$$

$$\eta^p_{\theta\theta z} = \nabla\varepsilon^p_{\theta z,\theta} + \nabla\varepsilon^p_{\theta z,\theta} - \nabla\varepsilon^p_{\theta\theta,z} = 2\nabla\varepsilon^p_{\theta z,\theta} - \nabla\varepsilon^p_{\theta\theta,z} \tag{B.11-23}$$

$$\eta^p_{\theta zz} = \nabla\varepsilon^p_{\theta z,z} + \nabla\varepsilon^p_{zz,\theta} - \nabla\varepsilon^p_{\theta z,z} = \nabla\varepsilon^p_{zz,\theta} \tag{B.11-24}$$

$$\eta^p_{zrz} = \nabla\varepsilon^p_{zz,r} + \nabla\varepsilon^p_{rz,z} - \nabla\varepsilon^p_{zr,z} = \nabla\varepsilon^p_{zz,r} \tag{B.11-25}$$

$$\eta^p_{z\theta z} = \nabla\varepsilon^p_{zz,\theta} + \nabla\varepsilon^p_{\theta z,z} - \nabla\varepsilon^p_{z\theta,z} = \nabla\varepsilon^p_{zz,\theta} \tag{B.11-26}$$

$$\eta^p_{zzz} = \nabla\varepsilon^p_{zz,z} + \nabla\varepsilon^p_{zz,z} - \nabla\varepsilon^p_{zz,z} = \nabla\varepsilon^p_{zz,z} \tag{B.11-27}$$

For plane problem, we only consider the 6 in-plane components:

$$\eta^p_{rrr} = \nabla\varepsilon^p_{rr,r} = \frac{\partial\varepsilon^p_{rr}}{\partial r} \tag{B.12-1}$$

$$\eta^p_{\theta\theta r} = 2\nabla\varepsilon^p_{\theta r,\theta} - \nabla\varepsilon^p_{\theta\theta,r} = 2\left(\frac{\partial\varepsilon^p_{r\theta}}{r\partial\theta} + \frac{\varepsilon^p_{rr}-\varepsilon^p_{\theta\theta}}{r}\right) - \frac{\partial\varepsilon^p_{\theta\theta}}{\partial r} \tag{B.12-2}$$

$$\eta^p_{r\theta r} = \nabla\varepsilon^p_{rr,\theta} = \frac{\partial\varepsilon^p_{rr}}{r\partial\theta} - \frac{2\varepsilon^p_{r\theta}}{r} \tag{B.12-3}$$

$$\eta^p_{rr\theta} = 2\nabla\varepsilon^p_{r\theta,r} - \nabla\varepsilon^p_{rr,\theta} = 2\frac{\partial\varepsilon^p_{r\theta}}{\partial r} - \left(\frac{\partial\varepsilon^p_{rr}}{r\partial\theta} - \frac{2\varepsilon^p_{r\theta}}{r}\right) \tag{B.12-4}$$

$$\eta^p_{\theta\theta\theta} = \nabla\varepsilon^p_{\theta\theta,\theta} = \frac{\partial\varepsilon^p_{\theta\theta}}{r\partial\theta} + \frac{2\varepsilon^p_{r\theta}}{r} \tag{B.12-5}$$

$$\eta^p_{r\theta\theta} = \nabla\varepsilon^p_{\theta\theta,r} = \frac{\partial\varepsilon^p_{\theta\theta}}{\partial r} \tag{B.12-6}$$

Finally, we can have **Eq.**(2.3.54).

# Appendix C: Numerical protocols

The essential knowledge required to advance the ***PRB*** model-based evaluation is the calculation of the ***GSFE*** curve, which should be a function of the hydrogen concentration. Without loss of generality, here I conduct the ***GSFE*** calculations of common metals (*e.g.*, $\alpha$-Fe and FeNiCr medium-entropy alloy (MEA)) by using the Large-scale Atomic/Molecular Massively Parallel Simulator (LAMMPS) code [236]. The interatomic interaction between Fe and H atoms in α-Fe system is described by using the Finnis-Sinclair type modification [25] of the embedded-atom-method (EAM) potential originally developed by Ramasubramaniam *et al*. [237] and modified by Song and Curtin [25]. The interatomic interaction in MEA system is described by using the EAM potential developed by Zhou *et al*. [238]. The time step for the velocity-Verlet integration is set as 1 fs. Periodic conditions are applied along *x*-, *y*- and *z*-axis directions. The simulation system of a cubic box is initially equilibrated at 300 K in the NPT ensemble. For FeNiCr MEA, additional relaxation is applied by using Monte Carlo method, then equilibrated within the NPT ensemble again. Subsequentially, hydrogen atoms are inserted in the simulation box randomly. The ***GSFE*** curve is evaluated by rigidly displacing two halves of the crystalline specimen, with the stacking sequence can be found in Ref.s [239, 240] for bcc and Ref.s [241, 242] for fcc lattice, respectively. **Fig. C1**-(a) shows the ***GSFE*** of the $\{112\}\langle 111\rangle$ slip system of the bcc Fe, with the Burgers vector $b = a\langle 111\rangle/2$, where $a$ is the lattice constant of Fe. **Fig. C1**-(b) shows that that the ***USF*** energy $\gamma_{usf}$ almost linearly increases with the increasing hydrogen concentration $c_H$.

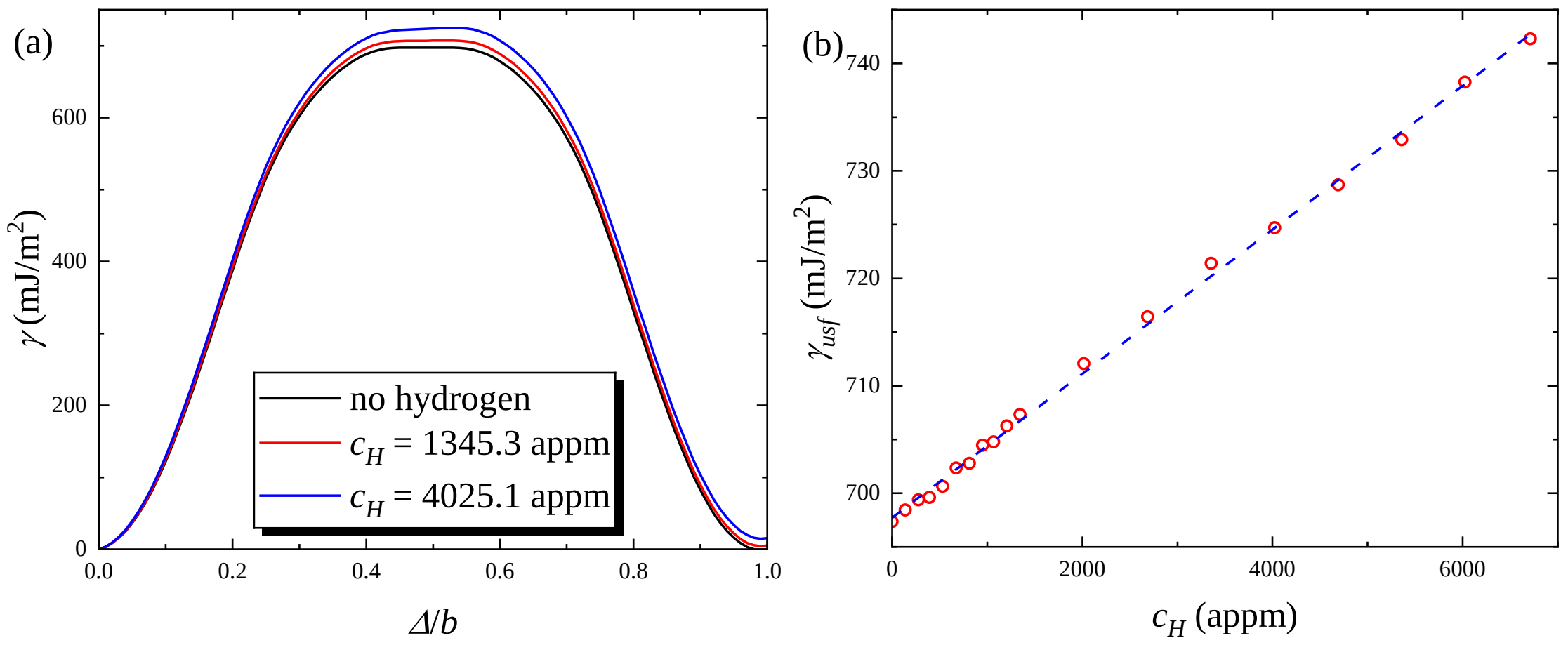

**Fig. C1**. (a) ***GSFE*** curves of the bcc Fe under various hydrogen concentrations, $c_H$ = 0, 139.3, 673.1, 1345.3, 2687.0, and 4025.1 appm, $b/a = \langle 111\rangle/2$; (b) $\gamma_{usf}$ as a function of the hydrogen concentration $c_H$, the blue dash line is fitted to the red open circles linearly with the slope $k$ = 0.0067 (mJ/m$^2$)/appm.

**Fig. C2**-(a) shows the ***GSFE*** curves of the $\{111\}\langle112\rangle$ slip system of the fcc FeNiCr MEA under different hydrogen concentrations, with the Burgers vector $b_p = a\langle11\bar{2}\rangle/6$, where $a$ is the lattice constant of FeNiCr MEA. **Fig. C2**-(b) shows that four parameters $\gamma_{usf}$, $\gamma_{ssf}$, $\gamma_{utf}$, and $\gamma_{stf}$ are linearly scaled with the hydrogen concentration $c_H$.

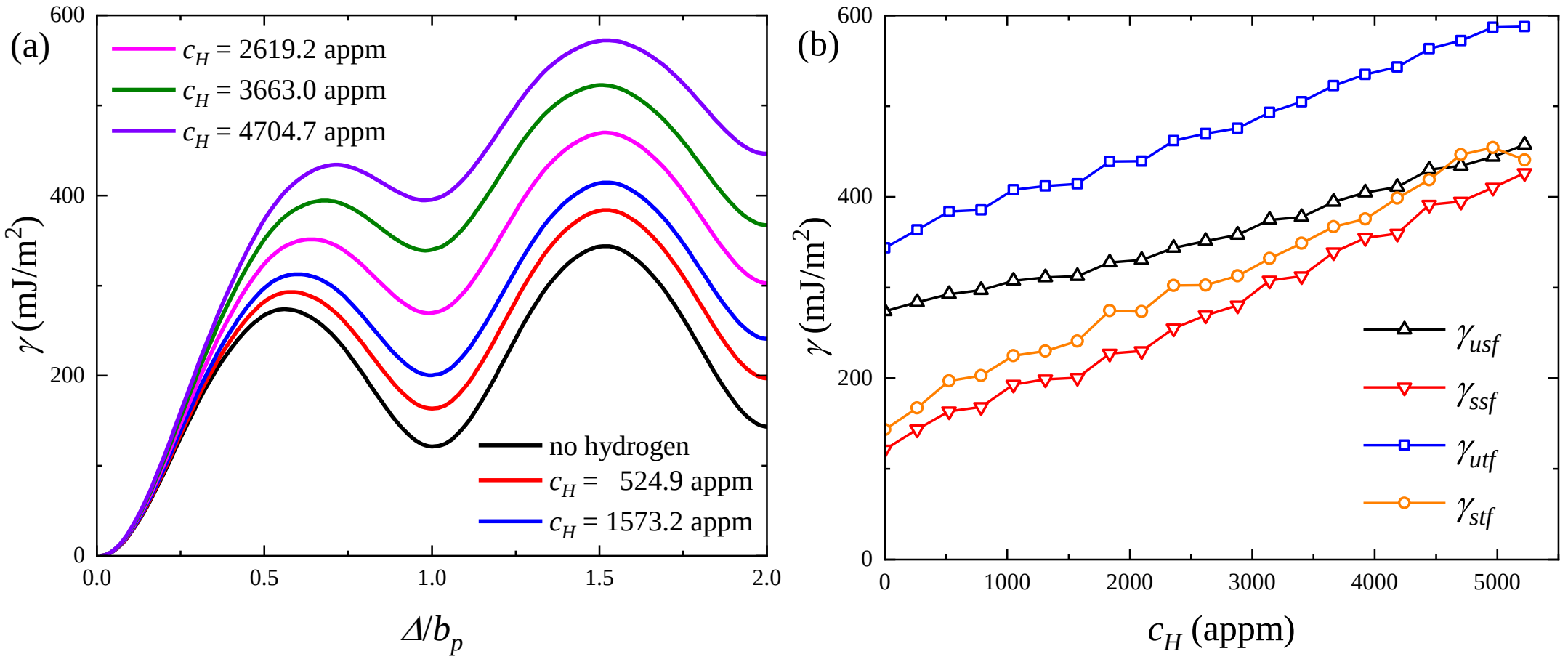

**Fig. C2**. (a) ***GSFE*** curves of the FeNiCr MEA under various hydrogen concentrations, $b_p/a = \langle11\bar{2}\rangle/6$; (b) $\gamma_{usf}$, $\gamma_{ssf}$, $\gamma_{utf}$ and $\gamma_{stf}$ as a function of the hydrogen concentration $c_H$.

## Data availability statement

Data will be made available on reasonable request.

# References

[1] I.M. Robertson, P. Sofronis, A. Nagao, M.L. Martin, S. Wang, D.W. Gross, K.E. Nygren, Hydrogen Embrittlement Understood, Metall. Mater. Trans. A 46 (2015) 2323-2341.
[2] H. Yu, A. Díaz, X. Lu, B. Sun, Y. Ding, M. Koyama, J. He, X. Zhou, A. Oudriss, X. Feaugas, Z. Zhang, Hydrogen Embrittlement as a Conspicuous Material Challenge - Comprehensive Review and Future Directions, Chem. Rev. 124 (2024) 6271-6392.
[3] A. Campari, F. Ustolin, A. Alvaro, N. Paltrinieri, A review on hydrogen embrittlement and risk-based inspection of hydrogen technologies, Int. J. Hydrog. Energy 48 (2023) 35316-35346.
[4] F.F. Dear, G.C.G. Skinner, Mechanisms of hydrogen embrittlement in steels: discussion, Philos. Trans. R. Soc. A Math. Phys. Eng. Sci. 375 (2017) 20170032.
[5] Y.-S. Chen, C. Huang, P.-Y. Liu, H.-W. Yen, R. Niu, P. Burr, K.L. Moore, E. Martínez-Pañeda, A. Atrens, J.M. Cairney, Hydrogen trapping and embrittlement in metals – A review, Int. J. Hydrog. Energy 136 (2025) 789-821.
[6] O. Barrera, D. Bombac, Y. Chen, T.D. Daff, E. Galindo-Nava, P. Gong, D. Haley, R. Horton, I. Katzarov, J.R. Kermode, C. Liverani, M. Stopher, F. Sweeney, Understanding and mitigating hydrogen embrittlement of steels: a review of experimental, modelling and design progress from atomistic to continuum, J. Mater. Sci. 53 (2018) 6251-6290.
[7] J. Zheng, X. Liu, P. Xu, P. Liu, Y. Zhao, J. Yang, Development of high pressure gaseous hydrogen storage technologies, Int. J. Hydrog. Energy 37 (2012) 1048-1057.
[8] P.C. Okonkwo, E.M. Barhoumi, I. Ben Belgacem, I.B. Mansir, M. Aliyu, W. Emori, P.C. Uzoma, W.H. Beitelmal, E. Akyüz, A.B. Radwan, R.A. Shakoor, A focused review of the hydrogen storage tank embrittlement mechanism process, Int. J. Hydrog. Energy 48 (2023) 12935-12948.
[9] W. Liu, J. Zhang, L. Qu, B. Wei, M. Niu, C. Sun, Hydrogen embrittlement of notched X80 steel under high-pressure gaseous hydrogen: insights from hollow specimen testing and finite element analysis, Int. J. Hydrog. Energy 188 (2025) 152125.

[10] W.H. Johnson, On Some Remarkable Changes Produced in Iron and Steel by the Action of Hydrogen and Acids, Nature 11 (1875) 393-393.
[11] M. Koyama, C.C. Tasan, E. Akiyama, K. Tsuzaki, D. Raabe, Hydrogen-assisted decohesion and localized plasticity in dual-phase steel, Acta Mater. 70 (2014) 174-187.
[12] A. Tehranchi, X. Zhou, W.A. Curtin, A decohesion pathway for hydrogen embrittlement in nickel: Mechanism and quantitative prediction, Acta Mater. 185 (2020) 98-109.
[13] A.R. Troiano, The Role of Hydrogen and Other Interstitials in the Mechanical Behavior of Metals, Trans. Am. Soc. Met. 52 (1960) 54-80.
[14] R.A. Oriani, P.H. Josephic, Equilibrium aspects of hydrogen-induced cracking of steels, Acta Metall. 22 (1974) 1065-1074.
[15] C.D. Beachem, A new model for hydrogen-assisted cracking (hydrogen “embrittlement”), Metall. Mater. Trans. B 3 (1972) 441-455.
[16] H.K. Birnbaum, P. Sofronis, Hydrogen-enhanced localized plasticity - a mechanism for hydrogen-related fracture, Mater. Sci. Eng. A 176 (1994) 191-202.
[17] S.M. Myers, M.I. Baskes, H.K. Birnbaum, J.W. Corbett, G.G. DeLeo, S.K. Estreicher, E.E. Haller, P. Jena, N.M. Johnson, R. Kirchheim, S.J. Pearton, M.J. Stavola, Hydrogen interactions with defects in crystalline solids, Rev. Mod. Phys. 64 (1992) 559-617.
[18] S. Laliberté-Riverin, J. Bellemare, F. Sirois, M. Brochu, Determination of hydrogen embrittlement stress intensity threshold by fractography, Materialia 12 (2020) 100759.
[19] S.P. Lynch, Hydrogen embrittlement and liquid-metal embrittlement in nickel single crystals, Scr. Metall. 13 (1979) 1051-1056.
[20] T. Neeraj, R. Srinivasan, J. Li, Hydrogen embrittlement of ferritic steels: Observations on deformation microstructure, nanoscale dimples and failure by nanovoiding, Acta Mater. 60 (2012) 5160-5171.
[21] S. Li, Y. Li, Y.-C. Lo, T. Neeraj, R. Srinivasan, X. Ding, J. Sun, L. Qi, P. Gumbsch, J. Li, The interaction of dislocations and hydrogen-vacancy complexes and its importance for deformation-induced proto nano-voids formation in α-Fe, Int. J. Plast. 74 (2015) 175-191.
[22] M. Nagumo, K. Takai, The predominant role of strain-induced vacancies in hydrogen embrittlement of steels: Overview, Acta Mater. 165 (2019) 722-733.
[23] I.H. Katzarov, A.T. Paxton, Hydrogen embrittlement II. Analysis of hydrogen-enhanced decohesion across (111) planes in $\ensuremath{\alpha}$-Fe, Phys. Rev. Mater. 1 (2017) 033603.
[24] P. Gong, A. Turk, J. Nutter, F. Yu, B. Wynne, P. Rivera-Diaz-del-Castillo, W. Mark Rainforth, Hydrogen embrittlement mechanisms in advanced high strength steel, Acta Mater. 223 (2022) 117488.
[25] J. Song, W.A. Curtin, Atomic mechanism and prediction of hydrogen embrittlement in iron, Nat. Mater. 12 (2013) 145-151.
[26] P. Sofronis, H.K. Birnbaum, Mechanics of the hydrogen-dislocation-impurity interactions - I. Increasing shear modulus, J. Mech. Phys. Solids 43 (1995) 49-90.
[27] I.M. Robertson, The effect of hydrogen on dislocation dynamics, Eng. Fract. Mech. 68 (2001) 671-692.
[28] M.L. Martin, M. Dadfarnia, A. Nagao, S. Wang, P. Sofronis, Enumeration of the hydrogen-enhanced localized plasticity mechanism for hydrogen embrittlement in structural materials, Acta Mater. 165 (2019) 734-750.
[29] V. Bulatov, W. Cai, One dislocation at a time, Nat. Mater. 22 (2023) 679-680.
[30] T. Richeton, J. Weiss, F. Louchet, Dislocation avalanches: Role of temperature, grain size and strain hardening, Acta Mater. 53 (2005) 4463-4471.
[31] T. Richeton, P. Dobron, F. Chmelik, J. Weiss, F. Louchet, On the critical character of plasticity in metallic single crystals, Mater. Sci. Eng. A 424 (2006) 190-195.
[32] J.R. Greer, J.-Y. Kim, M.J. Burek, The in-situ mechanical testing of nanoscale single-crystalline nanopillars, JOM 61 (2009) 19-25.
[33] R. Peierls, The size of a dislocation, Proceedings of the Physical Society 52 (1940) 34.
[34] F.R.N. Nabarro, Dislocations in a simple cubic lattice, Proceedings of the Physical Society 59 (1947) 256.
[35] B. Joós, Q. Ren, M.S. Duesbery, Peierls-Nabarro model of dislocations in silicon with generalized stacking-fault restoring forces, Phys. Rev. B 50 (1994) 5890-5898.
[36] B. Joós, M.S. Duesbery, The Peierls Stress of Dislocations: An Analytic Formula, Phys. Rev. Lett. 78 (1997) 266-269.
[37] G. Lu, The Peierls—Nabarro Model of Dislocations: A Venerable Theory and its Current Development, in: S. Yip (Ed.), Handbook of Materials Modeling: Methods, Springer Netherlands, Dordrecht, 2005, pp. 793-811.
[38] S. Patrizi, T. Sangsawang, From the Peierls–Nabarro model to the equation of motion of the dislocation continuum, Nonlinear Analysis 202 (2021) 112096.
[39] G. Schoeck, The Peierls model: Progress and limitations, Mater. Sci. Eng. A 400-401 (2005) 7-17.
[40] Y. Yao, T.C. Wang, The modified Peierls–Nabarro model of interfacial misfit dislocation, Acta Mater. 47 (1999) 3063-3068.
[41] H.B. Huntington, Modification of the Peierls-Nabarro Model for Edge Dislocation Core, Proceedings of the Physical Society. Section B 68 (1955) 1043.
[42] Y. Xiang, H. Wei, P. Ming, W. E, A generalized Peierls–Nabarro model for curved dislocations and core structures of dislocation loops in Al and Cu, Acta Mater. 56 (2008) 1447-1460.
[43] G. Schoeck, The generalized Peierls–Nabarro model, Philos. Mag. A 69 (1994) 1085-1095.
[44] Y.-H. Li, H.-B. Zhou, F. Gao, G. Lu, G.-H. Lu, F. Liu, Hydrogen induced dislocation core reconstruction in bcc tungsten, Acta Mater. 226 (2022) 117622.
[45] P. Yu, Y. Cui, G.-z. Zhu, Y. Shen, M. Wen, The key role played by dislocation core radius and energy in hydrogen interaction with dislocations, Acta Mater. 185 (2020) 518-527.
[46] M.S. Hasan, M.F. Kapci, B. Bal, M. Koyama, H. Bayat, W. Xu, An atomistic study on the HELP mechanism of hydrogen embrittlement in pure metal Fe, Int. J. Hydrog. Energy 57 (2024) 60-68.
[47] M.S. Hasan, H. Bayat, C. Delaney, C. Foronda, W. Xu, Hydrogen-Induced Transformation of Dislocation Core in Fe and Its Effect on Dislocation Mobility, TMS 2024 153rd Annual Meeting & Exhibition Supplemental Proceedings, Springer Nature Switzerland, Cham, 2024, pp. 1000-1007.
[48] C. Nowak, X.W. Zhou, An interplay between a hydrogen atmosphere and dislocation characteristics in BCC Fe from time-averaged molecular dynamics, Phys. Chem. Chem. Phys. 25 (2023) 8369-8375.
[49] G.P.M. Leyson, B. Grabowski, J. Neugebauer, Multiscale description of dislocation induced nano-hydrides, Acta Mater. 89 (2015) 50-59.
[50] L. Zhang, Y. Xiang, J. Han, D.J. Srolovitz, The effect of randomness on the strength of high-entropy alloys, Acta Mater. 166 (2019) 424-434.
[51] G. Stenerud, R. Johnsen, J.S. Olsen, J. He, A. Barnoush, Effect of hydrogen on dislocation nucleation in alloy 718, Int. J. Hydrog. Energy 42 (2017) 15933-15942.
[52] A. Barnoush, H. Vehoff, Recent developments in the study of hydrogen embrittlement: Hydrogen effect on dislocation nucleation, Acta Mater. 58 (2010) 5274-5285.

[53] D. Wang, X. Lu, Y. Deng, X. Guo, A. Barnoush, Effect of hydrogen on nanomechanical properties in Fe-22Mn-0.6C TWIP steel revealed by in-situ electrochemical nanoindentation, Acta Mater. 166 (2019) 618-629.
[54] R. Kirchheim, Reducing grain boundary, dislocation line and vacancy formation energies by solute segregation. I. Theoretical background, Acta Mater. 55 (2007) 5129-5138.
[55] R. Kirchheim, Reducing grain boundary, dislocation line and vacancy formation energies by solute segregation: II. Experimental evidence and consequences, Acta Mater. 55 (2007) 5139-5148.
[56] R. Kirchheim, Revisiting hydrogen embrittlement models and hydrogen-induced homogeneous nucleation of dislocations, Scr. Mater. 62 (2010) 67-70.
[57] R. Kirchheim, Solid solution softening and hardening by mobile solute atoms with special focus on hydrogen, Scr. Mater. 67 (2012) 767-770.
[58] G.P.M. Leyson, B. Grabowski, J. Neugebauer, Multiscale modeling of hydrogen enhanced homogeneous dislocation nucleation, Acta Mater. 107 (2016) 144-151.
[59] P.J. Ferreira, I.M. Robertson, H.K. Birnbaum, Hydrogen effects on the interaction between dislocations, Acta Mater. 46 (1998) 1749-1757.
[60] L. Huang, D. Chen, D. Xie, S. Li, Y. Zhang, T. Zhu, D. Raabe, E. Ma, J. Li, Z. Shan, Quantitative tests revealing hydrogen-enhanced dislocation motion in α-iron, Nat. Mater. 22 (2023) 710-716.
[61] Y. Ogawa, M. Tanaka, T. Fujita, A. Shibata, Thermally activated dislocation motion in hydrogen-alloyed Fe–Cr–Ni austenitic steel revisited via Haasen plot, Int. J. Hydrog. Energy 74 (2024) 170-182.
[62] D. Xie, S. Li, M. Li, Z. Wang, P. Gumbsch, J. Sun, E. Ma, J. Li, Z. Shan, Hydrogenated vacancies lock dislocations in aluminium, Nat. Commun. 7 (2016) 13341.
[63] M.F. Baltacioglu, M.F. Kapci, J.C. Schön, J. Marian, B. Bal, A phenomenological hydrogen induced edge dislocation mobility law for bcc Fe obtained by molecular dynamics, Int. J. Hydrog. Energy 87 (2024) 917-927.
[64] M.F. Kapci, J.C. Schön, B. Bal, The role of hydrogen in the edge dislocation mobility and grain boundary-dislocation interaction in α-Fe, Int. J. Hydrog. Energy 46 (2021) 32695-32709.
[65] K.-S. Kim, Q.-J. Li, J. Li, C.C. Tasan, Hydrogen can both move or pin dislocations in body-centered cubic metals, Nat. Commun. 16 (2025) 3936.
[66] I.H. Katzarov, D.L. Pashov, A.T. Paxton, Hydrogen embrittlement I. Analysis of hydrogen-enhanced localized plasticity: Effect of hydrogen on the velocity of screw dislocations in $\ensuremath{\alpha}$-Fe, Phys. Rev. Mater. 1 (2017) 033602.
[67] Y. Murakami, T. Kanezaki, Y. Mine, Hydrogen Effect against Hydrogen Embrittlement, Metall. Mater. Trans. A 41 (2010) 2548-2562.
[68] A.M.Z. Tan, Z. Li, Y. Zhao, U. Ramamurty, H. Gao, Modeling the improved hydrogen embrittlement tolerance of twin boundaries in face-centered cubic complex concentrated alloys, J. Mech. Phys. Solids 188 (2024) 105657.
[69] S.S. Shishvan, G. Csányi, V.S. Deshpande, Hydrogen induced fast-fracture, J. Mech. Phys. Solids 134 (2020) 103740.
[70] P. Novak, R. Yuan, B.P. Somerday, P. Sofronis, R.O. Ritchie, A statistical, physical-based, micro-mechanical model of hydrogen-induced intergranular fracture in steel, J. Mech. Phys. Solids 58 (2010) 206-226.
[71] M. Dadfarnia, M.L. Martin, A. Nagao, P. Sofronis, I.M. Robertson, Modeling hydrogen transport by dislocations, J. Mech. Phys. Solids 78 (2015) 511-525.
[72] D. Di Stefano, M. Mrovec, C. Elsässer, First-principles investigation of hydrogen trapping and diffusion at grain boundaries in nickel, Acta Mater. 98 (2015) 306-312.
[73] S. Wang, M.L. Martin, I.M. Robertson, P. Sofronis, Effect of hydrogen environment on the separation of Fe grain boundaries, Acta Mater. 107 (2016) 279-288.
[74] D. Di Stefano, R. Nazarov, T. Hickel, J. Neugebauer, M. Mrovec, C. Elsässer, First-principles investigation of hydrogen interaction with TiC precipitates in α-Fe, Phys. Rev. B 93 (2016) 184108.
[75] Y.-S. Chen, H. Lu, J. Liang, A. Rosenthal, H. Liu, G. Sneddon, I. McCarroll, Z. Zhao, W. Li, A. Guo, J.M. Cairney, Observation of hydrogen trapping at dislocations, grain boundaries, and precipitates, Science 367 (2020) 171.
[76] L. Vandewalle, T. Depover, K. Verbeken, Current state-of-the-art of hydrogen trapping by carbides: From theory to experiment, Int. J. Hydrog. Energy 136 (2025) 888-901.
[77] T. Depover, K. Verbeken, The detrimental effect of hydrogen at dislocations on the hydrogen embrittlement susceptibility of Fe-C-X alloys: An experimental proof of the HELP mechanism, Int. J. Hydrog. Energy 43 (2018) 3050-3061.
[78] Y. Wang, B. Sharma, Y. Xu, K. Shimizu, H. Fujihara, K. Hirayama, A. Takeuchi, M. Uesugi, G. Cheng, H. Toda, Switching nanoprecipitates to resist hydrogen embrittlement in high-strength aluminum alloys, Nat. Commun. 13 (2022) 6860.
[79] A. Nagao, M.L. Martin, M. Dadfarnia, P. Sofronis, I.M. Robertson, The effect of nanosized (Ti,Mo)C precipitates on hydrogen embrittlement of tempered lath martensitic steel, Acta Mater. 74 (2014) 244-254.
[80] V.S. Krasnikov, P.A. Bezborodova, A.E. Mayer, Effect of hydrogen accumulation on θ' precipitates on the shear strength of Al-Cu alloys, Int. J. Plast. (2022) 103475.
[81] C.-T. Santos Maldonado, A. Zafra, E. Martínez Pañeda, P. Sandmann, R. Morana, M.-S. Pham, Influence of dislocation cells on hydrogen embrittlement in wrought and additively manufactured Inconel 718, Communications Materials 5 (2024) 223.
[82] W. Chen, W. Zhao, P. Gao, F. Li, S. Kuang, Y. Zou, Z. Zhao, Interaction between dislocations, precipitates and hydrogen atoms in a 2000 MPa grade hot-stamped steel, J. Mater. Res. Technol. 18 (2022) 4353-4366.
[83] J. Li, C. Lu, L. Pei, C. Zhang, R. Wang, Hydrogen-modified interaction between lattice dislocations and grain boundaries by atomistic modelling, Int. J. Hydrog. Energy 45 (2020) 9174-9187.
[84] M. Koyama, S.M. Taheri-Mousavi, H. Yan, J. Kim, B.C. Cameron, S.S. Moeini-Ardakani, J. Li, C.C. Tasan, Origin of micrometer-scale dislocation motion during hydrogen desorption, Sci. Adv. 6 (2020) eaaz1187.
[85] Y.G. Li, W.H. Zhou, R.H. Ning, L.F. Huang, Z. Zeng, X. Ju, A Cluster Dynamics Model For Accumulation Of Helium In Tungsten Under Helium Ions And Neutron Irradiation, Communications in Computational Physics 11 (2012) 1547-1568.
[86] S.I. Golubov, A.M. Ovcharenko, A.V. Barashev, B.N. Singh, Grouping method for the approximate solution of a kinetic equation describing the evolution of point-defect clusters, Philos. Mag. A 81 (2001) 643-658.
[87] K. Zhao, J. He, A.E. Mayer, Z. Zhang, Effect of hydrogen on the collective behavior of dislocations in the case of nanoindentation, Acta Mater. 148 (2018) 18-27.
[88] J.M. Tarp, L. Angheluta, J. Mathiesen, N. Goldenfeld, Intermittent Dislocation Density Fluctuations in Crystal Plasticity from a Phase-Field Crystal Model, Phys. Rev. Lett. 113 (2014) 265503.
[89] Y. Lu, Y.-H. Zhang, E. Ma, W.-Z. Han, Relative mobility of screw versus edge dislocations controls the ductile-to-brittle transition in metals, Proc. Natl. Acad. Sci. U.S.A. 118 (2021) e2110596118.

[90] I. Finnie, R.A. Mayville, Historical Aspects in Our Understanding of the Ductile-Brittle Transition in Steels, Journal of Engineering Materials and Technology 112 (1990) 56-60.
[91] J.R. Rice, R. Thomson, Ductile Versus Brittle Behavior of Crystals, Philos. Mag. 29 (1974) 73-97.
[92] J.R. Rice, Dislocation Nucleation from a Crack Tip - an Analysis Based on the Peierls Concept, J. Mech. Phys. Solids 40 (1992) 239-271.
[93] J.R. Rice, G.E. Beltz, The activation energy for dislocation nucleation at a crack, J. Mech. Phys. Solids 42 (1994) 333-360.
[94] P.B. Hirsch, S.G. Roberts, J. Samuels, The brittle-ductile transition in silicon. II. Interpretation, Proc. R. Soc. Lond. A: Math. Phys. Sci. 421 (1989) 25-53.
[95] A. Giannattasio, S.G. Roberts, Strain-rate dependence of the brittle-to-ductile transition temperature in tungsten, Philos. Mag. 87 (2007) 2589-2598.
[96] P. Gumbsch, J. Riedle, A. Hartmaier, H.F. Fischmeister, Controlling Factors for the Brittle-to-Ductile Transition in Tungsten Single Crystals, Science 282 (1998) 1293-1295.
[97] Y. Sun, G.E. Beltz, Dislocation nucleation from a crack tip: A formulation based on anisotropic elasticity, J. Mech. Phys. Solids 42 (1994) 1905-1932.
[98] L. L. Fischer, G. E. Beltz, The effect of crack blunting on the competition between dislocation nucleation and cleavage, J. Mech. Phys. Solids 49 (2001) 635-654.
[99] S.J. Zhou, A.E. Carlsson, R. Thomson, Crack blunting effects on dislocation emission from cracks, Phys. Rev. Lett. 72 (1994) 852-855.
[100] G. Xu, A.S. Argon, M. Ortiz, Nucleation of dislocations from crack tips under mixed modes of loading: Implications for brittle against ductile behaviour of crystals, Philos. Mag. A 72 (1995) 415-451.
[101] P. Andric, W.A. Curtin, New theory for Mode I crack-tip dislocation emission, J. Mech. Phys. Solids 106 (2017) 315-337.
[102] X. Li, W. Li, D.L. Irving, L.K. Varga, L. Vitos, S. Schönecker, Ductile and brittle crack-tip response in equimolar refractory high-entropy alloys, Acta Mater. 189 (2020) 174-187.
[103] K. Zhao, F. Zhao, Q. Lin, X. Li, J. Xiao, Y. Gu, Q. Chen, Effect of loading rate on the dislocation emission from crack-tip under hydrogen environment, Mater. Today Commun. 37 (2023) 107269.
[104] K. Zhao, Y. Ding, H. Yu, J. He, Z. Zhang, An extended Rice model for intergranular fracture, Int. J. Mech. Sci. 286 (2025) 109891.
[105] T. Zhu, W. Yang, T. Guo, Quasi-cleavage processes driven by dislocation pileups, Acta Mater. 44 (1996) 3049-3058.
[106] M.J. Lii, X.F. Chen, Y. Katz, W.W. Gerberich, Dislocation modeling and acoustic emission observation of alternating ductile/brittle events in Fe-3wt%Si crystals, Acta Metall. Mater. 38 (1990) 2435-2453.
[107] W. Zielinski, M.J. Lii, W.W. Gerberich, Crack-tip dislocation emission arrangements for equilibrium —I. In situ TEM observations of Fe2wt%Si, Acta Metall. Mater. 40 (1992) 2861-2871.
[108] H. Huang, W.W. Gerberich, Crack-tip dislocation emission arrangements for equilibrium—II. Comparisons to analytical and computer simulation models, Acta Metall. Mater. 40 (1992) 2873-2881.
[109] P.G. Marsh, W. Zielinski, H. Huang, W.W. Gerberich, Crack-tip dislocation emission arrangements for equilibrium—III. Application to large applied stress intensities, Acta Metall. Mater. 40 (1992) 2883-2894.
[110] C.S. John, The brittle-to-ductile transition in pre-cleaved silicon single crystals, Philos. Mag. 32 (1975) 1193-1212.
[111] M.L. Martin, J.A. Fenske, G.S. Liu, P. Sofronis, I.M. Robertson, On the formation and nature of quasi-cleavage fracture surfaces in hydrogen embrittled steels, Acta Mater. 59 (2011) 1601-1606.
[112] M.L. Martin, I.M. Robertson, P. Sofronis, Interpreting hydrogen-induced fracture surfaces in terms of deformation processes: A new approach, Acta Mater. 59 (2011) 3680-3687.
[113] N.I. Muskhelishvili, Some Basic Problems of the Mathematical Theory of Elasticity, Springer Dordrecht, 1977.
[114] G.R. Irwin, Analysis of Stresses and Strains Near the End of a Crack Traversing a Plate, J. Appl. Mech. 24 (1957) 361-364.
[115] M. Creager, P.C. Paris, Elastic field equations for blunt cracks with reference to stress corrosion cracking, Int. J. Fract. Mech. 3 (1967) 247-252.
[116] M. Huang, Z. Li, Dislocation emission criterion from a blunt crack tip, J. Mech. Phys. Solids 52 (2004) 1991-2003.
[117] V. Lakshmanan, J.C.M. Li, Edge dislocations emitted along inclined planes from a mode I crack, Mater. Sci. Eng. A 104 (1988) 95-104.
[118] K. Zhou, A.A. Nazarov, M.S. Wu, Atomistic simulations of the tensile strength of a disclinated bicrystalline nanofilm, Philos. Mag. 88 (2008) 3181-3191.
[119] T. Shimokawa, M. Tsuboi, Atomic-scale intergranular crack-tip plasticity in tilt grain boundaries acting as an effective dislocation source, Acta Mater. 87 (2015) 233-247.
[120] D.H. Warner, W.A. Curtin, S. Qu, Rate dependence of crack-tip processes predicts twinning trends in f.c.c. metals, Nat. Mater. 6 (2007) 876-881.
[121] D.H. Warner, W.A. Curtin, Origins and implications of temperature-dependent activation energy barriers for dislocation nucleation in face-centered cubic metals, Acta Mater. 57 (2009) 4267-4277.
[122] T. Zhu, J. Li, A. Samanta, A. Leach, K. Gall, Temperature and Strain-Rate Dependence of Surface Dislocation Nucleation, Phys. Rev. Lett. 100 (2008) 025502.
[123] K. Zhao, I.G. Ringdalen, J.Y. Wu, J.Y. He, Z.L. Zhang, Ductile mechanisms of metals containing pre-existing nanovoids, Comput. Mater. Sci. 125 (2016) 36-50.
[124] S.M.O. Tavares, P.M.S.T. de Castro, Equivalent Stress Intensity Factor: The Consequences of the Lack of a Unique Definition, Applied Sciences, 2023, p. 4820.
[125] C.R. Weinberger, A.T. Jennings, K. Kang, J.R. Greer, Atomistic simulations and continuum modeling of dislocation nucleation and strength in gold nanowires, J. Mech. Phys. Solids 60 (2012) 84-103.
[126] Z. Zhao, Y. Wei, Intrinsic characteristics of grain boundary elimination induced by plastic deformation in front of intergranular microcracks in bcc iron, Int. J. Plast. 184 (2025) 104208.
[127] J. Chen, S. Takezono, The dislocation-free zone at a mode I crack tip, Eng. Fract. Mech. 50 (1995) 165-173.
[128] R. Thomson, Brittle-Fracture in a Ductile Material with Application to Hydrogen Embrittlement, J. Mater. Sci. 13 (1978) 128-142.
[129] D.M. Lipkin, G.E. Beltz, A simple elastic cell model of cleavage fracture in the presence of dislocation plasticity, Acta Mater. 44 (1996) 1287-1291.
[130] J.W. Hutchinson, Singular behaviour at the end of a tensile crack in a hardening material, J. Mech. Phys. Solids 16 (1968) 13-31.
[131] J.W. Hutchinson, Plastic stress and strain fields at a crack tip, J. Mech. Phys. Solids 16 (1968) 337-342.
[132] J.R. Rice, G.F. Rosengren, Plane strain deformation near a crack tip in a power-law hardening material, J. Mech. Phys. Solids 16 (1968) 1-12.

[133] C.F. Shih, J.W. Hutchinson, Fully Plastic Solutions and Large Scale Yielding Estimates for Plane Stress Crack Problems, Journal of Engineering Materials and Technology 98 (1976) 289-295.
[134] C.F. Shih, M.D. German, Requirements for a one parameter characterization of crack tip fields by the HRR singularity, Int. J. Fract. 17 (1981) 27-43.
[135] D. Hull, D.J. Bacon, Chapter 10 - Strength of Crystalline Solids, in: D. Hull, D.J. Bacon (Eds.), Introduction to Dislocations (Fifth Edition), Butterworth-Heinemann, Oxford, 2011, pp. 205-249.
[136] J.-S. Wang, The thermodynamics aspects of hydrogen induced embrittlement, Eng. Fract. Mech. 68 (2001) 647-669.
[137] J.P. Hirth, J.R. Rice, On the thermodynamics of adsorption at interfaces as it influences decohesion, Metall. Trans. A 11 (1980) 1501-1511.
[138] D.S. Lieberman, S. Zirinsky, A simplified calculation for the elatic constants of arbitrarily oriented single crystals, Acta Crystallogr. 9 (1956) 431-436.
[139] D.S. Lieberman, S. Zirinsky, A simplified calculation for the elastic constants of arbitrarily oriented single crystals: correction, Acta Crystallogr. 10 (1957) 242.
[140] M.I. Mendelev, S. Han, D.J. Srolovitz, G.J. Ackland, D.Y. Sun, M. Asta, Development of new interatomic potentials appropriate for crystalline and liquid iron, Philos. Mag. 83 (2003) 3977-3994.
[141] L. Zhang, G. Csányi, E. van der Giessen, F. Maresca, Atomistic fracture in bcc iron revealed by active learning of Gaussian approximation potential, npj Computational Materials 9 (2023) 217.
[142] G.C. Sih, P.C. Paris, G.R. Irwin, On cracks in rectilinearly anisotropic bodies, Int. J. Fract. Mech. 1 (1965) 189-203.
[143] A.P. Sutton, R.W. Balluffi, Interfaces in Crystalline Materials, Clarendon, Oxford, 1995.
[144] M. Lane, Interface Fracture, Annu. Rev. Mater. Res. 33 (2003) 29-54.
[145] J.J. Möller, E. Bitzek, Fracture toughness and bond trapping of grain boundary cracks, Acta Mater. 73 (2014) 1-11.
[146] E. Smith, The Nucleation and Growth of Cleavage Microcracks in Mild Steel, Proceedings of Physical Basis of Yield and Fracture (1966) 36-46.
[147] M.L. Jokl, V. Vitek, C.J. McMahon, A microscopic theory of brittle fracture in deformable solids: A relation between ideal work to fracture and plastic work, Acta Metall. 28 (1980) 1479-1488.
[148] C.J. McMahon, V. Vitek, The effects of segregated impurities on intergranular fracture energy, Acta Metall. 27 (1979) 507-513.
[149] A.H.M. Krom, R.W.J. Koers, A. Bakker, Hydrogen transport near a blunting crack tip, J. Mech. Phys. Solids 47 (1999) 971-992.
[150] J.P. Hirth, Effects of hydrogen on the properties of iron and steel, Metall. Trans. A 11 (1980) 861-890.
[151] P. Sofronis, R.M. McMeeking, Numerical analysis of hydrogen transport near a blunting crack tip, J. Mech. Phys. Solids 37 (1989) 317-350.
[152] A. Shibata, T. Yonemura, Y. Momotani, M.-h. Park, S. Takagi, Y. Madi, J. Besson, N. Tsuji, Effects of local stress, strain, and hydrogen content on hydrogen-related fracture behavior in low-carbon martensitic steel, Acta Mater. 210 (2021) 116828.
[153] W.A. Jesser, The role of frictional stress on the generation of misfit dislocations, South African Journal of Science 104 (2008) 379-382.
[154] F.R.N. Nabarro, The influence of elastic strain on the shape of particles segregating in an alloy, Proceedings of the Physical Society 52 (1940) 90.
[155] F.R.N. Nabarro, The strains produced by precipitation in alloys, Proceedings of the Royal Society of London. Series A. Mathematical and Physical Sciences 175 (1940) 519-538.
[156] W.A. Jesser, On the theory of loss of coherency by spherical precipitates, Philos. Mag. 19 (1969) 993-999.
[157] I. Chattoraj, S.B. Tiwari, A.K. Ray, A. Mitra, S.K. Das, Investigation on the mechanical degradation of a steel line pipe due to hydrogen ingress during exposure to a simulated sour environment, Corros. Sci. 37 (1995) 885-896.
[158] C.F. Dong, Z.Y. Liu, X.G. Li, Y.F. Cheng, Effects of hydrogen-charging on the susceptibility of X100 pipeline steel to hydrogen-induced cracking, Int. J. Hydrog. Energy 34 (2009) 9879-9884.
[159] Y. Wang, X. Wu, Z. Zhou, X. Li, Numerical analysis of hydrogen transport into a steel after shot peening, Results in Physics 11 (2018) 5-16.
[160] T. Lin, A.G. Evans, R.O. Ritchie, A statistical model of brittle fracture by transgranular cleavage, J. Mech. Phys. Solids 34 (1986) 477-497.
[161] W. Weibull, A statistical theory of the strength of materials, Ing. Vetenskap. Akad. Handl. 12 (1939) 153.
[162] A.M. Cuitiño, M. Ortiz, Ductile fracture by vacancy condensation in f.c.c. single crystals, Acta Mater. 44 (1996) 427-436.
[163] W. Yang, Macro and Micro-scale Fracture Mechanics, Defence Industry Press, 1995.
[164] T.C. Wang, S.H. Chen, Advanced Fracture Mechanics, Science Press, 2009.
[165] V. Shlyannikov, A. Tumanov, Characterization of crack tip stress fields in test specimens using mode mixity parameters, Int. J. Fract. 185 (2014) 49-76.
[166] C.F. Shih, Small-Scale Yielding Analysis of Mixed Mode Plane-Strain Crack Problems, in: G.R. Irwin (Ed.) Fracture Analysis: Proceedings of the 1973 National Symposium on Fracture Mechanics, Part II, ASTM International, 1974, p. 0.
[167] V.N. Shlyannikov, Mixed-mode crack behavior under plane stress and plane strain small scale yielding, in: V.N. Shlyannikov (Ed.), Elastic-Plastic Mixed-Mode Fracture Criteria and Parameters, Springer Berlin Heidelberg, Berlin, Heidelberg, 2003, pp. 1-72.
[168] H. Gao, Y. Huang, W.D. Nix, J.W. Hutchinson, Mechanism-based strain gradient plasticity— I. Theory, J. Mech. Phys. Solids 47 (1999) 1239-1263.
[169] N.A. Fleck, J.W. Hutchinson, Strain Gradient Plasticity, in: J.W. Hutchinson, T.Y. Wu (Eds.), Advances in Applied Mechanics, Elsevier1997, pp. 295-361.
[170] M. Isfandbod, E. Martínez-Pañeda, A mechanism-based multi-trap phase field model for hydrogen assisted fracture, Int. J. Plast. 144 (2021) 103044.
[171] A. Arsenlis, D.M. Parks, Crystallographic aspects of geometrically-necessary and statistically-stored dislocation density, Acta Mater. 47 (1999) 1597-1611.
[172] M.X. Shi, Y. Huang, H. Gao, The J-integral and geometrically necessary dislocations in nonuniform plastic deformation, Int. J. Plast. 20 (2004) 1739-1762.
[173] K.L. Murty, K. Detemple, O. Kanert, J.T.M. Dehosson, In-situ nuclear magnetic resonance investigation of strain, temperature, and strain-rate variations of deformation-induced vacancy concentration in aluminum, Metall. Mater. Trans. A 29 (1998) 153-159.
[174] T. Ungár, E. Schafler, P. Hanák, S. Bernstorff, M. Zehetbauer, Vacancy production during plastic deformation in copper determined by in situ X-ray diffraction, Mater. Sci. Eng. A 462 (2007) 398-401.
[175] M.J. Zehetbauer, Effects of Non-Equilibrium Vacancies on Strengthening, Key Engineering Materials 97-98 (1995) 287-306.
[176] P.J. Noell, R.B. Sills, A.A. Benzerga, B.L. Boyce, Void nucleation during ductile rupture of metals: A review, Prog. Mater. Sci. 135 (2023) 101085.
[177] G. Saada, Sur la nature des défauts ponctuels crées par le croisement des dislocations, Acta Metall. 9 (1961) 965-966.

[178] G. Saada, Production de défauts ponctuels par écrouissage dans un métal cubique à faces centrées, Physica 27 (1961) 657-660.
[179] S.J. Zhou, D.L. Preston, F. Louchet, Investigation of vacancy formation by a jogged dissociated dislocation with large-scale molecular dynamics and dislocation energetics, Acta Mater. 47 (1999) 2695-2703.
[180] M.J. Buehler, A. Hartmaier, H. Gao, M.A. Duchaineau, F.F. Abraham, The dynamical complexity of work-hardening: a large-scale molecular dynamics simulation, Acta Mech. Sin. 21 (2005) 103-111.
[181] M. Militzer, W.P. Sun, J.J. Jonas, Modelling the effect of deformation-induced vacancies on segregation and precipitation, Acta Metall. Mater. 42 (1994) 133-141.
[182] T. G.I., Plastic strain in metals, J. Inst. Met. 62 (1938) 307-324.
[183] G.I. Taylor, The mechanism of plastic deformation of crystals. Part I.—Theoretical, Proc. R. soc. Lond. Ser. A Contain. Pap. Math. Phys. Character 145 (1934) 362-387.
[184] D.O. Kharchenko, V.O. Kharchenko, A.I. Bashtova, A study of void size growth in nonequilibrium stochastic systems of point defects, Eur. Phys. J. B 89 (2016) 123.
[185] S. Harada, S. Yokota, Y. Ishii, Y. Shizuku, M. Kanazawa, Y. Fukai, A relation between the vacancy concentration and hydrogen concentration in the Ni–H, Co–H and Pd–H systems, J. Alloys Compd. 404-406 (2005) 247-251.
[186] H. Mecking, Y. Estrin, The effect of vacancy generation on plastic deformation, Scr. Metall. 14 (1980) 815-819.
[187] E.M. Lifshitz, L.P. Pitaevski, CHAPTER XII - KINETICS OF PHASE TRANSITIONS, in: E.M. Lifshitz, L.P. Pitaevski (Eds.), Physical Kinetics, Pergamon, Amsterdam, 1981, pp. 427-447.
[188] G.R. Stewart, Measurement of low‐temperature specific heat, Rev. Sci. Instrum. 54 (1983) 1-11.
[189] C.Y. Ho, R.W. Powell, P.E. Liley, Thermal Conductivity of the Elements, J. Phys. Chem. Ref. Data 1 (2009) 279-421.
[190] J. Langner, J.R. Cahoon, Increase in the Alpha to Gamma Transformation Temperature of Pure Iron upon Very Rapid Heating, Metall. Mater. Trans. A 41 (2010) 1276-1283.
[191] A.V. Savikovskii, A.S. Semenov, Calculation of mixed-mode stress intensity factors for orthotropic materials in the plane stress state, St. Petersburg State Polytechnical University Journal. Physics and Mathematics 15 (2022) 102–123.
[192] A.V. Savikovskii, A.S. Semenov, Computation of fracture parameters for cracks in materials with cubic symmetry in the plane strain state, St. Petersburg State Polytechnical University Journal. Physics and Mathematics 16 (2023) 131–149.
[193] T.L. Anderson, Fracture Mechanics: Fundamentals and Applications, 3rd ed., CRC Press, 2005.
[194] A. Sarmah, S. Asqardoust, M.K. Jain, H. Yuan, 3D microstructure-based modelling of ductile damage at large plastic strains in an aluminum sheet, Int. J. Plast. 181 (2024) 104088.
[195] T. Zhu, J. Li, S. Yip, Atomistic Study of Dislocation Loop Emission from a Crack Tip, Phys. Rev. Lett. 93 (2004) 025503.
[196] Y. Dai, Y. Liu, Y.J. Chao, Higher order asymptotic analysis of crack tip fields under mode II creeping conditions, Int. J. Solids Struct. 125 (2017) 89-107.
[197] V. Radu, A. Ion, A. Nitu, The Influence of Hydrogen Concentration on the Ramberg-Osgood Constitutive Equation for Zr-2.5%Nb Pressure Tube Alloy, The 11th Annual International Conference on Sustainable Development through Nuclear Research and Education, Institute for Nuclear Research, Pitesti (Romania), 2018, pp. 16-20.
[198] H. Park, S. Moon, K. Kang, The effect of atomic hydrogen on the behavior of a single dislocation of 〈111〉{112} in bcc tungsten: Atomistic study, J. Nucl. Mater. 589 (2024) 154842.
[199] Y. Sun, G.E. Beltz, J.R. Rice, Estimates from atomic models of tension-shear coupling in dislocation nucleation from a crack tip, Mater. Sci. Eng. A 170 (1993) 67-85.
[200] F. Shuang, L. Laurenti, P. Dey, Standard deviation in maximum restoring force controls the intrinsic strength of face-centered cubic multi-principal element alloys, Acta Mater. 282 (2025) 120508.
[201] S.J. Wang, H. Wang, K. Du, W. Zhang, M.L. Sui, S.X. Mao, Deformation-induced structural transition in body-centred cubic molybdenum, Nat. Commun. 5 (2014) 3433.
[202] Y.N. Gornostyrev, M.I. Katsnelson, Misfit stabilized embedded nanoparticles in metallic alloys, Phys. Chem. Chem. Phys. 17 (2015) 27249-27257.
[203] S.S. Quek, Y. Xiang, D.J. Srolovitz, Loss of interface coherency around a misfitting spherical inclusion, Acta Mater. 59 (2011) 5398-5410.
[204] Y. Xiang, D.J. Srolovitz, Dislocation climb effects on particle bypass mechanisms, Philos. Mag. 86 (2006) 3937-3957.
[205] Y. Xiang, D.J. Srolovitz, L.T. Cheng, W. E, Level set simulations of dislocation-particle bypass mechanisms, Acta Mater. 52 (2004) 1745-1760.
[206] V. Shlyannikov, E. Martínez-Pañeda, A. Tumanov, R. Khamidullin, Mode I and mode II stress intensity factors and dislocation density behaviour in strain gradient plasticity, Theor. Appl. Fract. Mech. 116 (2021) 103128.
[207] Y. Huang, S. Qu, K.C. Hwang, M. Li, H. Gao, A conventional theory of mechanism-based strain gradient plasticity, Int. J. Plast. 20 (2004) 753-782.
[208] S. Qu, Y. Huang, H. Jiang, C. Liu, P.D. Wu, K.C. Hwang, Fracture analysis in the conventional theory of mechanism-based strain gradient (CMSG) plasticity, Int. J. Fract. 129 (2004) 199-220.
[209] E. Martínez-Pañeda, N.A. Fleck, Mode I crack tip fields: Strain gradient plasticity theory versus J2 flow theory, Eur. J. Mech. A/Solids 75 (2019) 381-388.
[210] H. Jiang, Y. Huang, Z. Zhuang, K.C. Hwang, Fracture in mechanism-based strain gradient plasticity, J. Mech. Phys. Solids 49 (2001) 979-993.
[211] S. Nazar, S. Lipiec, E. Proverbio, FEM modelling of hydrogen embrittlement in API 5L X65 steel for safe hydrogen transportation, Journal of Materials Science: Materials in Engineering 20 (2025) 9.
[212] H. Yu, J.S. Olsen, A. Alvaro, L. Qiao, J. He, Z. Zhang, Hydrogen informed Gurson model for hydrogen embrittlement simulation, Eng. Fract. Mech. 217 (2019) 106542.
[213] M. Lin, H. Yu, Y. Ding, G. Wang, V. Olden, A. Alvaro, J. He, Z. Zhang, A predictive model unifying hydrogen enhanced plasticity and decohesion, Scr. Mater. 215 (2022) 114707.
[214] R. Depraetere, W. De Waele, S. Hertelé, Fully-coupled continuum damage model for simulation of plasticity dominated hydrogen embrittlement mechanisms, Comput. Mater. Sci. 200 (2021) 110857.
[215] S. Chroeun, Y. Charles, J. Mougenot, M. Gaspérini, Finite element modeling of hydrogen instantaneous trapping in porosities, Int. J. Hydrog. Energy 160 (2025) 150607.
[216] A.A. Benzerga, J.-B. Leblond, Ductile Fracture by Void Growth to Coalescence, in: H. Aref, E.v.d. Giessen (Eds.), Advances in Applied Mechanics, Elsevier2010, pp. 169-305.

[217] K. Zhao, J.Y. He, I.G. Ringdalen, Z.L. Zhang, Effect of amorphization-mediated plasticity on the hydrogen-void interaction in ideal lattices under hydrostatic tension, J. Appl. Phys. 123 (2018) 245101.
[218] V.A. Lubarda, M.S. Schneider, D.H. Kalantar, B.A. Remington, M.A. Meyers, Void growth by dislocation emission, Acta Mater. 52 (2004) 1397-1408.
[219] S. Traiviratana, E.M. Bringa, D.J. Benson, M.A. Meyers, Void growth in metals: Atomistic calculations, Acta Mater. 56 (2008) 3874-3886.
[220] E.M. Bringa, S. Traiviratana, M.A. Meyers, Void initiation in fcc metals: Effect of loading orientation and nanocrystalline effects, Acta Mater. 58 (2010) 4458-4477.
[221] Y. Tang, E.M. Bringa, B.A. Remington, M.A. Meyers, Growth and collapse of nanovoids in tantalum monocrystals, Acta Mater. 59 (2011) 1354-1372.
[222] Y. Tang, E.M. Bringa, M.A. Meyers, Ductile tensile failure in metals through initiation and growth of nanosized voids, Acta Mater. 60 (2012) 4856-4865.
[223] Q.-z. Chen, W.-y. Chu, C.-m. Hsiao, The in-situ observations of microcrack nucleation and bluntness in ductile fracture, Scr. Metall. Mater. 30 (1994) 1355-1358.
[224] Y. Zhang, Y.-B. Wang, W.-Y. Chu, C.-M. Hsiao, The in-situ TEM observation of microcrack nucleation in titanium aluminide, Scr. Metall. Mater. 31 (1994) 279-283.
[225] M. Wasim, M.B. Djukic, T.D. Ngo, Influence of hydrogen-enhanced plasticity and decohesion mechanisms of hydrogen embrittlement on the fracture resistance of steel, Engineering Failure Analysis 123 (2021) 105312.
[226] L.M. Martyushev, V.D. Seleznev, Maximum entropy production principle in physics, chemistry and biology, Phys. Rep. 426 (2006) 1-45.
[227] H. Ziegler, An Introduction to Thermomechanics, North Holland, 2012.
[228] L.M. Martyushev, The maximum entropy production principle: two basic questions, Philos. Trans. R. Soc. B Biol. Sci. 365 (2010) 1333-1334.
[229] M. Moshtaghi, M. Safyari, M.M. Khonsari, Hydrogen-enhanced entropy (HEENT): A concept for hydrogen embrittlement prediction, Int. J. Hydrog. Energy 53 (2024) 434-440.
[230] M. Safyari, N. Takata, D. Kim, M.M. Khonsari, M. Moshtaghi, Effect of hydrogen on dynamic precipitation in additively manufactured aluminum alloys: Evidence for HEENT mechanism, J. Alloys Compd. 1027 (2025) 180395.
[231] A. Zajkani, M. Khonsari, A unified thermomechanically-consistent framework for fatigue failure entropy, Mech. Mater. 207 (2025) 105379.
[232] O. Gaede, A. Karrech, K. Regenauer-Lieb, Anisotropic damage mechanics as a novel approach to improve pre- and post-failure borehole stability analysis, Geophysical Journal International 193 (2013) 1095-1109.
[233] P.W. Whaley, Entropy Production during Fatigue as a Criterion for Failure. A Local Theory of Fracture in Engineering Materials, 1984.
[234] J. Zhao, D. Pedroso, Strain gradient theory in orthogonal curvilinear coordinates, Int. J. Solids Struct. 45 (2008) 3507-3520.
[235] S. Swaddiwudhipong, K.K. Tho, J. Hua, Z.S. Liu, Mechanism-based strain gradient plasticity in C0 axisymmetric element, Int. J. Solids Struct. 43 (2006) 1117-1130.
[236] S. Plimpton, Fast parallel algorithms for short-range molecular dynamics, J. Comput. Phys. 117 (1995) 1-19.
[237] A. Ramasubramaniam, M. Itakura, E.A. Carter, Interatomic potentials for hydrogen in $\alpha$-iron based on density functional theory, Phys. Rev. B 79 (2009) 174101.
[238] X.W. Zhou, C. Nowak, R.S. Skelton, M.E. Foster, J.A. Ronevich, C. San Marchi, R.B. Sills, An Fe–Ni–Cr–H interatomic potential and predictions of hydrogen-affected stacking fault energies in austenitic stainless steels, Int. J. Hydrog. Energy 47 (2022) 651-665.
[239] A. Ojha, H. Sehitoglu, L. Patriarca, H.J. Maier, Twin nucleation in Fe-based bcc alloys - modeling and experiments, Model. Simul. Mater. Sci. Eng. 22 (2014) 075010.
[240] J.J. Möller, M. Mrovec, I. Bleskov, J. Neugebauer, T. Hammerschmidt, R. Drautz, C. Elsässer, T. Hickel, E. Bitzek, ${110}$ planar faults in strained bcc metals: Origins and implications of a commonly observed artifact of classical potentials, Phys. Rev. Mater. 2 (2018) 093606.
[241] S. Kibey, J.B. Liu, M.J. Curtis, D.D. Johnson, H. Sehitoglu, Effect of nitrogen on generalized stacking fault energy and stacking fault widths in high nitrogen steels, Acta Mater. 54 (2006) 2991-3001.
[242] X. Liu, H. Lee, Y. Li, L. Myhill, D. Rodney, P.-A. Geslin, N.C. Admal, G. Po, E. Martinez, Y. Cui, Atomistically informed partial dislocation dynamics of multi-principal element alloys, J. Mech. Phys. Solids 208 (2026) 106478.